\documentclass[aps, floatfix,
  showpacs,
  amssymb,
  preprint,
  preprintnumbers,
  nofootinbib,
  superscriptaddress,
]{revtex4-2}

\usepackage[T1]{fontenc}
\usepackage{amsmath,amssymb,amsfonts}
\usepackage{amsmath}
\usepackage{graphicx}
\usepackage{bm}
\usepackage{slashed}
\usepackage{bbm}
\usepackage{latexsym}
\usepackage{color,xcolor}
\usepackage{textcomp}
\usepackage{cancel}
\usepackage{placeins}
\usepackage{epstopdf}
\graphicspath{{Figures/}}
\usepackage{soul}

\usepackage[colorlinks=true]{hyperref}  
\usepackage{cleveref}                   

\begin{document}

\title{Black Hole Thermodynamics via Tsallis Statistical Mechanics and Phase Transitions Probed by Optical Characteristics}

\author{Phuwadon Chunaksorn \footnote{Email: maxwelltle@gmail.com}}
\affiliation{The Institute for Fundamental Study, Naresuan University, Phitsanulok, 65000, Thailand}

\author{Ratchaphat Nakarachinda \footnote{Email: ratchaphat.n@rumail.ru.ac.th}}
\affiliation{Quantum and Gravity Theory Research Group, Department of Physics, Faculty of Science, Ramkhamhaeng University, 282 Ramkhamhaeng Road, Hua mak, Bang Kapi, Bangkok 10240, Thailand}

\author{Pitayuth Wongjun \footnote{Email: pitbaa@gmail.com}}
\affiliation{The Institute for Fundamental Study, Naresuan University, Phitsanulok, 65000, Thailand}

\begin{abstract}
\par
We develop a non-extensive thermodynamic framework for Reissner--Nordstr\"om black holes based on a near-horizon photon-gas model within Tsallis statistics. We derive the generalized Bekenstein--Hawking entropy based on such an approach, consistent with the Bekenstein--Hawking area law in the extensive limit, $q \rightarrow 1$. The induced deformation gives rise to a rich thermodynamic structure consisting of small, intermediate, and large black-hole branches, exhibiting Van der Waals-like phase transitions characterized by mean-field critical exponents. We further establish an optical--thermodynamic analogy by relating photon-sphere observables, including orbital periods and Lyapunov exponents, to thermodynamic variables. These optical signatures qualitatively track the thermodynamic critical behavior and phase structure, suggesting their potential relevance as observational probes in future high-resolution measurements. These results may shed light on a conceptual connection between non-extensive entropy, black-hole critical phenomena, and strong-gravity optics.
\end{abstract}

\maketitle{}

\newpage

\section{Introduction}\label{sec:intro}
\par
Black holes are identified through the analysis of exact solutions to Einstein's field equations in general relativity. Characterized by the presence of an event horizon---a null hypersurface that defines a causal boundary—they forbid the escape of matter, radiation, and information once these cross into the interior region. When quantum field theory is formulated on a curved spacetime background, black holes are predicted to emit thermal radiation at the Hawking temperature~\cite{Hawking1975}. They also possess an entropy proportional to the horizon area, $S_{\rm BH} \propto A_{h}$, where $A_{h}$ denotes the area of the event horizon, known as the Bekenstein--Hawking entropy~\cite{Bekenstein1973,Hawking1976,York1986}. The interplay between general relativity, quantum field theory, and thermodynamics~\cite{BardeenCarterHawking1973,Wald:1995yp} therefore elevates black holes to genuine thermodynamic systems and provides an important theoretical window into the quantum nature of gravity.
\par
Despite these achievements, the statistical origin of black hole entropy remains elusive. In conventional statistical mechanics, entropy is defined as the logarithm of the number of microscopic configurations consistent with fixed macroscopic constraints. In the black hole setting, the event horizon acts as an absolute causal barrier that prevents an external observer from accessing the internal microstates. This obstruction complicates any direct statistical interpretation of the entropy and indicates that the standard statistical-mechanical framework must be applied with care or suitably generalized.
\par
To address this difficulty, several theoretical approaches have been developed to replace the inaccessible interior degrees of freedom with near-horizon or externally measurable quantities. By considering a composite system formed by a black hole and a surrounding gas shell, one can compute the entropy of the gas in the near-horizon region, obtaining an area-law behavior analogous to the black hole surface entropy. Motivated by this idea, the familiar Bekenstein--Hawking entropy has been reproduced within a near-horizon thermodynamic framework~\cite{Kolekar:2011ideal,Mirza:2012condensation,Bhattacharya:2017entropy,Li:2021shell,Sourtzinou:2023quantum}, where the goal is to derive the entropy solely from observables accessible outside the horizon.
\par
These considerations naturally raise the question of whether the Gibbs--Boltzmann (GB) statistical framework is adequate for systems with intrinsically non-additive entropy such as black holes. Since the black hole entropy does not satisfy standard additivity, generalized entropy formalisms offer a more appropriate statistical description. Examples include the Tsallis entropy, which incorporates non-extensivity through the deformation parameter $q$~\cite{Tsallis1988,Tsallis_2004,BiroVan2011}; the Rényi entropy, which is additive and logarithmically related to the Tsallis form~\cite{Renyi:1961,ABE1997326,BiroVan2011}; and the Barrow entropy, inspired by fractal geometric corrections~\cite{Barrow:2020tzx,Saridakis:2020zol,Capozziello:2025axh}. These generalized entropies have found broad applications in gravitational and cosmological models~\cite{Mamon2021GSLBarrow,Yasir:2024wir,Yasir:2024eyv,Ladghami:2024yjn,Ladghami:2024sen,Zafar:2025sxl,Capozziello:2025axh}, while multi-parameter extensions such as the Sharma--Mittal entropy~\cite{sharma1975new,Ghaffari_2019} interpolate smoothly between the Tsallis and Rényi limits. These frameworks have also attracted interest in holography and quantum gravity~\cite{Nojiri2003,Biro:2013cra,AlonsoSerrano2021,Nojiri:2022dark,Nojiri:2022holographic,Odintsov:2023bounce,Cimidiker2023}. For its minimal deformation and tractability, we adopt the Tsallis entropy as the basis for our analysis.
\par
The deformation parameter $q$ is introduced within the framework of Tsallis entropy, which characterizes deviations from extensivity and is well suited for describing the strong long-range correlations inherent in gravitational systems~\cite{Campa_2009,Padmanabhan:1989gm,tsallis2012blackhole}. It has been successfully applied in astrophysics~\cite{PLASTINO1993384,Lima2001,Taruya:2001mv,SilvaLima2005,Chavanis:2002pw,Du2007,LouisMartinez2011,LiDuGuo2011,Zheng2017}, kinetic theory~\cite{Chunaksorn:2024gwo}, and generalized uncertainty principles (GUPs), where the presence of a minimal length scale induces non-extensive statistical features~\cite{Luciano:2021ndh}, as well as in cosmology~\cite{Tsallis:1995,Sheykhi2018,Lymperis2018,Saridakis:2018unr,Sheykhi2020,Jizba2023,Nakarachinda:2023jko,Dehpour2024}. Although it is natural to identify \(S_{\rm BH} \equiv S_q\), Tsallis entropy generally does not satisfy the zeroth law of thermodynamics because of the non-additivity of entropy. A consistent empirical temperature can instead be defined through a logarithmic mapping to the additive Rényi entropy, which has therefore been widely employed in black hole thermodynamics~\cite{TsallisCirto2013,Czinner2016,Czinner2017,Maleki:2024nonextensive,Tannukij2020,Promsiri2020,Promsiri2021,Nojiri:2021czz,Nakarachinda2021,Hirunsirisawat2022,Chunaksorn2022,Nakarachinda:2022gsb,Anusonthi:2025dup}. Furthermore, investigations of black hole thermodynamics associated with non-extensive entropy have been presented in Refs.~\cite{Nojiri:2022sfd,Elizalde:2025iku,Czinner:2025koi,Czinner:2025zbr,Elizalde:2025iku}.
\par
Recent investigations have shown that applying Tsallis statistics to ideal gases near black hole horizons yields entropy functionals that reduce to the Bekenstein--Hawking entropy in the extensive limit while remaining consistent with the first law and the Smarr relation~\cite{Chunaksorn:2025nsl}. Extensions to interacting bosonic systems have further supported these results~\cite{Maleki:2024nonextensive}, reinforcing Tsallis statistics as a coherent and predictive framework for black hole thermodynamics.
\par
Motivated by these developments, we extend the Tsallis formalism to a photon gas confined within the near-horizon region. Photon trajectories provide a useful probe of black hole geometries, as their dynamics reflect both thermodynamic and geometric properties of the underlying spacetime. In particular, photon orbits near the photon sphere encode structural information that can be observed by distant detectors. Our aim is to construct a non-extensive statistical model of black hole thermodynamics capable of (i) reproducing standard black hole thermodynamics in the limit \(q \to 1\) and (ii) investigating observable imprints of non-extensive effects on photon-sphere dynamics
\par
Photon-sphere observables—including the angular velocity, radius, and characteristic decay rates—are known to correlate strongly with black hole thermodynamic properties~\cite{He:2010,Liu:2014,Chabab:2016,Zou:2017}. The advent of high-resolution observational facilities such as LIGO and the Event Horizon Telescope (EHT)~\cite{LIGOScientific:2016a,Akiyama:2019a,Akiyama:2019b} further motivates exploring these connections, enabling direct empirical access to black hole dynamics and near-horizon structure. Recent studies have revealed critical scaling behavior in photon-sphere observables near thermodynamic phase transitions~\cite{Wei:2018,Zhang:2019,Zhang:2020,Belhaj:2020,Du:2023,Guo:2022,Yang:2023,Lyu:2024,Kumara:2024,Hale:2024lzh,Shukla:2024tkw,Anand:2025vfj}, for instance in charged anti-de Sitter (AdS) black holes~\cite{Mahapatra:2016dae,Promsiri:2024hrl}, indicating a possible correspondence between gravitational optics and thermodynamic universality classes.
\par
In this work, we construct a Tsallis-deformed statistical framework for a photon gas confined in the near-horizon region of a black hole. Within this framework, the black hole entropy associated with Tsallis statistics, namely the $q$-generalized Bekenstein--Hawking entropy, is derived and shown to recover the standard Bekenstein--Hawking entropy in the limit $q \rightarrow 1$. Based on this construction, the corresponding thermodynamic quantities, including the black hole temperature, are obtained in a manner consistent with the generalized first law of thermodynamics and the Smarr relation. We then investigate the thermodynamic properties and phase structure of charged black holes within the Tsallis-deformed statistical framework. Furthermore, we explore the connection between black hole thermodynamic phase transitions and photon-sphere observables, allowing the phase structure of the system to be probed through optical signatures. In this way, potential observational imprints of the non-extensive effects may arise in photon-sphere characteristics, which could indirectly signal the presence of charge in black holes and provide a possible avenue for constraining the non-extensive parameter through optical observations.
\par
The remainder of this paper is organized as follows. Sec.~\ref{entropy} introduces the Tsallis-deformed entropy model for a near-horizon photon gas and contrasts it with the conventional GB framework. Sec.~\ref{thermo} develops the resulting thermodynamics, including critical behavior, heat capacities, and a generalized Smarr relation. Sec.~\ref{optics} investigates photon-sphere observables across the phase diagram and discusses their observational implications. Throughout this work, we employ units, setting \(c = h = k_{B} = G = 1\), except where dimensional consistency is explicitly required in Sec.~\ref{entropy}.

\section{Near-Horizon Thermodynamics and Black Hole Entropy}\label{entropy}
\par
In this section, we analyze the thermodynamic behavior of a composite system consisting of a static black hole and a surrounding photon gas, modeled as two subsystems in thermal contact undergoing a quasi-static, reversible process. The subsystems are assumed to exchange energy while maintaining thermodynamic equilibrium, and the photon gas is treated purely as a thermodynamic probe, without invoking quantum field theoretic effects such as particle creation. We neglect the back-reaction of the gas on the spacetime geometry and assume the black hole provides a fixed gravitational background. To specify the model, we consider the photon gas forming a thin, spherically symmetric shell of thickness \( H \) in the static spacetime around the black hole, with approximately uniform local temperature \( T := 1/(k_{B}\beta) \), where \( \beta \) denotes the inverse temperature. The spacetime geometry is described by the general static, spherically symmetric line element
\begin{equation}
\displaystyle
ds^{2} = -g(r)\, c^{2} dt^{2} + \frac{dr^{2}}{g(r)} + r^{2}\left(d\theta^{2} + \sin^{2}\theta\, d\phi^{2}\right),
\label{ds}
\end{equation}
where \( g(r) \) is the horizon function. The condition \( g_{00} = g_{11}^{-1} \) holds when the energy-momentum tensor satisfies \( T^{0}_{\,\,\,0} = T^{1}_{\,\,\,1} \), a criterion fulfilled in a broad class of static, spherically symmetric spacetimes, including vacuum and many matter-dominated solutions.  
\par
Given this background geometry, we next construct the invariant phase-space measure for particles propagating in such spacetime. The invariant phase-space volume element for a single photon with energy \( E \) is given by
\begin{equation}
\displaystyle
d\mathcal{V} = \frac{(4\pi)^{2} E^{3}}{3 h^{3} c^{3}}\frac{r^{2}}{g^{2}(r)} dr,
\label{phase 2}
\end{equation}
which provides the differential phase-space volume accessible to photons in a static gravitational field.
This expression forms the starting point for evaluating the phase-space structure of a photon gas surrounding a black hole. 
\par
To proceed, we confine the photon gas to a thin spherical shell in the radial interval \( r_{bh} + L < r < r_{bh} + L + H \), where \( r_{bh} \) is the black hole horizon radius and \( L \) is the coordinate distance from the horizon to the shell’s inner boundary. Focusing on the near-horizon regime (\( L \ll r_{bh}, H \)), the shell remains entirely outside the event horizon, ensuring photons are in a well-defined thermal state at fixed radius. Applying Eq.~\eqref{phase 2} to this configuration requires an expansion of the horizon function near \(r = r_{bh}\), yielding
\begin{equation}
\displaystyle
g(r) \approx \left. \frac{d g(r)}{dr} \right|_{r = r_{bh}} (r - r_{bh}) = \frac{2\kappa}{c^{2}} (r - r_{bh}),
\label{approx g}
\end{equation}
where $\kappa := c^2dg(r)/(2dr)\big|_{r = r_{bh}}$ is the surface gravity~\cite{poisson2009relativists}.
Note that the above expression is valid only for non-extremal black holes (\(\kappa \neq 0\)). 
In case of extremal black holes, the higher-order terms of \((r-r_{bh})\) must be retained. As a result, the total phase-space volume at near-horizon regime (i.e., $L \ll H$ and $r\approx r_{bh}$) becomes
\begin{equation}
\displaystyle
\mathcal{V} \simeq \frac{\pi c E^{3} A_{h}}{3 h^{3} \kappa^{2} L},
\label{phase3}
\end{equation}
where $A_{h} := 4\pi r_{bh}^{2}$ the horizon area.
The divergence of the above volume when \(L \to 0\)  reflects the accumulation of high-energy modes near the horizon, in agreement with ’t Hooft’s brick-wall model~\cite{tHooft:1984kcu}, where \(L\) acts as a short-distance regulator. This regulated volume will be used to construct the photon-gas partition function and to extract the corresponding thermodynamic quantities in the next subsection.

\subsection{Black Hole Entropy within the Gibbs--Boltzmann Framework}\label{subsec: S GB}
\par 
We now turn to the statistical description of the photon gas within the standard GB framework. Let us consider photons capable of occupying a quantum state \(s\) with occupation number $n_{s}$. Due to their bosonic nature, there is no restriction on the number of photons that can simultaneously occupy a single quantum state. Therefore, the total number of photons is not conserved in thermal equilibrium. The photons may be freely emitted or absorbed by the thermal reservoir provided by the confining shell. 
Consequently, the chemical potential vanishes, implying that the appropriate thermodynamic description is furnished by the grand canonical ensemble at zero chemical potential. 
\par
Within this grand canonical formulation, the total partition function \(Z\) factorizes over the complete set of photon states, as statistical independence ensures that each mode contributes multiplicatively:
\begin{equation}
\displaystyle
Z = \sum_{n_{1}, n_{2}, n_{3}, \ldots = 0}^{\infty} \exp \left[-\beta \sum_{s = 1}^{\infty} n_{s} E_{s} \right] 
= \prod_{s=1}^{\infty} \left[\sum_{n_{s}=0}^{\infty} \exp (- \beta n_{s} E_{s})\right] 
\equiv \prod_{s=1}^{\infty} Z^{(s)},
\label{Z sum}
\end{equation}
where \(Z^{(s)}\) denotes the single-mode partition function of the \(s\)-th quantum state with energy $E_s$.
It is explicitly defined as the sum over all possible occupation numbers of that mode, each term weighted by the corresponding Boltzmann factor:
\begin{equation}
\displaystyle
Z^{(s)} 
= \sum_{n_{s} = 0}^{\infty} \exp (-\beta n_{s} E_{s})
= \frac{1}{1 - \exp(-\beta E_{s})}.
\label{partition GB s}
\end{equation}
To obtain the closed form in the last equality, we have employed the geometric series: $\sum_{k=0}^{\infty} u^{k} = 1/(1 - u)$ for $u=\exp(-\beta E_{s}) < 1$.
Because the chemical potential vanishes, the grand canonical and canonical ensembles become thermodynamically equivalent for the photon gas, particularly for energy-based observables. 
This equivalence allows us, in the next step, to combine the regulated phase-space measure with the GB formalism to construct explicit expressions for the thermodynamic quantities of interest.
\par
In the high-temperature or large-volume limit, the photon population spreads over a dense spectrum of high-energy states, making the discrete energy level spacing negligible compared with thermal fluctuations. The logarithmic function of the total partition function is found to be approximated as 
\begin{eqnarray}
\displaystyle
\ln Z =-\sum_{s=1}^{\infty} \ln \big|1 - \exp(-\beta E_{s})\big| \simeq -\int_{0}^{\infty} d\mathcal{V} \ln \left|1 - \exp (-\beta E)\right|
\simeq\frac{\pi^{5} c}{45 h^{3}} \left(\frac{A_{h}}{\beta^{3} \kappa^{2} L}\right).
\label{phase 4}
\end{eqnarray}
where \(d\mathcal{V}\) is precisely the phase-space volume computed in Eq.~\eqref{phase3}.
In the second step, the discrete sum over quantum states can be converted into an integral over the energy spectrum, with the density of states encoded by the differential phase-space volume element. It is also noticed from the final result of Eq.~\eqref{phase 4} that the inverse-cubic temperature scaling reflects the ultra-relativistic nature of massless radiation in three spatial dimensions, modulated by the redshift and horizon area as determined from the geometry. 
\par
Remarkably, the continuum-limit partition function derived in Eq.~\eqref{phase 4} includes all photon modes, in particular the zero-energy ground state \(E_0 = 0\). Owing to the massless nature of photons and the non-conservation of their total number, this mode contributes only a constant additive term to \(\ln Z\), which drops out of all thermodynamic observables obtained from temperature derivatives, such as the internal energy or entropy~\cite{Huang1987,LandauLifshitz1980,PathriaBeale2011,KittelKroemer1980}. Therefore, the lower limit of the integral in Eq.~\eqref{phase 4} may be taken to be zero without affecting any physical predictions.
\par
To express thermodynamic quantities in terms of locally measurable observables, we relate the coordinate separation \(L\) of the photon shell to the proper radial distance \(l_{\mathrm{loc}} = \int_{r_{bh}}^{r_{bh}+L} dr/\sqrt{g(r)}\), which measures the physical distance from the horizon to the inner boundary of the shell as observed locally. 
This local observer at the radial coordinate \(r\) also measures the temperature, $\beta_{\rm loc}$ which is related to the asymptotic temperature $\beta$ via Tolman’s law~\cite{PhysRev.35.904,tolman1934relativity,PhysRev.36.1791,landau1975classical,Wald:1984rg} as $\beta_{\rm loc}(r) = \beta \sqrt{-g_{00}(r)}$. Under the thin-shell approximation, the redshift factor varies negligibly across the shell's thickness \(H\), so \(\beta_{\rm loc}\) can be treated as approximately uniform.
Consequently, the temperature of the gas is supposed to be that evaluated at the inner edge of the shell $r=r_{bh}+L$.
In the near-horizon regime, the logarithmic function of the total partition function in Eq.~\eqref{phase 4} is therefore expressed approximately as
\begin{equation}
\displaystyle
\ln Z \simeq \frac{2 \pi^{5}}{45 h^{3} c^{3}} \left(\frac{A_{h} l_{\mathrm{loc}}}{\beta_{\mathrm{loc}}^{3}}\right),
\label{phase GB 1}
\end{equation}
which exhibits the correct scaling with proper local quantities and provides a consistent thermodynamic description for static observers situated just outside the event horizon.
Note that the expression in Eq.~\eqref{phase GB 1} should recover the flat-space blackbody expression in terms of locally measurable quantities in order to confirm that conventional statistical mechanics remains valid when properly localized in curved spacetime.
To achieve this issue, we define the effective local volume of the photon gas shell as
\begin{equation}
\displaystyle
V_{\mathrm{loc}} := \frac{A_{h} l_{\mathrm{loc}}}{4},\label{V loc}
\end{equation}
accounting for confinement to the thin, curved region near the horizon and ensuring dimensional consistency with the standard blackbody volume. It is important to emphasize that the partition function \(\ln Z(\beta_{\mathrm{loc}}, V_{\mathrm{loc}})\) is defined with respect to a static observer co-moving with the shell, using the proper volume and local inverse temperature evaluated at a fixed radial position (the shell’s inner boundary). Time-translation invariance, ensured by the time-like Killing vector of the static and spherically symmetric spacetime, guarantees that these local quantities are stationary. This highlights that, in static curved spacetimes, thermodynamic observables are naturally referenced to proper local measurements, thereby maintaining consistency with the underlying geometry. 
\par
From this locally defined partition function, the standard canonical ensemble relations yield the photon gas thermodynamic quantities as measured by the static observer:
\begin{eqnarray}
\displaystyle
U_{\mathrm{loc}} &=& -\frac{\partial}{\partial \beta_{\mathrm{loc}}} \ln Z = \frac{8 \pi ^{5}}{15 h^{3} c^{3}} \frac{V_{\mathrm{loc}}}{\beta_{\mathrm{loc}}^{4}}, \label{U GB}\\
\displaystyle
P_{\mathrm{loc}} &=& \frac{1}{\beta_{\mathrm{loc}}} \frac{\partial}{\partial V_{\mathrm{loc}}} \ln Z = \frac{8 \pi^{5}}{45 h^{3} c^{3}} \frac{1}{\beta_{\mathrm{loc}}^{4}}, \label{P GB} \\
\displaystyle
S_{\mathrm{loc}} &=& k_{B} \left(1 - \beta_{\mathrm{loc}} \frac{\partial}{\partial \beta_{\mathrm{loc}}}\right) \ln Z = k_{B} \frac{32 \pi^{5}}{45 h^{3} c^{3}} \frac{V_{\mathrm{loc}}}{\beta_{\mathrm{loc}}^{3}}.
\label{S GB}
\end{eqnarray}
These relations demonstrate that, once expressed in terms of proper local quantities \(V_{\mathrm{loc}}\) and \(T_{\mathrm{loc}} = 1/(k_B \beta_{\mathrm{loc}})\), the thermodynamic scaling laws
\[
U_{\mathrm{loc}} \propto V_{\mathrm{loc}} T_{\mathrm{loc}}^{4}, \qquad 
P_{\mathrm{loc}} \propto T_{\mathrm{loc}}^{4}, \qquad 
S_{\mathrm{loc}} \propto V_{\mathrm{loc}} T_{\mathrm{loc}}^{3},
\]
naturally emerge, reproducing standard blackbody behavior. This is consistent with the local validity of conventional thermodynamic laws in strongly curved near-horizon regions, provided that all quantities are expressed in terms of properly and physically measurable observables, thereby establishing a clear link between near-horizon geometry, phase-space structure, and localized thermodynamics. The thermodynamic quantities for the static observer derived in Eqs.~\eqref{U GB}--\eqref{S GB} also satisfy the first law of thermodynamics
\begin{equation}
\displaystyle
dU_{\mathrm{loc}} = \frac{1}{k_{B}\beta_{\mathrm{loc}}}\, dS_{\mathrm{loc}} - P_{\mathrm{loc}}\, dV_{\mathrm{loc}}. 
\end{equation}
This indicates that the locally regularized partition function \(\ln Z(\beta_{\mathrm{loc}}, V_{\mathrm{loc}})\) functions as a well-defined thermodynamic quantity in the near-horizon regime. 
\par
Assuming thermal contact between the black hole and the photon-gas shell, the composite system attains thermal equilibrium, with the temperature of the photon gas identified with the Hawking (inverse) temperature of the black hole~\cite{ISRAEL1976107,York1986}:
\begin{equation}
\displaystyle
\beta_{\mathrm{H}} = \frac{4\pi^{2} c }{\hbar\, \kappa}.
\end{equation}
Substituting this equilibrium temperature into the local entropy formula~\eqref{S GB}, it yields the entropy of the photon gas near the black hole's  horizon:
\begin{equation}
\displaystyle
S_{\mathrm{loc}}^{(0)} = k_{B} \frac{A_{h}}{360 \pi l_{\mathrm{loc}}^{2}},
\label{GB entropy 1}
\end{equation}
where the superscript ``$(0)$'' of the above entropy denotes the quantity of the gas system being in equilibrium with the black hole. Obviously, this entropy depends on the proper distance \(l_{\mathrm{loc}}\) from the horizon. Within this near-horizon setup, the resulting entropy is therefore proportional to the horizon area. The entropy reproduces the well-known Bekenstein--Hawking one,
\(S^{(0)}_{\mathrm{loc}} \to S_{\mathrm{BH}}:=k_{B} A_{h}/(4l_P^2)\) where $l_{P}$ is the Planck length~\cite{Bekenstein1973,Hawking1976,damour1982surface,giddings1992black,mathur2005fuzzball}, when the proper radial distance is chosen as
\begin{equation}
\displaystyle
l_{*} = \frac{1}{3 \sqrt{10 \pi}} \, l_{P} \approx 0.0595 \, l_{P},
\label{proper length}
\end{equation}
introducing a UV cut-off at a length scale of the order of the Planck length. 
Let us write the entropy of a black hole as such $S^{(0)}_{bh}=S^{(0)}_{\rm loc}\big|_{l=l_{*}}$ for convenience.
In this setup, the divergence of the near-horizon phase-space volume
in Eq.~\eqref{phase3} is regulated by \(l_{*}\), yielding a finite local entropy proportional to the horizon area and
consistent with the semiclassical Bekenstein--Hawking result, without
presupposing the microscopic origin of black hole entropy.
\par
Beyond this general result, the proper thickness required for a near-horizon photon gas is slightly smaller than that of a classical massive ideal gas shell, with a representative value \(l_{\mathrm{loc}} \approx 0.0932\, l_{P}>l_*\)~\cite{Chunaksorn:2025nsl}. 
This reduction is physically expected: owing to the radiation equation of state, a photon gas possesses a higher entropy density at a fixed local temperature and can therefore accommodate the same total entropy within a smaller proper volume, approximately \(31.91\%\) that required in the massive ideal gas case. 
Accordingly, the near-horizon photon gas reproduces \(S_{\mathrm{BH}}\) with an enhanced local entropy density \(S^{(0)}_{bh}/V_{\mathrm{loc}}\). The appearance of a finite proper thickness \(l_{\mathrm{loc}}\) may thus be viewed as an effective UV scale below which continuum statistical mechanics becomes inadequate, in qualitative agreement with quantum-gravity expectations of minimal length arising from discreteness, generalized uncertainty principles, or metric fluctuations~\cite{garay1995quantum,amati1989can,maggiore1993generalized,rovelli1995discreteness}.
\par
Given this minimal-length rationale, it is natural to identify the geometric horizon area \(A_{h}\) with the effective surface area of the photon gas shell located at a proper distance \(l_{*}\) outside the horizon. Because of the extreme gravitational red-shift and the resulting causal disconnection, this thin shell is operationally indistinguishable from the true event horizon and may therefore be regarded as a \textit{stretched horizon}~\cite{thorne1986black,susskind1993stretched}. In this picture, the stretched horizon acts as a thermodynamically well-defined time-like boundary endowed with a finite red-shifted temperature and entropy carried by microscopic degrees of freedom, while the null event horizon itself remains causally inaccessible. 
Accordingly, black hole thermodynamics can be consistently formulated on this effective surface, which encapsulates the effective screening of near-horizon microscopic information by the event horizon and supports the interpretation of \(S_{\mathrm{BH}}\) as an intrinsic geometric quantity, robust across a wide class of classical and quantum statistical descriptions~\cite{Jacobson:1995ab,Carlip:1999cy,Solodukhin:2011gn}. This means that the recent approach furnishes an operationally consistent framework for black hole thermodynamics by replacing the ill-defined notion of entropy on a null surface with a well-defined statistical ensemble on a time-like hypersurface.
\par
While the black hole entropy can be consistently accounted for using a
near-horizon photon gas, other thermodynamic quantities—such as pressure,
volume, and enthalpy—are determined by the underlying spacetime geometry. It should be noted that the entropy $S^{(0)}_{bh}$ scales with the black hole surface area rather than the volume. This implies that the entropy is non-extensive quantity.
In contrast to conventional thermodynamic systems, the entropy of the governed by GB statistics is extensive. This issue motivates the exploration of generalized entropy frameworks, such as the Tsallis formalism, which are tailored to describe non-additive systems with long-range correlations or strong interactions. 
Within this framework, the area law of entropy can be reconciled with a statistical description of near-horizon matter fields. In the following subsection, we present a detailed analysis of a near-horizon photon gas using Tsallis non-extensive statistical mechanics, leading to a generalized black hole entropy expression consistent with the Tsallis deformation and offering insight into possible microscopic origins and deformations of black hole entropy.
This proposed approach to determining a black hole entropy is quite general and can then be extended directly to an investigation based on Tsallis statistics.

\subsection{Black Hole Entropy within the Tsallis Statistical Framework}
\label{subsec:Tsallis}
\par
We now consider a generalized statistical framework capable of capturing the non-extensive character of black hole entropy. 
The Bekenstein--Hawking entropy, scaling with the event horizon area rather than the spatial volume, signals the presence of long-range gravitational correlations and horizon-localized microscopic degrees of freedom.
This motivates the adoption of non-extensive statistical mechanics, in particular the Tsallis entropy formalism, which is well-suited to systems exhibiting strong correlations or non-additivity~\cite{Tsallis1988,TsallisBook:2009}.
\par
To formalize this perspective, we follow a methodology analogous to that employed in the standard GB framework, expressing the black hole entropy within the Tsallis non-extensive statistical mechanics context. 
The Tsallis entropy, or \(q\)-entropy, generalizes the standard GB entropy by introducing a real deformation parameter \(q \in \mathbb{R}\), which quantifies the degree of non-extensivity:
\begin{equation}
\displaystyle
S_{q} = -k_{B} \sum_{i=1}^{\Omega} p_{i}^{q} \ln_{q} p_{i},
\label{Tsallis entropy}
\end{equation}
where \( \Omega \) is the number of accessible microstates, \( p_{i} \)
denotes the probability of the \(i\)-th microstate, and \( \ln_{q} p_{i} \) is the \(q\)-logarithm. This function and its inverse function are, respectively, defined as 
\begin{equation}
\displaystyle
\ln_{q} X = \frac{X^{1 - q} - 1}{1 - q}, \qquad
\exp_{q} X = \left[1 + (1 - q) X\right]^{\frac{1}{1 - q}},
\label{Eq-ln-q-exp-q}
\end{equation}
for \( X \in \mathbb{R} \).
The domain restriction \( 1 + (1 - q) X > 0 \) ensures that the probabilities remain physically admissible.
It can be checked that the aforementioned functions can be reduced to the natural logarithm and exponential functions as \( q \to 1 \). Consequently, in the limit \( q \to 1 \), the entropy defined in Eq.~\eqref{Tsallis entropy} recovers the standard GB entropy, providing a smooth connection between the extensive and non-extensive regimes.
\par
A central feature of Tsallis entropy is its generalized non-additive composition rule. For a composite system of two statistically independent subsystems, labeled by \( (1) \) and \( (2) \), the total entropy obeys the pseudo-additivity relation~\cite{TsallisBook:2009}:
\begin{equation}
\displaystyle
S_{q(1+2)} = S_{q(1)} + S_{q(2)} + \frac{(1-q)}{k_{B}} S_{q(1)} S_{q(2)},
\label{Tsallis-composition}
\end{equation}
which reduces to the additive rule for the GB entropy when \( q \to 1 \) as expected. 
The sign and magnitude of $(1-q)$ determine the deviation from extensivity. 
For $q>1$, the entropy is \textit{sub-extensive}, reflecting suppressed growth due to long-range correlations or constraints. For $q<1$, it is \textit{super-extensive}, capturing enhanced statistical weight of collective or multi-fractal microscopic structures.
\par
In statistical mechanics, equilibrium distributions are obtained by maximizing a specified entropy functional subject to appropriate physical constraints. 
Within the Tsallis framework, the entropy in Eq.~\eqref{Tsallis entropy}, is maximized under the normalization condition and constraints on macroscopic observables, such as the mean internal energy. In the grand canonical ensemble, the average number of particles is also constrained. Unlike the standard GB case, the non-extensive Tsallis formalism allows multiple definitions of internal energy, including unnormalized and normalized \(q\)-expectation values (escort averages), each leading to a different equilibrium distribution and corresponding thermodynamic relations~\cite{Curado:1991jc,Tsallis:1998ws,Coraddu:1998yb,Martinez:2000,Lenzi:2000,Abe:2001,Taruya:2002,Tsallis:2002tp,Wilk:2002uf,Taruya:2003a,Taruya:2003b,Sakagami:2004,Jiulin:2004a,Jiulin:2004b,Ferri:2005yf,Zavala:2006,Chakrabarti:2010,Chandrashekar:2011,Wilk:2014zka}. This flexibility makes the formalism well-suited for modeling systems with long-range correlations and non-local interactions, such as the near-horizon thermodynamics of the gas system.
\par
Despite the existence of multiple inequivalent constraint schemes in Tsallis statistical mechanics, these formulations can often be related through appropriate transformations~\cite{FerriMartinezPlastino2005}, allowing one to select the most convenient scheme for analytical or interpretive purposes. In the context of near-horizon thermodynamics, where the photon gas serves as a natural microscopic probe, we adopt the standard normalization condition
\begin{equation}
\displaystyle
\sum_{i=1}^{\Omega} p_{i} = 1,\label{q norm cond}
\end{equation}
together with the deformed, normalized \(q\)-expectation value for the internal energy
\begin{equation}
\displaystyle
\sum_{i=1}^{\Omega} p_{i}^{q} E_{i} = U_{q},
\label{mean U}
\end{equation}
as originally proposed in Ref.~\cite{Tsallis:1998ws}. This choice preserves the Legendre structure of \(q\)-thermodynamics and ensures internal consistency, providing a clear and physically meaningful framework for systems in which particle number is not conserved and the internal energy serves as the primary thermodynamic constraint. Consequently, this formulation provides a consistent derivation of equilibrium distributions appropriate for near-horizon black hole physics.
\par
Alternative formulations, such as those based on escort averages or generalized entropic measures, yield physically equivalent results under appropriate variable redefinitions and rescalings~\cite{Ferri:2005yf}. This equivalence motivates focusing on the normalized \(q\)-expectation scheme, which preserves the Legendre structure and allows a transparent variational treatment. With these constraints, the equilibrium distribution \(\{p_i\}\) is obtained by maximizing \(S_q\) using the method of Lagrange multipliers. Introducing multipliers \(a_1\) and \(a_2\) corresponding to the normalization and \(q\)-expectation internal energy constraints, the variational principle reads:
\begin{eqnarray}
\displaystyle
\delta \left[S_{q} - a_{1} \left(\sum_{i=1}^{\Omega} p_{i} - 1\right) - a_{2} \left(\sum_{i=1}^{\Omega} p_{i}^{q} E_{i} - U_{q}\right)\right] &=& 0.
\label{extremize Sq}
\end{eqnarray}
Substituting Eq.~\eqref{Tsallis entropy} into the above equation, the variation with respect to $p_i$ yields the condition:
\begin{equation}
\displaystyle
\frac{k_{B} q}{1-q} p_{i}^{q-1} - a_{1} - a_{2} q E_{i} p_{i}^{q-1} = 0,
\end{equation}
which can be solved for the generalized equilibrium distribution $p_i$. 
As a result, the distribution is taken in the form of
\begin{equation}
\displaystyle
p_{i} = \frac{1}{Z_q}\exp_q(-\beta_qE_i),\label{q-prob}
\end{equation}
where $Z_q=(1-q)/(k_Ba_1q)$ and $\beta_q=a_2/k_B$.
The factor $1/Z_q$ can be thought of as the normalized factor satisfied condition~\eqref{q norm cond}.
This leads to defining the \(q\)-generalized partition function:
\begin{equation}
\displaystyle
Z_{q} = \sum_{i=1}^{\Omega} \exp_{q} (-\beta_{q} E_{i}),
\label{q-partition 1}
\end{equation}
It is noticed that the distribution in Eq.~\eqref{q-prob} will be reduced to the GB version in the limit $q\to1$ with an additional requirement: $\beta_q\to\beta$.
Remarkably, the inverse temperature must be promoted to be a $q$-dependent quantity for consistency in black hole thermodynamics, as will be discussed in the next section.
\par
We now apply the framework to black hole thermodynamics. 
The $s$-th state of the photon gas is characterized by an occupation number \(n_s\) and a discrete energy \(E_s\). 
To follow the standard GB statistics, let the partition function be required to be factorized into single-mode contributions $Z_{q}^{(s)}$ as follows: 
\begin{eqnarray}
\displaystyle
Z_{q} = \prod_{s=1}^{N} Z_{q}^{(s)}, \quad \text{with} \quad
\displaystyle
Z_{q}^{(s)} = \sum_{n_{s} = 0}^{\infty} \exp_q(-\beta_{q} n_{s} E_{s}).
\end{eqnarray}
Due to the fact that the $q$-exponential function differs from the exponential one (see Eq.~\eqref{Eq-ln-q-exp-q}), the microscopic energy is not strictly additive. Instead, the Tsallis formalism requires a pseudo-additive composition rule that explicitly incorporates multi-particle correlations~\cite{wang2000nonextensivedistributionfactorizationjoint}:
\begin{equation}
\label{q-total energy}
\displaystyle
E_{\mathrm{tot}} = \sum_{i=1}^{N} n_{i} E_{i} + \sum_{j=2}^{N} \left[(q-1) \beta_{q}\right]^{j-1} \sum_{n_{1} < n_{2} < \cdots < n_{j}}^{N} \prod_{l = 1}^{j} n_{l} E_{l}.
\end{equation} 
Obviously, the non-additive parts vanish in the extensive limit \(q \to 1\), thereby recovering the additive rule for standard GB statistics. To reconcile this microscopic non-additivity with macroscopic extensivity, one introduces a logarithmic mapping that defines an \emph{additive} energy variable~\cite{ABE1997326,BiroVan2011}. This mapping ensures that the Legendre structure of thermodynamics is preserved while the deformation of the microscopic viewpoint due to the Tsallis statistics is still captured. 
\par
It is interestingly found that, under a condition \(\beta_q E_s \ll 1\), the single-mode partition function can be approximated by a geometric-like series,
\begin{equation}
\label{partition of s 2}
\displaystyle
Z_{q}^{(s)} \simeq \sum_{n_{s} = 0}^{\infty} \left[\exp_{q} (-\beta_{q} E_{s})\right]^{n_{s}} = \frac{1}{1 - \exp_{q}(-\beta_{q} E_{s})},
\end{equation}
which is valid under the convergence condition \(|\exp_{q}(-\beta_{q} E_{s})| < 1\). This resembles the standard form expressed in Eq.~\eqref{partition GB s}. Note that the condition $\beta_{q} E_{s} \ll 1$ refers to the system being in the high-temperature regime (or i.e., very small $\beta_{q}$). It is the condition that is considered in other studies for the classical gas~\cite{Prato1995GeneralizedSM}. As a result, in the thermodynamic limit (i.e., \(N \to \infty\)), the total partition function in Eq.~\eqref{q-partition 1} within the aforementioned regime, can be obtained as
\begin{equation}
\label{Zq final}
\displaystyle
Z_{q} \simeq \prod_{s=1}^{\infty} \frac{1}{1 - \exp_{q} (-\beta_{q} E_{s})}.
\end{equation}
This provides a non-extensive generalization of the grand canonical partition function for an ideal bosonic gas, suitable for near-horizon regions of strongly gravitating systems where conventional extensivity may break down. Regarding the convergence condition: $\left| \exp_{q} (-\beta_{q} E_{s}) \right| < 1$, this becomes real and positive \(q\)-exponential when we simply require $\exp_{q} (-\beta_{q} E_{s}) < 1$. The consequences of this convergence condition depend critically on the sign of \((1-q)\), giving rise to two qualitatively distinct regimes:
$i$) \(q < 1\) (super-extensive), the convergence condition yields the limitation of the energy as $E_{s} < E_{\mathrm{max}} \equiv 1/[(1-q) \beta_{q}]$.
Thus, the admissible energy spectrum is bounded within \(0 < E_{s} < E_{\mathrm{max}}\), introducing a natural UV cut-off characteristic of super-extensive systems. $ii)$ \(q > 1\) (sub-extensive): the convergence condition does not constrain \(E_{s} > 0\). Note also that the upper bound of the energy of the super-extensive case approaches infinity in the high-temperature regime.
\par
It has been found that the $q$-deformation of the partition function in Eq.~\eqref{Zq final} near the black hole's horizon can be approximately obtained as
\begin{equation}
\displaystyle
Z_{q} \simeq \exp \left[\frac{4 \pi^{5} c^{3}}{45 h^{3}} \left( \frac{V_{\mathrm{loc}}}{(\beta_q \kappa l_\textrm{loc})^{3}} \, \Gamma_q(3) \right)\right],
\label{partition 5 q}
\end{equation}
where $\Gamma_{q}(s)$ is the $q$-deformed gamma function defined as
\begin{equation}
\displaystyle
\Gamma_{q}(s) = \int_{0}^{\infty} du \, u^{s-1} \exp_q(-u),\label{def q Gamma}
\end{equation}
generalizing the classical gamma function to non-extensive statistics and converging under appropriate conditions on \(s\) and \(q\). For specific values of \(q\), the $q$-gamma function takes the explicit form~\cite{naik2016qLaplace}:
\begin{equation}
\displaystyle
\Gamma_{q}(s) = 
\begin{cases}
\displaystyle 
\frac{1}{(1-q)^{s}} \frac{\Gamma(s)\, \Gamma\left(\displaystyle \frac{1}{1-q} + 1 \right)}{\Gamma\left(\displaystyle \frac{1}{1-q} + s + 1 \right)}, & q<1, \\[1.5ex]
\displaystyle \Gamma(s), & q=1, \\[1.5ex]
\displaystyle \frac{1}{(q-1)^{s}} \frac{\Gamma(s)\, \Gamma\left(\displaystyle \frac{1}{q-1} - s \right)}{\Gamma\left(\displaystyle \frac{1}{q-1} \right)}, & q>1,
\end{cases}
\end{equation}
valid for all \(s>0\) and \(q \in \mathbb{R}^{+} \setminus \{1\}\), provided the gamma functions are defined.
The detail of obtaining the result in Eq.~\eqref{partition 5 q} can be seen in Appendix~\ref{Integral}. This result can be thought of as modifying the standard Stefan–Boltzmann law by replacing the numerical factor with the $q$-gamma correction \( \Gamma_{q}(3) \), which encodes the non-extensive deformation effects. In addition, for practical evaluations, an explicit rational form of \( \Gamma_{q}(3) \) is often employed:
\begin{equation}
\displaystyle
\Gamma_{q}(3) = \frac{2}{(2-q)(3-2q)(4-3q)},
\label{Gamma3}
\end{equation}
valid in both sub-extensive (\(q>1\)) and super-extensive (\(q<1\)) regimes. In the limit \( q \to 1 \), this smoothly reduces to the classical result \( \Gamma(3) = 2 \), restoring consistency with the extensive case.
\par
The $q$-logarithmic representation of the $q$-partition function given in Eq.~\eqref{partition 5 q} can be expressed in the form
\begin{equation}
\displaystyle
\ln_{q} Z_{q} = \frac{1}{1 - q} \left( Z_{q}^{1 - q} - 1 \right)
\simeq \frac{1}{1 - q} \left[ \exp\left( \frac{4 \pi^{5} (1-q)}{45 h^{3} c^{3}} \frac{V_{\mathrm{loc}}}{\beta_{q,\mathrm{loc}}^3} \Gamma_q(3) \right) - 1 \right],
\label{Zq 2}
\end{equation}
where we have consistently implemented the $q$-generalized Tolman's law, $\beta_{q,\mathrm{loc}} = \beta_q \kappa l_{\mathrm{loc}}$, in order to express all thermodynamic quantities as measured in a local frame within the framework of Tsallis-deformed statistics. Accordingly, thermodynamic observables within the Tsallis framework can be derived by differentiating its $q$-logarithm with respect to the appropriate variables. In particular, the $q$-local internal energy and $q$-local pressure are obtained as follows:
\begin{eqnarray}
\displaystyle
U_{q, \mathrm{loc}} &=& -\frac{\partial}{\partial \beta_{q, \mathrm{loc}}} \ln_{q} Z_{q} 
= \frac{4 \pi^{5}}{15 h^{3} c^{3}} \frac{V_{\mathrm{loc}}}{\beta_{q, \mathrm{loc}}^{4}} \Gamma_{q}(3) \,
\exp\!\Bigg[\frac{4\pi^{5} (1 - q)}{45 h^{3} c^{3}} \frac{V_{\mathrm{loc}}}{\beta_{q, \mathrm{loc}}^{3}} \Gamma_{q}(3)\Bigg], 
\label{q energy} \\
\displaystyle
P_{q, \mathrm{loc}} &=& \frac{1}{\beta_{q, \mathrm{loc}}} \frac{\partial}{\partial V_{\mathrm{loc}}} \ln_{q} Z_{q} 
= \frac{4 \pi^{5}}{45 h^{3} c^{3}} \frac{\Gamma_{q}(3)}{\beta_{q, \mathrm{loc}}^{4}} \,
\exp\!\Bigg[\frac{4\pi^{5} (1 - q)}{45 h^{3} c^{3}} \frac{V_{\mathrm{loc}}}{\beta_{q, \mathrm{loc}}^{3}} \Gamma_{q}(3)\Bigg].
\label{q pressure}
\end{eqnarray}
In addition, the $q$-local entropy follows directly from the non-extensive thermodynamic identity. Substituting Eq.~\eqref{Zq 2}, we obtain
\begin{eqnarray}
\displaystyle
S_{q, \mathrm{loc}} &=& k_{B} \left( 1 - \beta_{q, \mathrm{loc}} \frac{\partial}{\partial \beta_{q, \mathrm{loc}}} \right) \ln_{q} Z_{q} \nonumber \\
\displaystyle
&=& \frac{k_{B}}{1 - q} \Bigg[ 
\Bigg( 1 + 3 (1 - q) \frac{4 \pi^{5}}{45 h^{3} c^{3}} \frac{V_{\mathrm{loc}}}{\beta_{q, \mathrm{loc}}^{3}} \Gamma_{q}(3) \Bigg)
\exp\!\left(\frac{4 \pi^{5} (1 - q)}{45 h^{3} c^{3}} \frac{V_{\mathrm{loc}}}{\beta_{q, \mathrm{loc}}^{3}} \Gamma_{q}(3) \right)
- 1 \Bigg].
\label{q entropy}
\end{eqnarray}
One can verify that these $q$-deformed thermodynamic quantities consistently satisfy the generalized first law of thermodynamics in the local frame,
\begin{equation}
\displaystyle
dU_{q, \mathrm{loc}} = \frac{1}{k_{B} \beta_{q, \mathrm{loc}}} \, dS_{q, \mathrm{loc}} - P_{q, \mathrm{loc}} \, dV_{\mathrm{loc}},
\label{1st law of q}
\end{equation}
thereby demonstrating that the Legendre structure is preserved under non-extensive deformation, with the internal consistency of this $q$-generalized first law verified explicitly in Appendix~\ref{check 1st law}.
\par
Motivated by Ref.~\cite{Chunaksorn:2025nsl}, we introduce a mapping between the $q$-deformed inverse temperature and its conventional GB counterpart of the form
\begin{equation}
\displaystyle
\beta_{q} = \mathcal{J}(q, A_h) \, \beta,
\label{Jq 1}
\end{equation}
where $\mathcal{J}(q, A_{h})$ is assumed to be a smooth and positive-definite function satisfying a consistency condition $\lim_{q\to 1} \mathcal{J}(q, A_{h}) = 1$. Building upon this prescription, we adopt the standard physical input from the GB framework in order to recover the Bekenstein--Hawking entropy in the extensive limit. In particular, the local volume is taken to be $V_{\mathrm{loc}} = A_{h} \, l_{\mathrm{loc}}/4$, the inverse temperature is specified by $\beta = 4 \pi^{2} c/(h \kappa)$, and the proper distance from the horizon is fixed at $l_{*} = l_{P}/(3 \sqrt{10 \pi})$. Under these conditions, the entropy of the thermal system evaluated at the stretched horizon can be expressed as
\begin{equation}
\displaystyle
S^{(0)}_{q, \mathrm{loc}} 
= \frac{k_{B}}{1 - q} \left[ 
\left( 1 + 3 (1 - q) \frac{\Gamma_{q}(3)}{32 \mathcal{J}(q, A_{h})^{3}} \frac{A_{h}}{l_{P}^{2}} \right) 
\exp\!\left( (1 - q) \frac{\Gamma_{q}(3)}{32 \mathcal{J}(q, A_{h})^{3}} \frac{A_{h}}{l_{P}^{2}} \right) - 1 
\right],
\label{Sq 1}
\end{equation}
which provides an extension of the classical area law into the non-extensive regime.
\par
It is important to argue that black hole entropy should be unique and independent of the type of near-horizon gas system. Consequently, this requirement leads to a matching condition between the entropy given in Eq.~\eqref{Sq 1} and the $q$-generalized Bekenstein--Hawking entropy constructed in Ref.~\cite{Chunaksorn:2025nsl}, given by
\begin{equation}
\label{used Sq}
\displaystyle
S_{q, \mathrm{BH}} = \frac{k_{B}}{1-q} 
\left[ 
\left(1 + 3 (1-q) \frac{A_{h}}{l_{P}^{2}} \right) 
\exp \left(-\frac{11 (1-q) \, A_{h}}{4 l_{P}^{2}\left(1 + 3(1-q) \, \displaystyle \frac{A_{h}}{l_{P}^{2}}\right)} \right) - 1 
\right].
\end{equation}
This black hole entropy is a homogeneous function (see more details in the next section) by eliminating the $q$-Gamma function $\Gamma_{q}(3)$ through the imposition of the energy condition given by
\begin{equation}
\displaystyle
U_{q, \mathrm{loc}} = u_{1} \, \Gamma_{q}(3)^{\,u_{2}} \, U_{\mathrm{loc}},
\label{U condition}
\end{equation}
where the constants $u_{1}$ and $u_{2}$ are fixed by consistency requirements to be $u_{1} = 2^{1/3}$ and $u_{2} = -1/3$. The requirement \(U_{q, \mathrm{loc}} > 0\), consistent with its interpretation as the black hole mass, imposes the constraint
\begin{equation}
\label{bound q 1}
0 < q < \frac{4}{3},
\qquad
\frac{3}{2} < q < 2.
\end{equation}
By solving the entropy-matching condition, $S^{(0)}_{q, \mathrm{loc}} \rightarrow S_{q, \mathrm{BH}}$, one obtains a closed-form expression for the deformation function $\mathcal{J}(q, A_{h})$ through the introduction of the Lambert function $W(x)$, yielding
\begin{equation}
\displaystyle
\mathcal{J}(q, A_{h}) = \left[\frac{3(1-q) A_{h} \Gamma_{q}(3)}{32 l^{2}_{P} \left(W\left(K\right) - 1\right)}\right]^{1/3},
\label{JqSequal}
\end{equation}
where the argument $K$ of the Lambert function is given by
\begin{equation}
\displaystyle
K := \frac{e^{1/3}}{3} \left(1 + 3(1-q) \frac{A_{h}}{l^{2}_{P}}\right) \exp \left[-\frac{11 (1-q) A_{h}}{4 l_{P}^{2} \left(1 + 3(1-q) \displaystyle \frac{A_{h}}{l_{P}^{2}}\right)}\right].
\end{equation}
It is noted that Eq.~\eqref{JqSequal} is obtained from an equation of the form $j e^{j} = x$, for which the solution is $j = W(x)$. Moreover, the function $\mathcal{J}(q, A_{h})$ reduces to unity in the extensive limit, $\lim_{q \rightarrow 1} \mathcal{J}(q, A_{h}) = 1$. Importantly, the Lambert-domain constraint $x \geq -1/e$ is satisfied for all physically admissible values of $q$, ensuring that the Lambert function $W(x)$ remains real-valued on the principal branch $W_{0}$. Consequently, Eq.~\eqref{JqSequal} defines a real, continuous, and non-negative deformation function $\mathcal{J}(q, A_h)$. This, in turn, guarantees that the deformed local inverse temperature $\beta_{q,\mathrm{loc}} = \mathcal{J}(q, A_h)\beta_{\mathrm{loc}}$ remains non-negative whenever $\beta_{\mathrm{loc}} \geq 0$.
\par
Before concluding this section, we reconsider the physical interpretation of non-extensive effects in the near-horizon gas system. We begin with the $q$-entropy introduced in Eq.~\eqref{q entropy}, under the assumption that strong inter-mode correlations are absent. Within this approximation, the local entropy exhibits the scaling behavior
\begin{equation}
\label{Sq1}
\displaystyle
S_{q, \rm loc} \sim \alpha_{q} k_{B} \frac{V_{\rm loc}}{\beta_{q, \rm loc}^{3}},
\end{equation}
where $\alpha_{q} \equiv \frac{16 \pi^{5}}{45 h^{3} c^{3}} \Gamma_{q}(3)$. 
In this regime, the microcanonical and canonical ensembles remain effectively equivalent, so that the number of accessible microstates is related to the entropy via the $q$-exponential,
\begin{equation}
\displaystyle
Z_{q} \sim \Omega \sim \exp_{q}\!\left(\frac{S_{q,\mathrm{loc}}}{k_{B}}\right), \quad \text{or} \quad \Omega \sim \exp_{q}\!\left[\alpha_{q} \frac{V_{\mathrm{loc}}}{\beta_{q,\mathrm{loc}}^{3}}\right].
\end{equation}
\par
The physical implications of this structure follow from the non-additive nature of the $q$-exponential. Since $\exp_q(x+y) \neq \exp_q(x)\exp_q(y)$, the composition rule of microstates is modified, leading to deviations from extensivity. In the \textit{super-extensive regime} ($q < 1$),
\(
\Omega_{(1+2)} > \Omega_{(1)} \cdot \Omega_{(2)},
\)
reflecting an enhancement of accessible configurations. Conversely, in the \textit{sub-extensive regime} ($q > 1$),
\(
\Omega_{(1+2)} < \Omega_{(1)} \cdot \Omega_{(2)},
\)
indicating that correlations or long-range interactions effectively restrict the available configuration space, as expected in the presence of strong gravitational fields near the horizon. In both cases, the entropic index $q$ governs the departure from extensivity through the modified composition rule, rather than through the local entropy scaling itself.
\par
Having established the statistical origin, we now examine the implications of the deformation for the macroscopic thermodynamic structure. The equation of state for radiation within Tsallis-deformed statistics follows from Eqs.~\eqref{q energy} and~\eqref{q pressure} as
\begin{equation}
\displaystyle
P_{q, \mathrm{loc}} = \frac{1}{3} \frac{U_{q, \mathrm{loc}}}{V_{\mathrm{loc}}}.
\end{equation}
This form coincides with that of a relativistic radiation fluid in standard GB statistics and arises from kinetic theory via the covariant energy--momentum tensor,
\begin{equation}
\displaystyle
T^{\mu \nu} = \int \frac{d^{3} \mathbf{p}}{(2\pi \hbar)^{3}} \frac{p^{\mu} p^{\nu}}{p^{0}} f_{q}(\mathbf{r}, \mathbf{p}),
\end{equation}
where $f_{q}$ is the $q$-deformed distribution. In the local rest frame of an isotropic massless gas,
$\rho \sim c \int dp\, p^{3} f_{q}(p)$, and $P \sim \int dp\, p^{3} f_{q}(p)/(3c)$, so that $P = \rho/3 = U/(3V)$ holds independently of the specific form of $f_{q}$. This demonstrates that the equation of state is fixed by kinematic considerations, while non-extensive effects enter only through deformations of the thermodynamic quantities.
\par
This distinction between statistical deformation and kinematic structure motivates an examination of thermodynamic consistency. The Tsallis parameter $q$ governs the onset of non-classical thermodynamic behavior, providing a possible extension of the Bekenstein--Hawking entropy while preserving its geometric character. This perspective is consistent with Hawking’s area theorem,
\begin{equation}
\displaystyle
\Delta A_{h} \geq 0,
\end{equation}
which ensures the irreversibility of classical processes obeying the null energy condition~\cite{Hawking:1971tu}. At the semiclassical level, Hawking radiation~\cite{Hawking1975} allows the horizon area to decrease, motivating the generalized second law (GSL):
$\Delta \left(S_{\mathrm{BH}} + S_{\mathrm{matter}}\right) \geq 0$ %
which ensures the non-decrease of total entropy~\cite{Bekenstein1973,Wall:2011hj}.
\par
Within the Tsallis-deformed framework, these considerations suggest that the entropy should remain monotonic with respect to the horizon area,
\begin{equation}
\displaystyle
\frac{dS_{q, \mathrm{BH}}}{dA_{h}} \geq 0, \qquad A_{h} \geq 0.
\end{equation}
This condition is satisfied for all $A_h$ in the \textit{super-extensive regime} ($q < 1$), whereas in the \textit{sub-extensive regime} ($q > 1$) it holds only within a finite interval,
\begin{equation}
\label{max Ah}
\displaystyle
1 \leq \frac{A_{h}}{l_{P}^{2}} \leq \frac{1}{36(q - 1)},
\end{equation}
indicating a saturation of accessible microstates induced by strong correlations. Building on this monotonicity and semiclassical considerations, one is naturally led to a generalized form of the second law in the Tsallis framework,
\begin{equation}
\label{GSL}
\displaystyle
\Delta \left(S_{q, \mathrm{BH}} + S_{\mathrm{matter}}\right) \gtrsim 0,
\end{equation}
which reduces to the standard result in the extensive limit and provides a plausible extension of thermodynamic principles to non-extensive regimes.
\par
To distinguish these non-extensive corrections from those arising in quantum gravity, we note that in approaches such as loop quantum gravity and string theory, the Bekenstein--Hawking area law receives sub-leading logarithmic corrections,
\begin{equation}
\displaystyle
S_{\mathrm{bh}} = k_{B} \frac{A_{h}}{4 l_{P}^{2}} + \alpha \ln \left|\frac{A_{h}}{l_{P}^{2}}\right| + \hdots
\end{equation}
where $\alpha$ is a constant.
The leading term of the right hand side of the above expression is geometric in origin, whereas the logarithmic correction is commonly associated with quantum fluctuations. By contrast, the Tsallis deformation generates corrections of statistical origin that persist in the macroscopic regime. Expanding Eq.~\eqref{used Sq} around the extensive limit $q = 1$ yields
\begin{equation}
\label{Sq series}
\displaystyle
S_{q, \mathrm{BH}} = k_{B} \frac{A_{h}}{4 l_{P}^{2}} + \frac{121 k_{B}}{32} \left(\frac{A_{h}}{l_{P}^{2}}\right)^{2} (1 - q) + \mathcal{O}[(1 - q)^{2}],
\end{equation}
where the sub-leading contribution arises from non-extensive effects, highlighting a different physical origin from the logarithmic corrections typically encountered in quantum gravitational approaches.
\par
Having established the mathematical structure and physical interpretation of the Tsallis-deformed entropy, we are now in a position to formulate a generalized thermodynamic framework in which non-extensivity is systematically incorporated. In the following section, we construct this framework by deriving the first law of thermodynamics, identifying the relevant conjugate variables, analyzing the resulting equation of state, and investigating how the Tsallis parameter $q$ affects thermodynamic stability, response functions, and critical phenomena.

\section{Black Hole Thermodynamics with Tsallis Entropy}\label{thermo}
\par
Building on the non-extensive entropy formalism introduced earlier, we will construct a generalized framework in which black hole thermodynamics can be formulated in terms of the $q$-generalized Bekenstein--Hawking entropy. We also derive thermodynamically consistent quantities and analyze their physical properties, with particular emphasis of the effects of non-extensivity on stability and the emergence of critical phenomena.
\par
The analysis begins with the first law of black hole thermodynamics, treating the black hole mass as a thermodynamic potential that is homogeneous in its extensive variables. This homogeneity is a fundamental feature of gravitational thermodynamics, as it allows the application of Euler’s theorem to derive generalized Smarr relations. These relations ensure internal consistency among the thermodynamic quantities and clarify the interplay between entropy, energy, and their conjugate variables.
\par
For precision, recall that a real-valued function $f(x_{i})$, defined on variables $x_{i} \in \mathbb{R}$, is homogeneous of degree $n \in \mathbb{R}$ if it satisfies
\begin{equation}
\displaystyle
f(I x_{i}) = I^{n} f(x_{i}),
\end{equation}
for any scaling factor $I \in \mathbb{R}$. Under this condition, Euler’s theorem can be expressed as
\begin{equation}
\displaystyle
n f(x_{i}) = \sum_{j} x_{j} \frac{\partial f}{\partial x_{j}},
\end{equation}
which plays a central role in thermodynamics, provided the potential function maintains its homogeneity—i.e., scales consistently with the extensive variables. In black hole thermodynamics, particularly when generalized to include contributions such as a cosmological constant or non-extensive deformation parameters, Euler’s theorem provides a rigorous foundation for constructing Smarr-type relations. With this formalism established, we now apply it to the $q$-deformed black hole entropy $S_{q, \mathrm{BH}}(A_{h})$ introduced in Eq.~\eqref{used Sq}. This systematic extension of black hole thermodynamics beyond the extensive regime reveals new features emerging from non-extensive statistics.

\subsection{Extended Phase-Space Thermodynamic Formulation}
\par
As a concrete realization of the general framework, we develop a thermodynamic formulation based on the $q$-generalized entropy for the static, spherically symmetric charged black hole described by the Reissner--Nordström (RN) metric. The RN spacetime is characterized by the metric function
\begin{equation}
g(r) = 1 - \frac{2M}{r} + \frac{Q^{2}}{r^{2}},
\label{horizon function Q}
\end{equation}
where $M$ and $Q$ denote the black hole mass and electric charge, respectively. 
The event horizon radius is given by the largest root of $g(r)=0$, which yields
\begin{equation}
M = \frac{r_{bh}}{2} \left(1 + \frac{Q^{2}}{r_{bh}^{2}}\right).
\label{M_relation}
\end{equation}
The RN geometry admits two horizons: the outer (event) horizon 
$r_{+}=M + \sqrt{M^{2} - Q^{2}}$ and the inner (Cauchy) horizon 
$r_{-}=M - \sqrt{M^{2} - Q^{2}}$. 
In the extremal limit $M = Q$, they coincide and the associated Hawking temperature vanishes. Throughout this work, we identify the event horizon with $r_{bh} \equiv r_{+}$, as all thermodynamic quantities are defined there.
\par
Within the standard extensive framework, the Bekenstein--Hawking entropy is expressed as
\begin{equation}
\displaystyle
S_{\mathrm{BH}(\mathrm{RN})} = \pi r_{bh}^{2}.
\label{Sch}
\end{equation}
Substituting Eq.~\eqref{Sch} into Eq.~\eqref{M_relation} allows one to express the black hole mass explicitly in terms of entropy and charge:
\begin{equation}
\displaystyle
M = \frac{\sqrt{S_{\mathrm{BH}(\mathrm{RN})}}}{2\sqrt{\pi}} \left(1 + \frac{\pi Q^{2}}{S_{\mathrm{BH}(\mathrm{RN})}}\right).
\label{M 2}
\end{equation}
This relation exhibits homogeneity of degree $1/2$ under the rescaling
\[
S_{\mathrm{BH}(\mathrm{RN})} \to I S_{\mathrm{BH}(\mathrm{RN})}, \quad Q^{2} \to I Q^{2},
\]
which justifies the application of Euler’s theorem to yield the Smarr relation
\begin{equation}
\displaystyle
M = 2 T_{\mathrm{H}(\mathrm{RN})} S_{\mathrm{BH}(\mathrm{RN})} + \Phi_{(\mathrm{RN})} Q,
\label{Smarr charge}
\end{equation}
where the conjugate variables are
\begin{equation}
\displaystyle
T_{\mathrm{H}(\mathrm{RN})} = \left(\frac{\partial M}{\partial S_{\mathrm{BH}(\mathrm{RN})}}\right)_Q, \qquad \Phi_{(\mathrm{RN})} = \left(\frac{\partial M}{\partial Q}\right)_{S_{\mathrm{BH}(\mathrm{RN})}}.
\label{conjugate charge}
\end{equation}
These relations satisfy the standard first law of black hole thermodynamics,
\begin{equation}
\displaystyle
dM = T_{\mathrm{H}(\mathrm{RN})} \, dS_{\mathrm{BH}(\mathrm{RN})} + \Phi_{(\mathrm{RN})} \, dQ,
\label{1st law charge}
\end{equation}
serving as a baseline for comparison with non-extensive effects.
\par
To explore such non-extensive effects, we adopt the Tsallis-deformed entropy defined in Eq.~\eqref{used Sq}, introducing $\eta := 1 - q$:
\begin{equation}
\displaystyle
S_{\eta, \mathrm{BH}(\mathrm{RN})} = \frac{1}{\eta} \left[(1 + 3 \eta A_{h}) \exp \!\left(- \frac{11 \eta A_{h}}{4(1 + 3 \eta A_{h})}\right) - 1\right],
\label{charged Sq}
\end{equation}
with horizon area $A_{h} = 4\pi r_{bh}^{2} = 4\pi (M + \sqrt{M^{2} - Q^{2}})^{2}$, giving explicitly
\begin{eqnarray}
\displaystyle
S_{\eta, \mathrm{BH}(\mathrm{RN})} &=& \frac{1}{\eta} \Bigg[(1 + 12 \pi \eta (M + \sqrt{M^{2} - Q^{2}})^{2}) \nonumber \\
&& \times \exp \!\left(- \frac{11 \pi \eta (M + \sqrt{M^{2} - Q^{2}})^{2}}{1 + 12 \pi \eta (M + \sqrt{M^{2} - Q^{2}})^{2}} \right) - 1\Bigg].
\label{charged Sq 2}
\end{eqnarray}
The transcendental form of Eq.~\eqref{charged Sq 2} precludes an analytic inversion to $M(S_{\eta,\mathrm{BH}})$, motivating an analysis based on homogeneity in the variables $(M, Q, \eta)$. Under the scaling transformation $M^{2} \to I M^{2}$,
$Q^{2} \to I Q^{2}$, and $\eta^{-1} \to I \eta^{-1}$, the entropy scales as
\begin{equation}
S_{\eta, \mathrm{BH}(\mathrm{RN})}(I M^{2}, I Q^{2}, I \eta^{-1})
= I \, S_{\eta, \mathrm{BH}(\mathrm{RN})}(M^{2}, Q^{2}, \eta^{-1}),
\end{equation}
thereby showing that it is homogeneous of degree one.
\par
Applying Euler’s theorem to the Tsallis-deformed entropy leads to
\begin{equation}
\displaystyle
S_{\eta, \mathrm{BH}(\mathrm{RN})} = \frac{1}{2} M \left( \frac{\partial S_{\eta, \mathrm{BH}(\mathrm{RN})}}{\partial M} \right)_{Q,\eta} + \frac{1}{2} Q \left( \frac{\partial S_{\eta, \mathrm{BH}(\mathrm{RN})}}{\partial Q} \right)_{M,\eta} - \eta \left( \frac{\partial S_{\eta, \mathrm{BH}(\mathrm{RN})}}{\partial \eta} \right)_{M,Q},
\end{equation}
which naturally defines the Tsallis-deformed Smarr relation:
\begin{equation}
\displaystyle
M = 2 T_{\eta, \mathrm{H}(\mathrm{RN})} S_{\eta, \mathrm{BH}(\mathrm{RN})} - 2 \Phi_{\eta(\mathrm{RN})} \eta + \Phi_{(\mathrm{RN})} Q,
\label{q Smarr}
\end{equation}
with the generalized potentials
\begin{eqnarray}
\displaystyle
T_{\eta, \mathrm{H}(\mathrm{RN})} &=& \left( \frac{\partial M}{\partial S_{\eta, \mathrm{BH}(\mathrm{RN})}} \right)_{Q, \eta}, \label{Tq} \\
\Phi_{\eta(\mathrm{RN})} &=& \left( \frac{\partial M}{\partial \eta} \right)_{Q, S_{\eta, \mathrm{BH}(\mathrm{RN})}}, \\
\Phi_{(\mathrm{RN})} &=& \left( \frac{\partial M}{\partial Q} \right)_{\eta, S_{\eta, \mathrm{BH}(\mathrm{RN})}}. \label{Phi charge}
\end{eqnarray}
With these potentials, one can define the generalized first law in the Tsallis framework as
\begin{equation}
\displaystyle
dM = T_{\eta, \mathrm{H}(\mathrm{RN})} \, dS_{\eta, \mathrm{BH}(\mathrm{RN})} + \Phi_{(\mathrm{RN})} \, dQ + \Phi_{\eta(\mathrm{RN})} \, d\eta.
\label{q 1st law 1}
\end{equation}
In this section, we will consider thermodynamic processes at fixed $\eta$, so that the first law simplifies to
\begin{equation}
\displaystyle
dM = T_{\eta, \mathrm{H}(\mathrm{RN})} \, dS_{\eta, \mathrm{BH}(\mathrm{RN})} + \Phi_{(\mathrm{RN})} \, dQ.
\label{1st fixed eta}
\end{equation}
\begin{figure}[ht]\centering
\includegraphics[width=6cm]{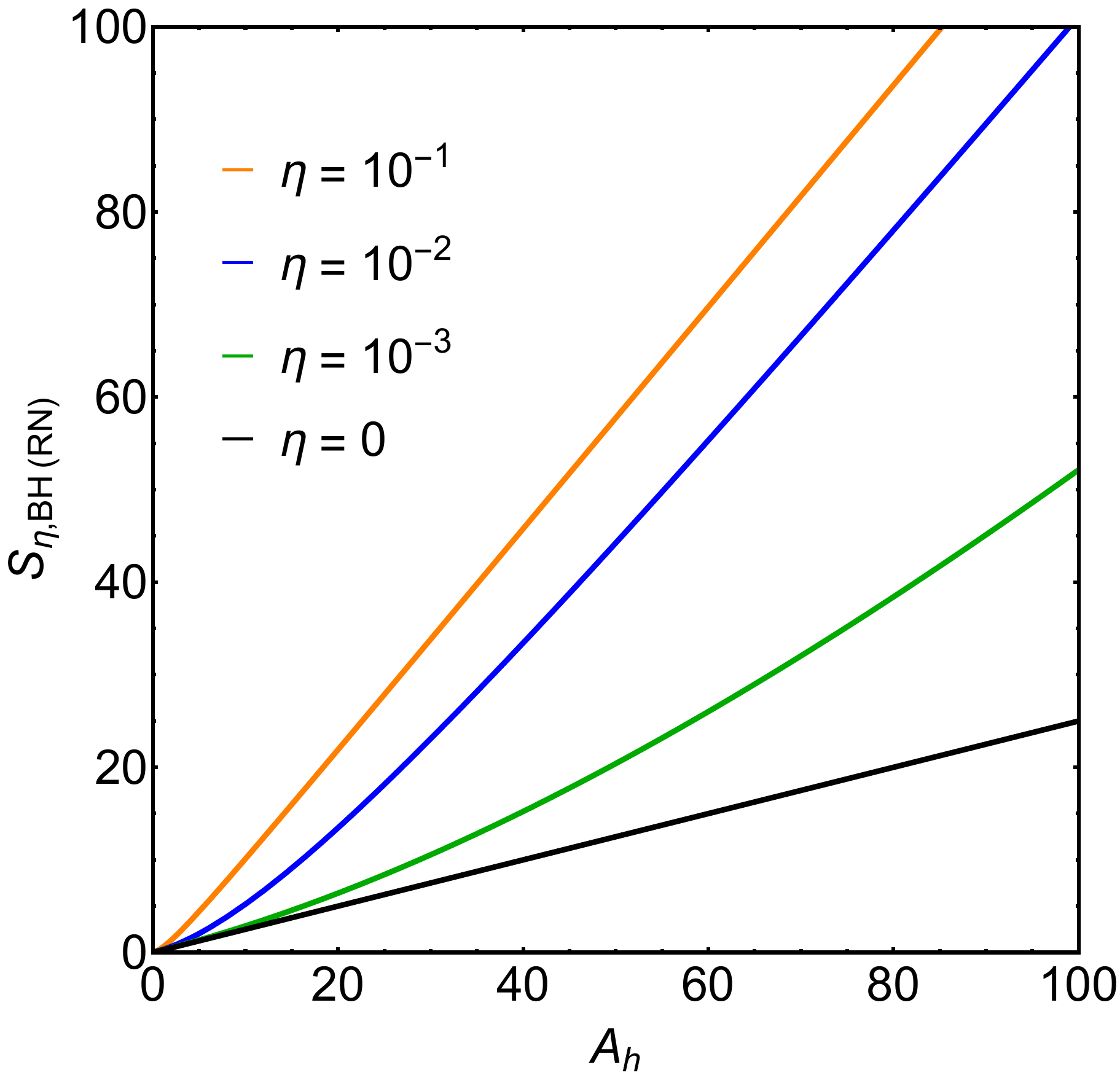}
\qquad
\includegraphics[width=6cm]{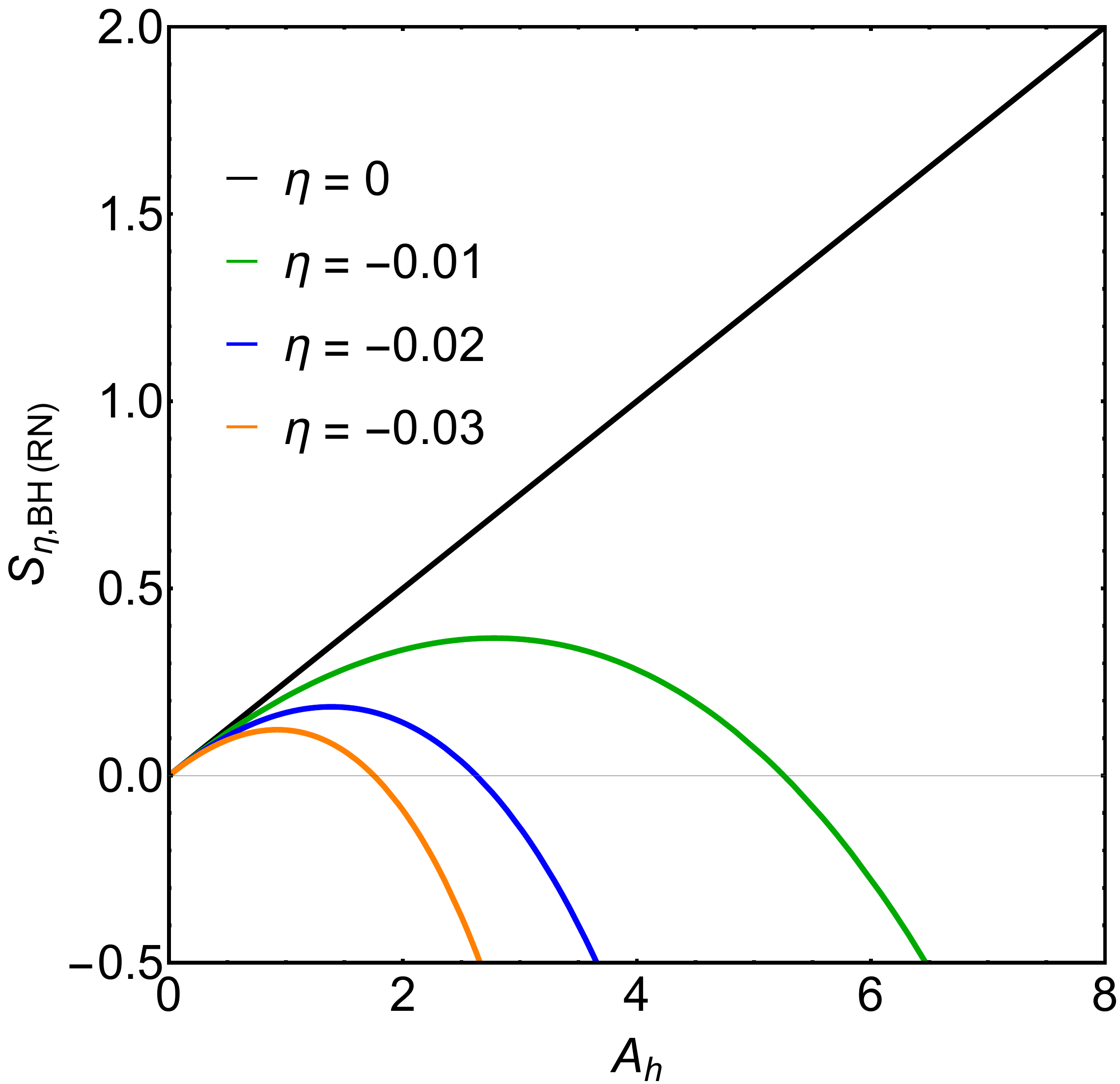}
\caption{Black hole entropy versus horizon area $A_{h}$ for positive $\eta$ (left) and negative $\eta$ (right).}
\label{Smax}
\end{figure}
\par
The behavior of the entropy $S_{\eta,\mathrm{BH(RN)}}$ as a function of the horizon area $A_h$, shown in Fig.~\ref{Smax}, exhibits two distinct regimes determined by the sign of $\eta$. In the left panel, the entropy is a monotonically increasing function of $A_h$. In contrast, in the right panel it increases only within the range $A_h < A_{\mathrm{max}}$, where the maximum occurs at
\begin{equation}
A_{\mathrm{max}} = -\frac{1}{36 \eta}.
\end{equation}
This turnover indicates a saturation of accessible microstates for $A_h > A_{\mathrm{max}}$. Furthermore, the physical relevance of the generalized entropy is assessed at the Planck scale. In Planck units, the minimal horizon area is $A_h = 1$, corresponding to a single Planck area, thereby excluding the unphysical regime $A_h \ll 1$. To remain within the semiclassical regime, we restrict attention to $A_h \ge 1$, which leads to the refined bound on the deformation parameter
\begin{equation}
-\frac{1}{36} < \eta < 1.
\label{bound q 2}
\end{equation}
This constraint is stronger than that in Eq.~\eqref{bound q 1} and ensures that the generalized entropy remains positive and monotonically increasing within the physical domain.
\par
Within the Tsallis statistical framework, the generalized Hawking temperature given by Eq.~\eqref{Tq} can be written explicitly as
\begin{equation}
T_{\eta, \mathrm{H}(\mathrm{RN})} = 
\frac{(1 + 3 \eta A_{h}) \exp \left[\dfrac{11 \eta A_{h}}{4 (1 + 3 \eta A_{h})} \right]}{2 \sqrt{\pi A_{h}} (1 + 36 \eta A_{h})} 
\left(1 - \frac{4 \pi Q^2}{A_{h}}\right)
= T_{\eta, \mathrm{H}(\mathrm{Sch})} \left(1 - \frac{4 \pi Q^2}{A_{h}}\right),
\label{charged temp 1}
\end{equation}
where $T_{\eta, \mathrm{H}(\mathrm{Sch})}$ denotes the Tsallis-deformed temperature of a Schwarzschild black hole. In the extensive limit $\eta \to 0$, this expression reduces smoothly to the standard Hawking temperature.
\begin{figure}[ht]\centering
\includegraphics[width=6.1cm]{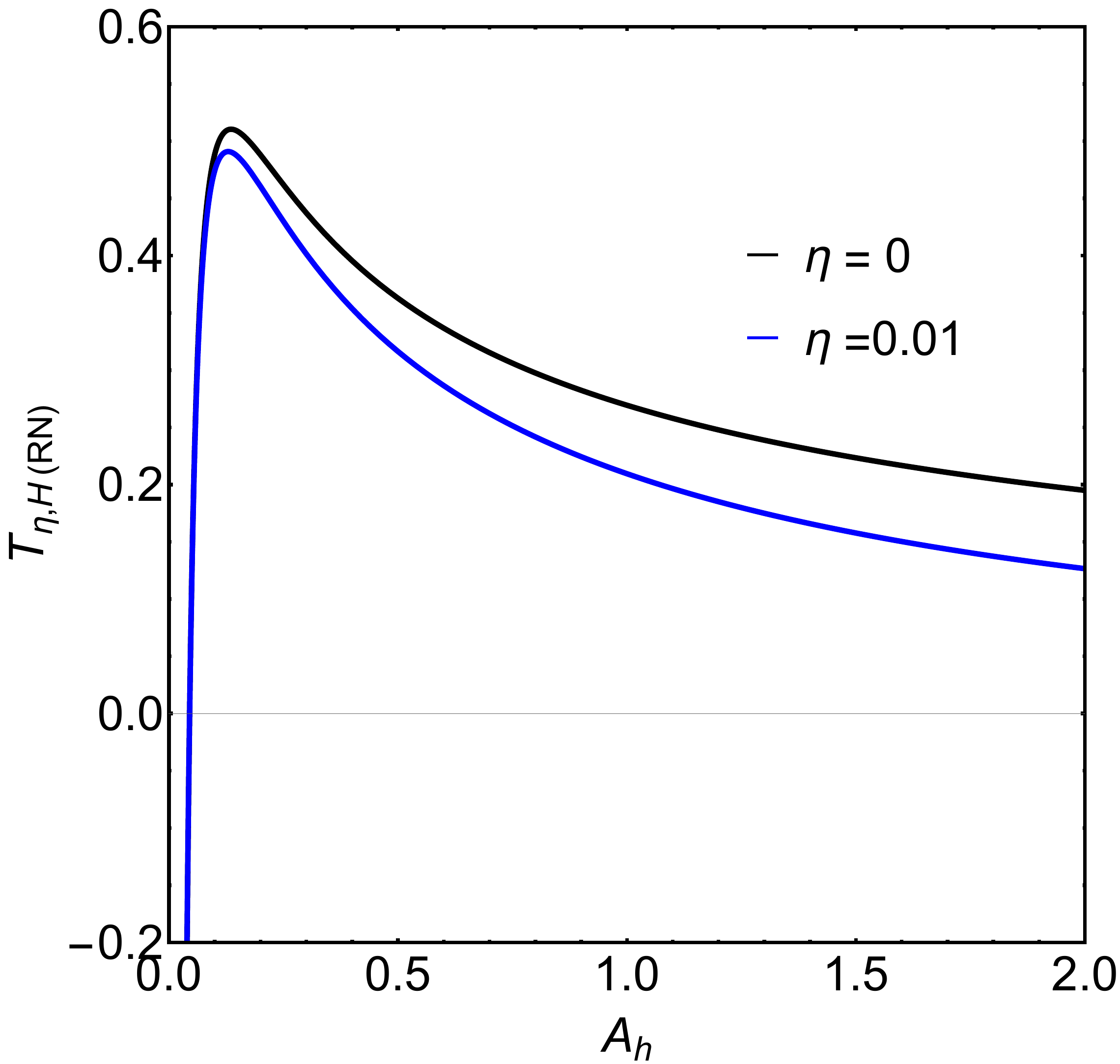}
\qquad
\includegraphics[width=6cm]{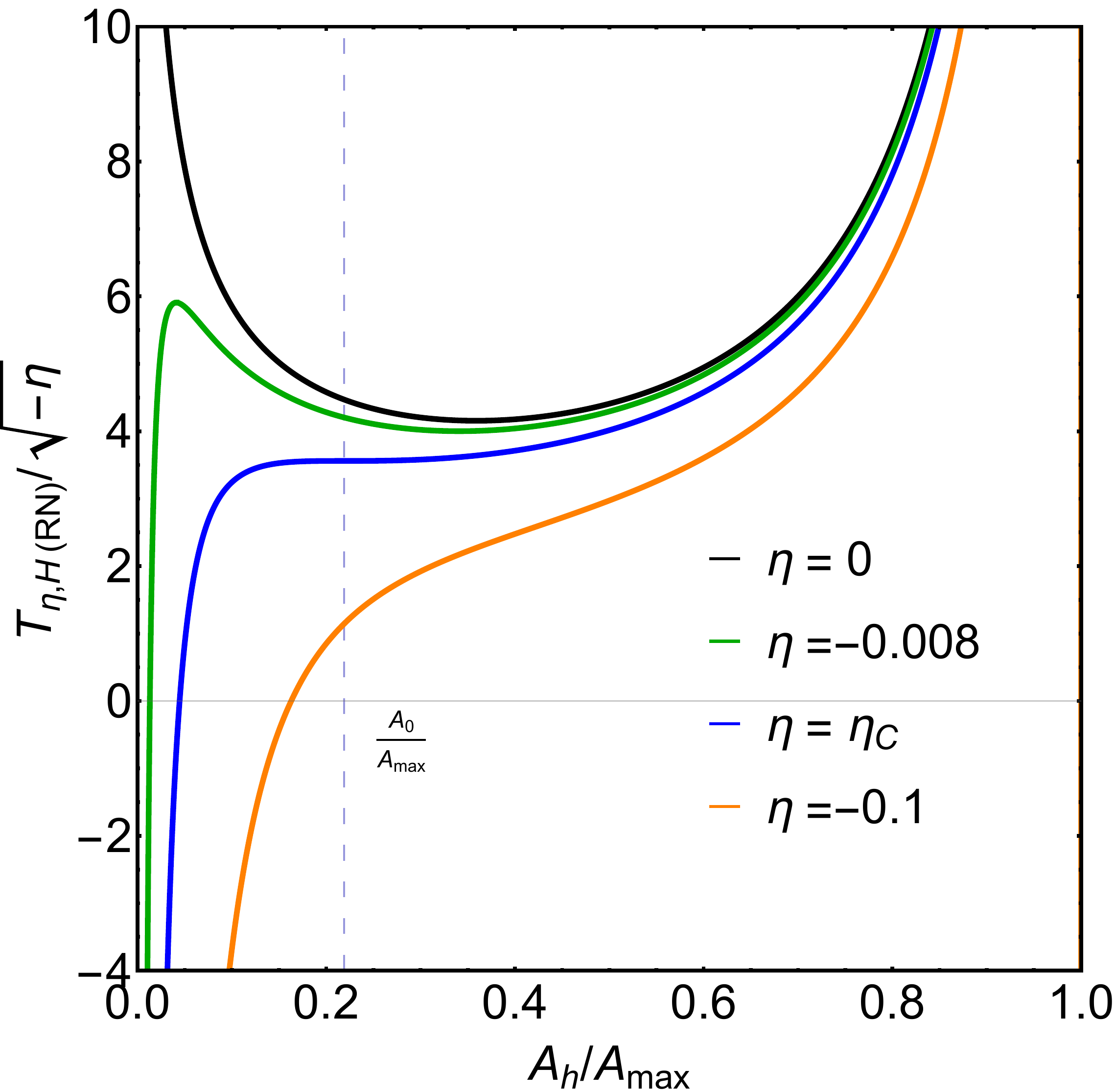}
\caption{Profiles of the black hole temperature for positive $\eta$ (left) and negative $\eta$ (right) at fixed electric charge $Q=0.06$.}
\label{T}
\end{figure}
\par
The behavior of the temperature is illustrated in Fig.~\ref{T}. In the \textit{super-extensive regime} ($\eta > 0$, left panel of Fig.~\ref{T}), the temperature vanishes as $A_h \rightarrow 4 \pi Q^{2}$, corresponding to the extremal limit $M = Q$. It then increases monotonically to a maximum and subsequently decreases gradually toward zero as $A_h \rightarrow \infty$, exhibiting a behavior qualitatively similar to that obtained from the standard Bekenstein--Hawking entropy. In this case, only two physical branches are present, and no additional phase structure emerges. Consequently, this regime is less relevant for investigating novel black hole phase transitions. In contrast, the \textit{sub-extensive regime} ($\eta < 0$, right panel of Fig.~\ref{T}) exhibits a qualitatively different, non-monotonic temperature profile. The temperature initially increases to a local maximum, then decreases to a local minimum, and finally diverges as $A_h \rightarrow A_{\mathrm{max}}^{-}$ from below, signaling the approach to a thermodynamic boundary.
\par
In general, the sign of the heat capacity is directly related to the slope of the temperature curve: a positive slope corresponds to positive heat capacity and hence local thermodynamic stability, whereas a negative slope indicates negative heat capacity and instability. Moreover, the heat capacity diverges at the extrema of the temperature profile, where the slope $\left(\partial T_{\eta,\mathrm{H}}/\partial A_h\right)_Q$ vanishes. Therefore, the structure of local stability is entirely encoded in the geometric properties of the temperature curve.
\par
Accordingly, from Fig.~\ref{T}, the black hole branch in the range $4\pi Q^{2} < A_{h} < A_{h(\rm max)}$ is locally stable, while the other branch with $A_{h} > A_{h(\rm max)}$ is locally unstable for $\eta > 0$, where $A_{h(\rm max)}$ denotes the area corresponding to the maximum of the temperature, obtained from $dT_{\eta, \rm H}/dA_{h} = 0$. Moreover, for $\eta < 0$, there exists a novel branch that is locally stable. This feature in the $\eta < 0$ case indicates that the Tsallis-deformed framework induces a richer thermodynamic structure, thereby motivating a more detailed analysis of stability and phase transitions.
\par
To assess local thermodynamic stability more explicitly, we analyze the behavior of the slope of the temperature profile $T_{\eta,\mathrm{H}}(A_h)$ at fixed charge, which can be expressed as
\begin{equation}
\frac{\partial}{\partial A_{h}} T_{\eta, \mathrm{H}(\mathrm{RN})}  
= \frac{(\mathcal{A}_{1} A_{h}^{4} + \mathcal{A}_{2} A_{h}^{3} + \mathcal{A}_{3} A_{h}^{2} + \mathcal{A}_{4} A_{h} + \mathcal{A}_{5}) 
\exp \left[\dfrac{11 \eta A_{h}}{4 (1 + 3\eta A_{h})} \right]}{8 A_{h}^{5/2} \sqrt{\pi} (1+3\eta A_{h}) (1+36 \eta A_{h})^{2}},
\label{dT1}
\end{equation}
with coefficients
\begin{align*}
\mathcal{A}_{1} &:= -648\, \eta^{3}, &
\mathcal{A}_{2} &:= -\eta^{2} \left(450 - 7776\, \pi \eta Q^{2}\right), \\
\mathcal{A}_{3} &:= -\eta \left(205 - 5400\, \pi \eta Q^{2}\right), &
\mathcal{A}_{4} &:= -2 + 1492\, \pi \eta Q^{2}, &
\mathcal{A}_{5} &:= 24\, \pi Q^{2}.
\end{align*}
\par
The critical points of local thermodynamic stability are determined by the real, positive roots of the quartic polynomial
\begin{equation}
\mathcal{P} (A_{h}) := \mathcal{A}_{1} A_{h}^{4} + \mathcal{A}_{2} A_{h}^{3} + \mathcal{A}_{3} A_{h}^{2} + \mathcal{A}_{4} A_{h} + \mathcal{A}_{5} = 0,
\label{poly 1}
\end{equation}
which we denote by $A_{C_Q} = A_{C_Q}(Q,\eta)$. The existence and multiplicity of these roots depend sensitively on both the deformation parameter $\eta$ and the electric charge $Q$, as they enter nonlinearly into the coefficients $\mathcal{A}_i$.
\begin{figure}[ht] 
\centering 
\includegraphics[width=6cm]{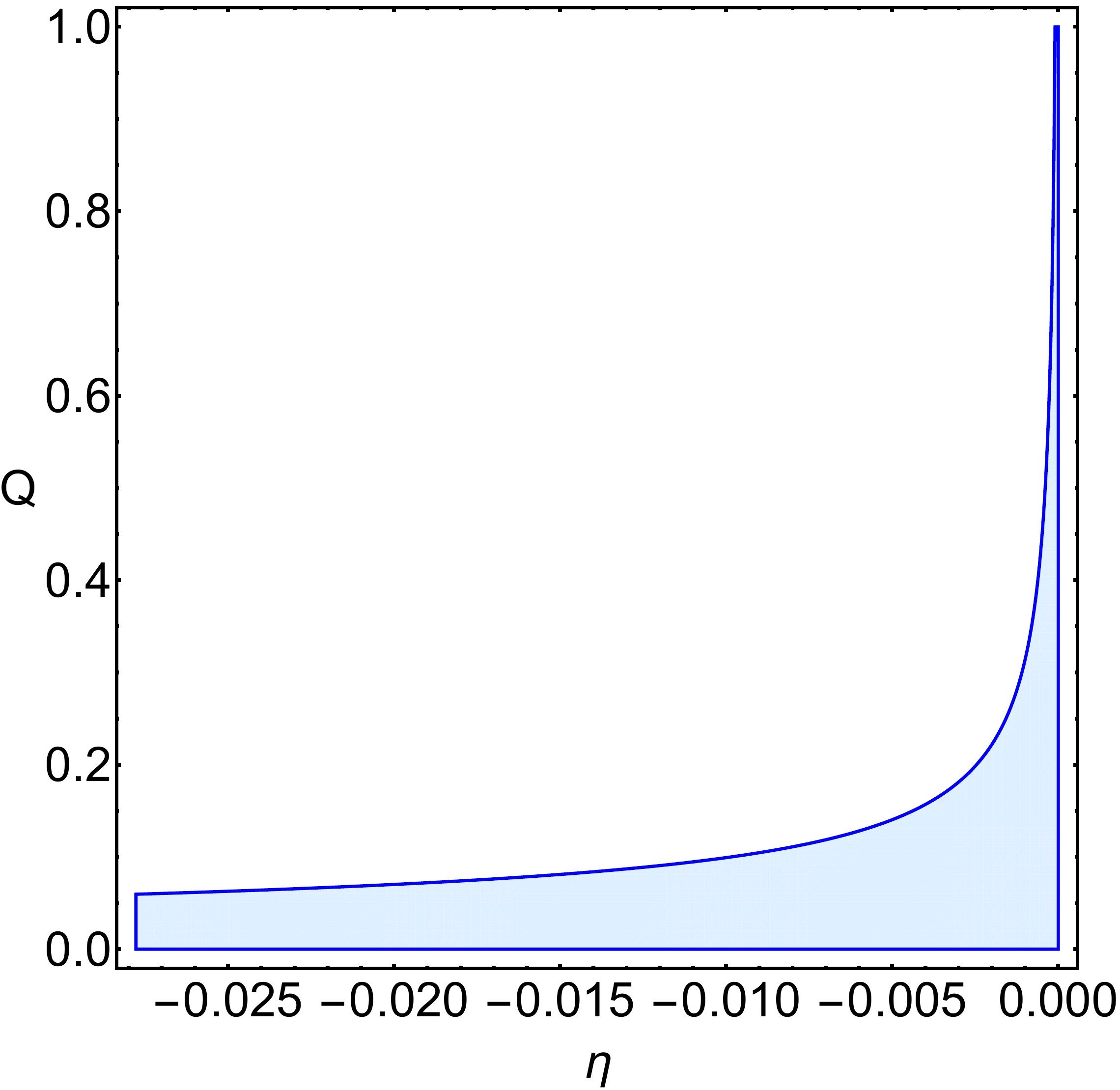} \caption{Admissible region in the $(\eta,Q)$-plane corresponding to the simultaneous existence of the two extrema $A_{C_{Q(-)}}$ and $A_{C_{Q(+)}}$.} 
\label{region local 3} \end{figure}
\par
The simultaneous existence of two distinct real, positive roots defines a non-trivial region in the $(\eta,Q)$-plane. Within this domain, the temperature develops two turning points, giving rise to multiple thermodynamic branches. The physically admissible region in which both extrema coexist is shown in Fig.~\ref{region local 3}.
\par
At a critical (degenerate) point, the two extrema merge, $A_{C_{Q(-)}} = A_{C_{Q(+)}} = A_{0}$, where both the first and second derivatives vanish:
\begin{equation}
\frac{\partial}{\partial A_{h}} T_{\eta, \mathrm{H(RN)}}(A_{0}) = 0,
\qquad
\frac{\partial^{2}}{\partial A_{h}^{2}} T_{\eta, \mathrm{H(RN)}}(A_{0}) = 0.
\end{equation}
Numerically, for $Q=0.06$, this occurs at $A_{0} \approx 0.222$ with $\eta_{C} \approx -0.0274$, as shown in the right panel of Fig.~\ref{T}.
\par
Having analyzed the temperature behavior and its critical structure under Tsallis deformation, we now introduce the electric potential given by Eq.~\eqref{Phi charge} as
\begin{equation}
\Phi_{(\mathrm{RN})} = 2 \sqrt{\pi} \frac{Q}{\sqrt{A_{h}}},
\label{charge Phi 1}
\end{equation}
which is independent of $\eta$. This relation will be used to eliminate the horizon area $A_h$, allowing the thermodynamic quantities to be reformulated in terms of $Q$ and $\Phi$ for the fluid reinterpretation to be developed below.
\par
With the generalized entropy and temperature at hand, local thermodynamic stability is determined by the heat capacity, which measures the response to small energy fluctuations at fixed charge. A positive heat capacity corresponds to a locally stable branch, whereas a negative value signals thermodynamic instability and runaway evaporation. Therefore, the extrema of the generalized temperature directly determine the horizon areas that separate stable and unstable branches.
\par
\begin{figure}[ht]
\centering
\includegraphics[width=6cm]{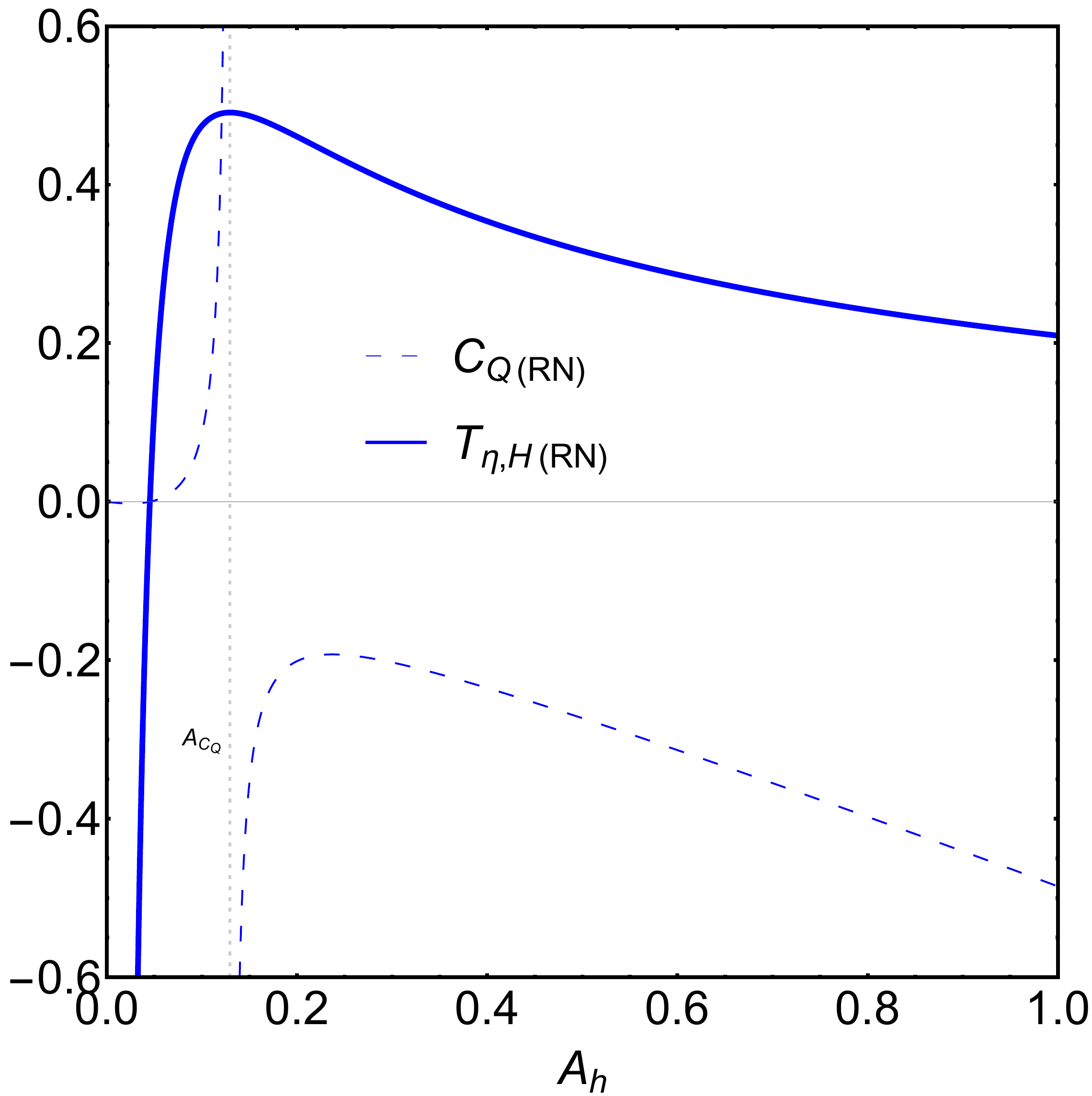}
\qquad
\includegraphics[width=6cm]{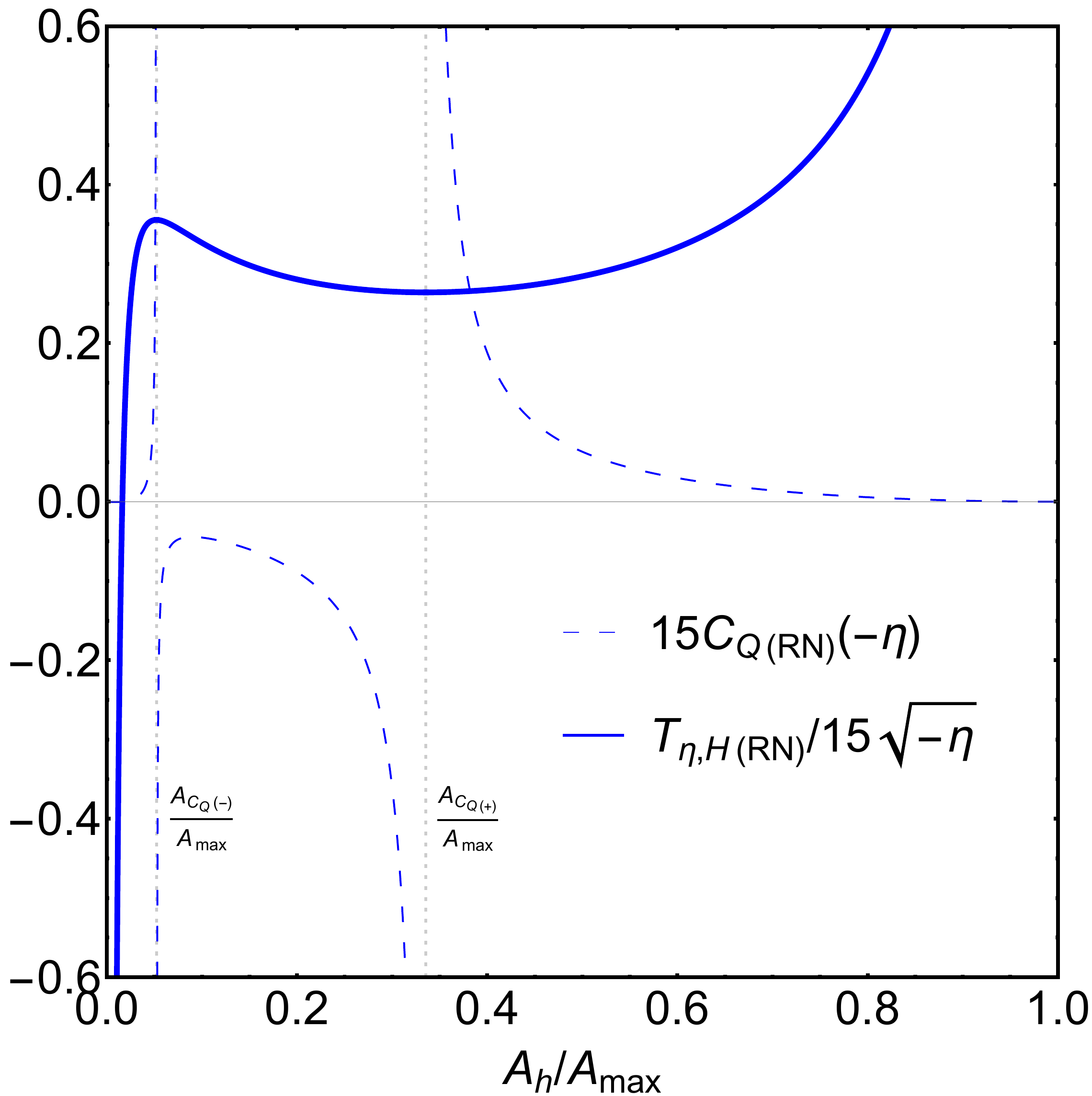}
\caption{Profiles of the heat capacity and temperature for fixed $Q = 0.06$ with $\eta = 0.01$ (left panel) and $\eta = -0.01$ (right panel).}
\label{T C}
\end{figure}

For fixed electric charge $Q$, the heat capacity in the Tsallis framework is given by
\begin{align}
\displaystyle
C_{Q(\mathrm{RN})} &= \left(\frac{\partial M}{\partial T_{\eta, \mathrm{H}(\mathrm{RN})}}\right)_{Q} 
= \left(\frac{\partial M}{\partial A_{h}}\right)_{Q} \left(\frac{\partial T_{\eta, \mathrm{H}(\mathrm{RN})}}{\partial A_{h}}\right)^{-1}_{Q} \nonumber \\
\displaystyle
&= \frac{A_{h}^{2} (1 + 3\eta A_{h}) (1 + 36 \eta A_{h})^{2} \exp \left[\displaystyle -\frac{11 \eta A_{h}}{4 (1 + 3 \eta A_{h})}\right]}{\mathcal{P} (A_{h})} \left(1 - \frac{4 \pi Q^{2}}{A_{h}}\right),
\label{heat cap 1}
\end{align}
where $\mathcal{P}(A_{h})$ is the quartic polynomial defined in Eq.~\eqref{poly 1}. The divergences and sign changes of $C_{Q(\mathrm{RN})}$ occur precisely at the extrema of $T_{\eta, \mathrm{H}(\mathrm{RN})}$; hence, the previously determined critical areas fully characterize local thermodynamic stability, as shown in Fig.~\ref{T C}.
\par
For $\eta > 0$, the charged black hole is locally stable near extremality ($A_{h} \gtrsim 4\pi Q^{2}$), but becomes unstable at larger horizon areas. In this regime, the generalized Hawking temperature develops a single maximum, so the heat capacity changes sign only once and remains negative thereafter, leading to a smooth thermodynamic response without multiple stability branches. In contrast, for $\eta < 0$, the suppression of entropy growth modifies the temperature profile, producing two extrema that divide the solutions into three branches: a small black-hole (SBH) branch and a large black-hole (LBH) branch, both locally stable, separated by an intermediate black-hole (IBH) branch that is locally unstable. This structure is analogous to the phase behavior of Van der Waals (VdW) fluids.
\par
Beyond local stability, global thermodynamic stability is determined by the Gibbs free energy, which compares the black hole configuration with alternative backgrounds—typically a thermal spacetime with vanishing free energy. The phase with lower Gibbs free energy is globally preferred. This criterion identifies globally stable phases and signals the possibility of first-order phase transitions, either between distinct black hole branches or between a black hole and the thermal background.
\par
For fixed deformation parameter $\eta$, the Gibbs free energy of the charged black hole is
\begin{align}
\displaystyle
\mathcal{G}_{\eta(\mathrm{RN})} &= M - T_{\eta, \mathrm{H}(\mathrm{RN})} \, S_{\eta, \mathrm{BH}(\mathrm{RN})} \nonumber \\
\displaystyle
&= \mathcal{G}_{\eta(\mathrm{Sch})} 
+ \frac{\sqrt{\pi} Q^{2} \Big[ 2 - 2 (1 + 3 \eta A_{h}) 
\exp \left[\displaystyle \frac{11 \eta A_{h}}{4 (1 + 3 \eta A_{h})}\right] 
+ \eta A_{h} (13 + 54 \eta A_{h}) \Big]}
{\eta A_{h}^{3/2} (1 + 36 \eta A_{h})},
\label{free energy 1}
\end{align}
where $\mathcal{G}_{\eta(\mathrm{Sch})}$ denotes the Gibbs free energy of the neutral Sch black hole:
\begin{equation}
\displaystyle
\mathcal{G}_{\eta(\mathrm{Sch})} 
= \frac{-2 - 11 \eta A_{h} + 18 \eta^{2} A_{h}^{2} 
+ 2 (1 + 3 \eta A_{h}) 
\exp \left[\displaystyle \frac{11 \eta A_{h}}{4 (1 + 3 \eta A_{h})}\right]}
{4 \eta \sqrt{\pi A_{h}} (1 + 36 \eta A_{h})}.
\end{equation}
In the limit $Q \to 0$, $\mathcal{G}_{\eta(\mathrm{RN})}$ smoothly reduces to 
$\mathcal{G}_{\eta(\mathrm{Sch})}$. The sign of $\mathcal{G}_{\eta(\mathrm{RN})}$ 
therefore determines whether the charged black hole is globally preferred 
relative to the reference thermal background in the canonical ensemble.
\par
To elucidate the thermodynamic role of the Gibbs free energy, 
we rewrite the first law in Eq.~\eqref{1st fixed eta} 
as the total differential of $\mathcal{G}_{\eta(\mathrm{RN})}$:
\begin{equation}
\displaystyle
d\mathcal{G}_{\eta(\mathrm{RN})} 
= d\!\left(M - T_{\eta, \mathrm{H}(\mathrm{RN})} 
S_{\eta, \mathrm{BH}(\mathrm{RN})}\right)
= - S_{\eta, \mathrm{BH}(\mathrm{RN})} \, dT_{\eta, \mathrm{H}(\mathrm{RN})} 
+ \Phi_{(\mathrm{RN})} \, dQ,
\end{equation}
from which the conjugate variables follow:
\begin{equation}
\displaystyle
\Phi_{(\mathrm{RN})} 
= \left( \frac{\partial \mathcal{G}_{\eta(\mathrm{RN})}}{\partial Q} \right)_{T_{\eta, \mathrm{H}(\mathrm{RN})}}, 
\qquad 
S_{\eta, \mathrm{BH}(\mathrm{RN})} 
= - \left( \frac{\partial \mathcal{G}_{\eta(\mathrm{RN})}}{\partial T_{\eta, \mathrm{H}(\mathrm{RN})}} \right)_Q.
\label{Gibbs S}
\end{equation}
\par
It must be emphasized that the thermodynamic interpretation of the reference state is more subtle in the charged case than in the neutral one. For asymptotically flat Schwarzschild black holes, the surrounding spacetime may be regarded as a thermal reservoir characterized solely by a fixed temperature $T$, so that a comparison with hot spacetime is well defined within the canonical ensemble. By contrast, in the charged asymptotically flat case, the background spacetime is assigned a temperature but no electric charge. It therefore cannot be consistently interpreted as a reservoir that simultaneously fixes $(T,Q)$, as required for canonical ensemble consistency.
\par
To address this issue, we introduce an idealized thermodynamic bath with fixed $(T,Q)$ as an auxiliary thermal system. Within this ensemble, global stability is determined by comparison among configurations that share the same temperature and conserved electric charge. Accordingly, a charged black hole should be compared with other configurations in the same $(T,Q)$ sector, rather than with a neutral thermal background~\cite{Burikham:2014gwa}. In particular, a charged black hole is globally stable if its free energy is minimal among all configurations at fixed temperature and electric charge.
\begin{figure}[ht]
\centering
\includegraphics[width=6cm]{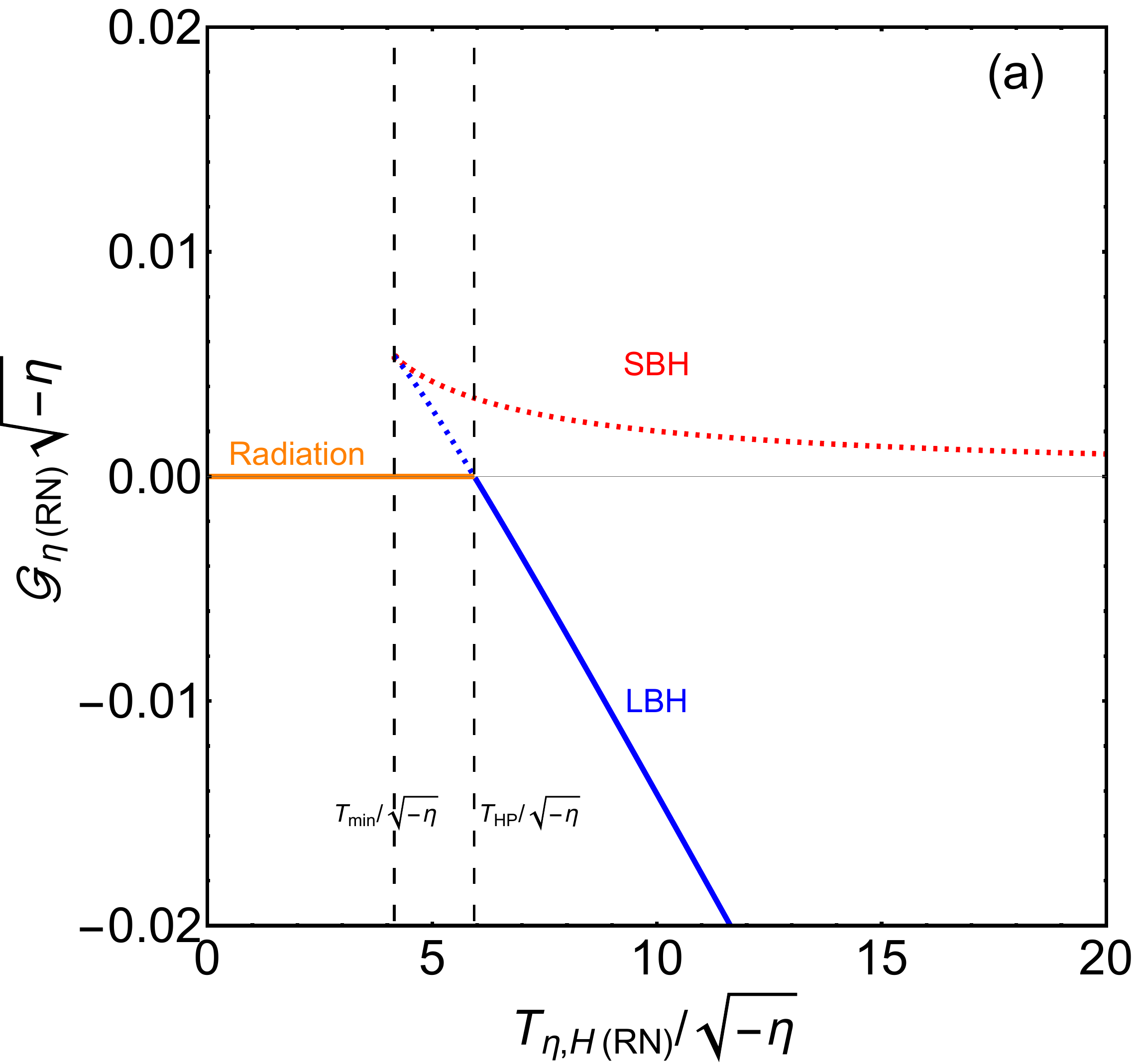}
\qquad
\includegraphics[width=6cm]{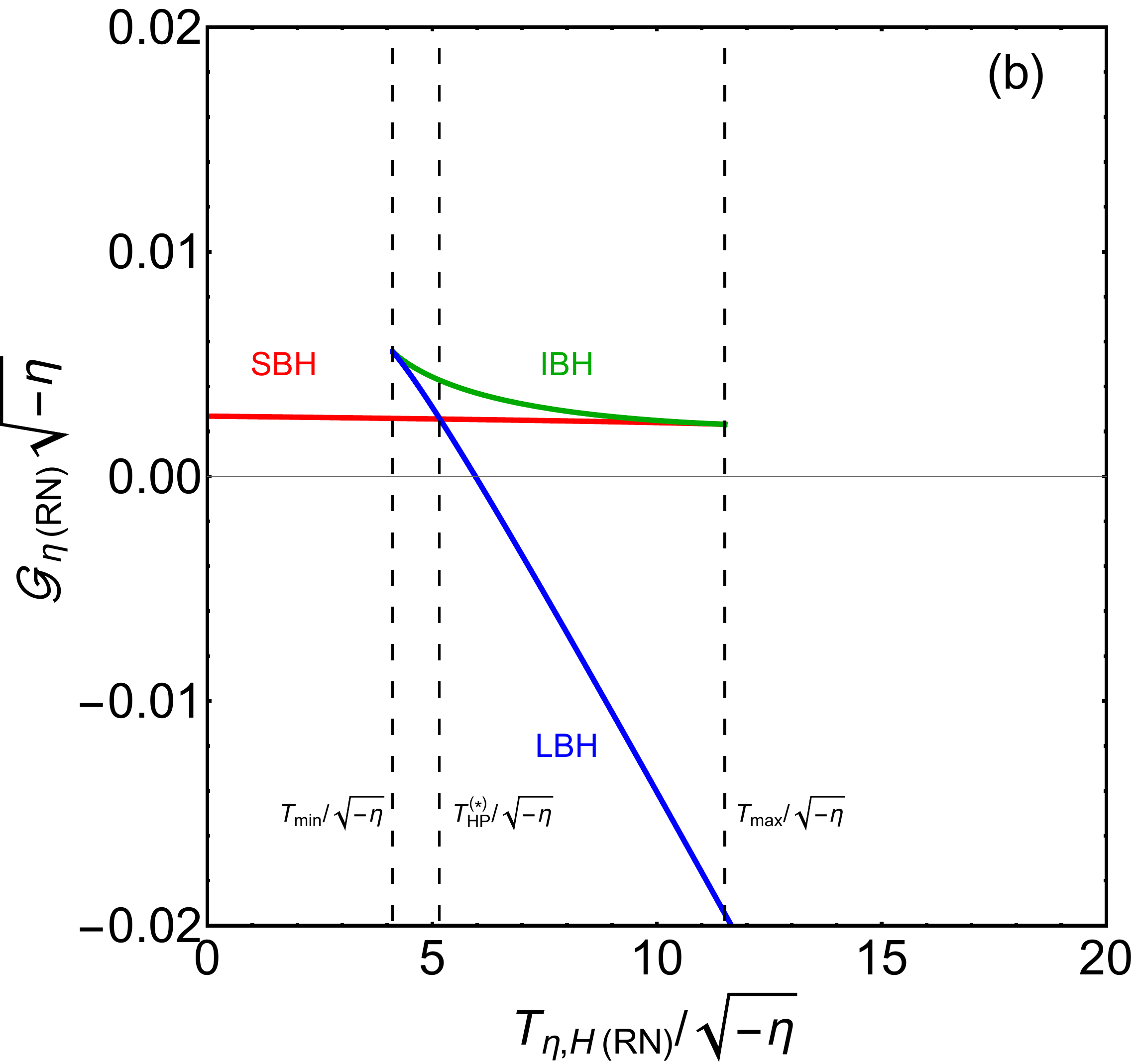}
\qquad
\includegraphics[width=6cm]{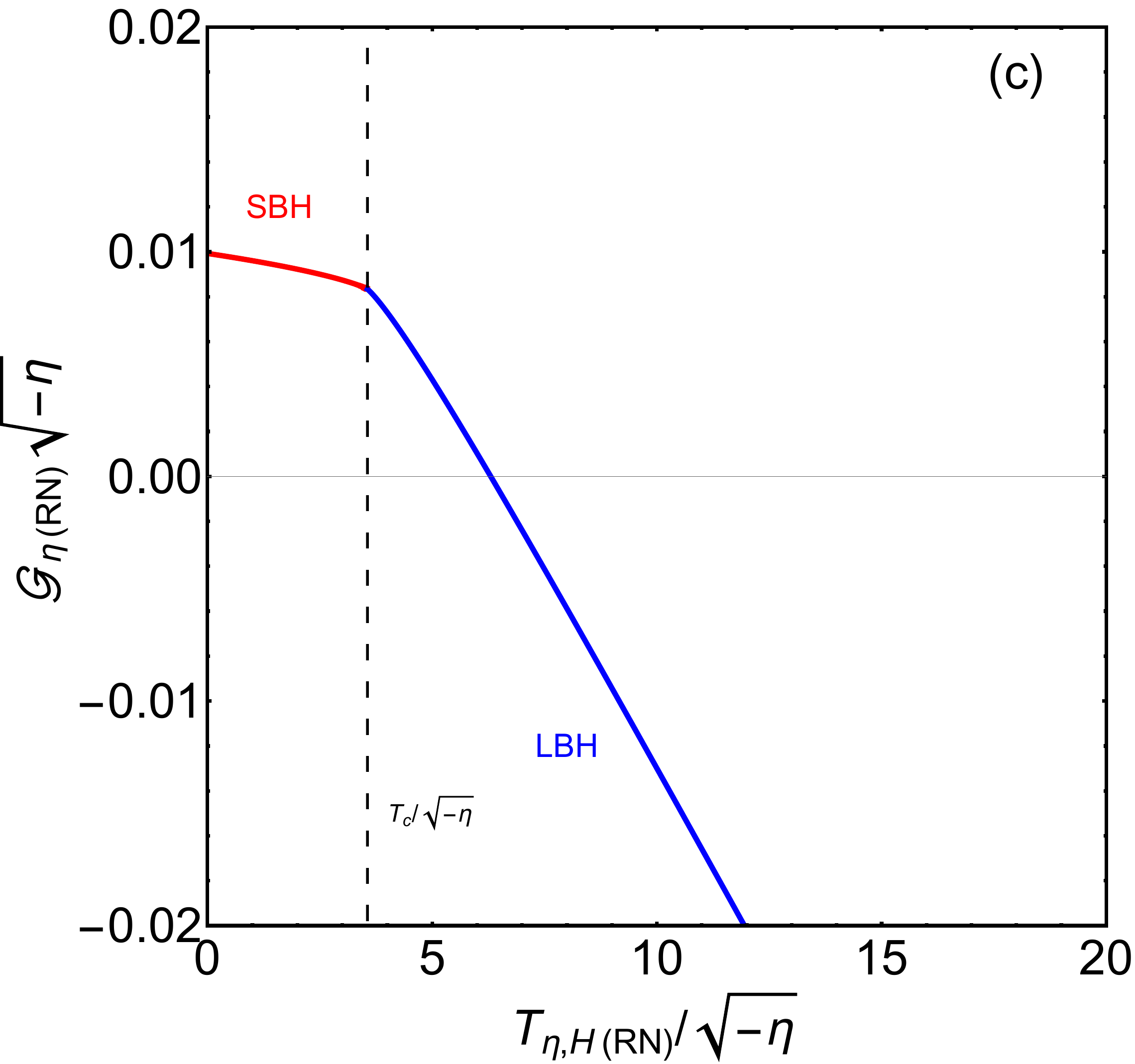}
\qquad
\includegraphics[width=6cm]{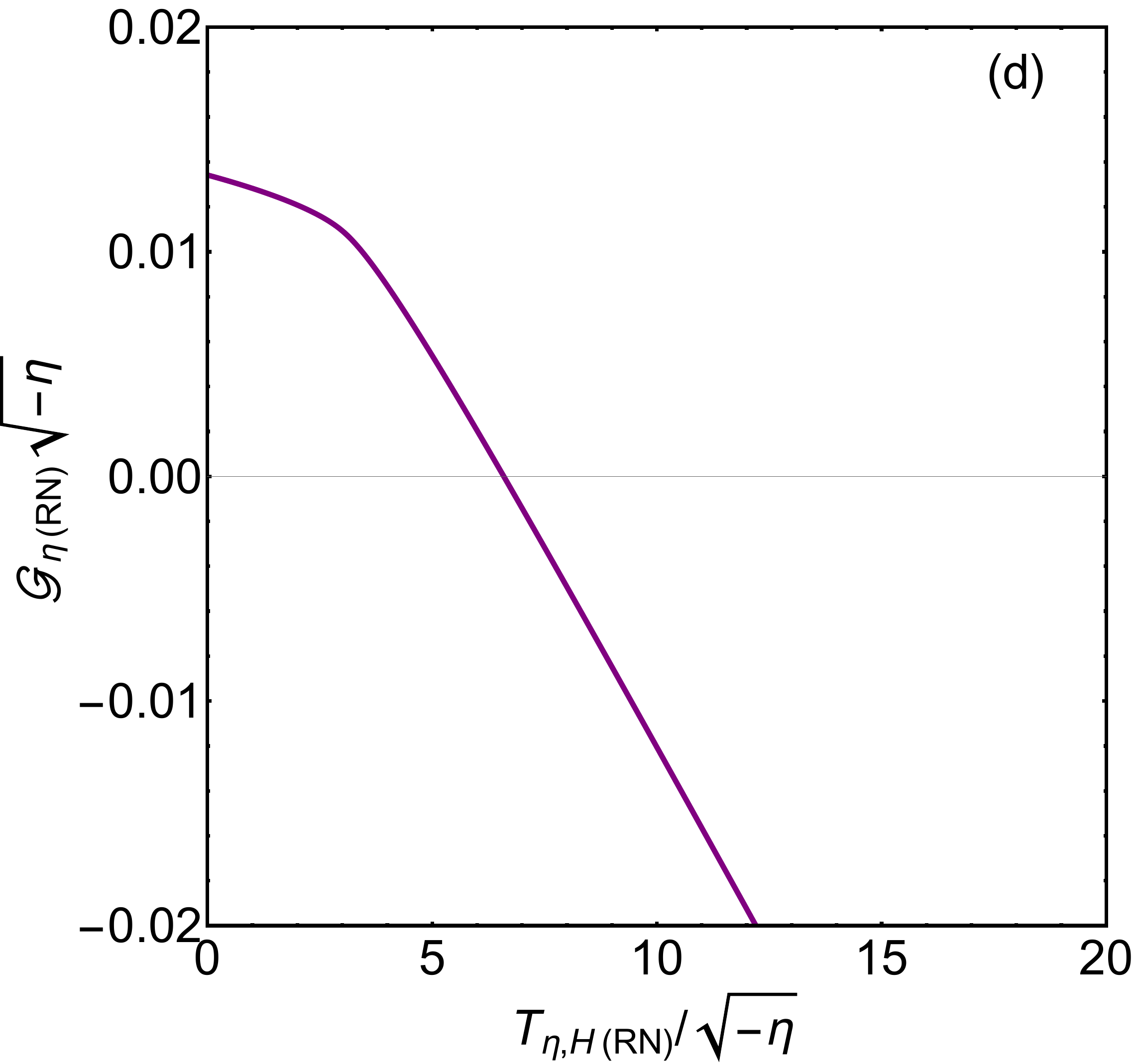}
\caption{Gibbs free energy as a function of temperature for varying $\eta$: (a) non-extensive Sch with $\eta = -0.01, Q=0$; (b)–(d) charged cases at fixed $\eta = -0.002$ with (b) $Q = 0.06 < Q_c$, (c) $Q = Q_c$, and (d) $Q = 0.3 > Q_c$, respectively.}
\label{GT 1}
\end{figure}
\par
It is instructive to analyze the Gibbs free energy $\mathcal{G}_{\eta}$ as a function of the generalized Hawking temperature $T_{\eta,\mathrm{H}}$, as illustrated in Fig.~\ref{GT 1}. In panel~(a), corresponding to the uncharged Schwarzschild case ($Q = 0$), a characteristic cusp structure appears at the critical horizon area $A_{C_Q}$, where the temperature attains a local minimum. This cusp signals the emergence of a non-trivial thermodynamic branch structure induced by the non-extensive deformation. Panel~(a) further distinguishes the thermodynamic branches and the associated Hawking--Page-type transition. The red dashed curve denotes locally unstable configurations with negative heat capacity, while the blue dashed curve represents locally stable but globally unfavorable black holes. Although these latter configurations possess positive heat capacity, their Gibbs free energy exceeds that of the thermal radiation phase (orange horizontal line). As the temperature increases, the system evolves toward states of lower Gibbs free energy and eventually undergoes a first-order phase transition at the Hawking--Page temperature $T_{\mathrm{HP}}$~\cite{Hawking:1982dh}, defined by the intersection of the black hole and radiation branches.
\par
Quantitatively, the deformation-induced Hawking--Page transition in the uncharged case is determined by the degeneracy condition $\mathcal{G}_{\eta(\mathrm{Sch})}=0$. 
Numerically, this yields $A_{\mathcal{G}_{\eta}} \simeq 0.695\,A_{\rm max}$ and, in the small-deformation regime ($|\eta A_{\mathcal{G}_{\eta}}|\ll1$),
\[
T_{\mathrm{HP}} \simeq 5.935\,\sqrt{-\eta}.
\]
As $\eta \to 0^{-}$, the critical temperature vanishes, consistently recovering the classical instability of asymptotically flat Sch black holes.
\par
For charged black holes ($Q \neq 0$), panels~(b)–(d) of Fig.~\ref{GT 1} demonstrate that the interplay between electric charge and non-extensive deformation substantially enriches the phase structure. In contrast to the uncharged case, global stability must now be determined by comparing Gibbs free energies of black hole configurations at fixed charge $Q$, rather than relative to hot spacetime. 
\par
For a richer structure in the case $\eta < 0$, $\mathcal{G}_\eta(T_{\eta,\mathrm{H}})$ develops two cusp-like structures associated with the local extrema of the temperature at $A_{C_{Q(-)}}$ and $A_{C_{Q(+)}}$. At these points, the heat capacity $C_Q$ diverges, signaling local thermodynamic criticality. The region between the extrema corresponds to an IBH branch with negative heat capacity, which separates the locally stable SBH and LBH phases and gives rise to a multi-valued Gibbs free energy profile.
\par
For $Q<Q_c$ [panel~(b)], the system undergoes a first-order \emph{black hole--black hole} phase transition at finite temperature $T_{\rm HP}^{(*)},$ determined by the degeneracy condition
\begin{equation}
\mathcal{G}_{\eta(\mathrm{SBH})}(T_{\rm HP}^{(*)}, Q)=
\mathcal{G}_{\eta(\mathrm{LBH})}(T_{\rm HP}^{(*)}, Q).
\end{equation}
For example, at $Q=0.06<Q_c$ and $\eta=-0.002$, this occurs at $T_{\rm HP}^{(*)}\sqrt{-\eta} \approx 5.16$, marking coexistence of two distinct black hole phases, analogous to a liquid--gas transition.
\par
As $Q\to Q_c$ [panel~(c)], the IBH branch shrinks and the two divergence points merge. 
At $Q=Q_c$, the SBH and LBH branches join smoothly at a single critical temperature $T_c\approx0.161$, indicating genuine second-order critical behavior characterized by a continuous Gibbs free energy and divergent heat capacity.
\par
In the supercritical regime ($Q > Q_c$, panel~(d)), the multi-branch structure disappears entirely, and the Gibbs free energy becomes single-valued and smooth, with the heat capacity remaining finite and no phase transition occurring. The resulting thermodynamic behavior is analogous to that of a supercritical fluid, where distinct phases cease to exist beyond the critical point. This behavior can be understood within the thermodynamic reinterpretation of the system, in which a black hole equation of state emerges with a structure closely analogous to that of ordinary fluids.
\par
In this framework, the electric potential $\Phi_{(\mathrm{RN})}$ and electric charge $Q$ are recast as analogues of thermodynamic volume and pressure, respectively~\cite{Kastor:2009,Dolan:2010ha,Banerjee:2011au,Kubiznak:2012wp,Gunasekaran:2012dq,Poshteh:2013}, while the generalized Hawking temperature $T_{\eta, \mathrm{H}(\mathrm{RN})}$ plays the role of the thermal control parameter governing equilibrium. To further develop this thermodynamic correspondence, one may reformulate the first law of black hole thermodynamics within the fixed-$\eta$ framework. It takes the form
\begin{equation}
\displaystyle
d(M - Q \Phi_{(\mathrm{RN})}) = T_{\eta, \mathrm{H}(\mathrm{RN})} \, dS_{\eta, \mathrm{BH}(\mathrm{RN})} - Q \, d\Phi_{(\mathrm{RN})},
\end{equation}
which directly parallels the standard thermodynamic relation
\begin{equation}
\displaystyle
dU = T \, dS - P \, dV,
\end{equation}
with the effective internal energy identified as
\begin{equation}
\displaystyle
U_{Q} = M - Q \Phi_{(\mathrm{RN})}.
\end{equation}
To formalize this mapping further, this correspondence yields a natural thermodynamic dictionary:  
\begin{equation}
\displaystyle
Q \leftrightarrow P, \qquad \Phi_{(\mathrm{RN})} \leftrightarrow V,
\end{equation}
where \( Q \) plays the role of an effective pressure and \( \Phi_{(\mathrm{RN})} \) functions as a thermodynamic volume. This mapping allows the black hole system to be studied using familiar techniques from classical fluid thermodynamics, including the identification of critical points, phase coexistence, and first- or second-order transitions.
\begin{figure}[ht]
\centering
\includegraphics[width=6cm]{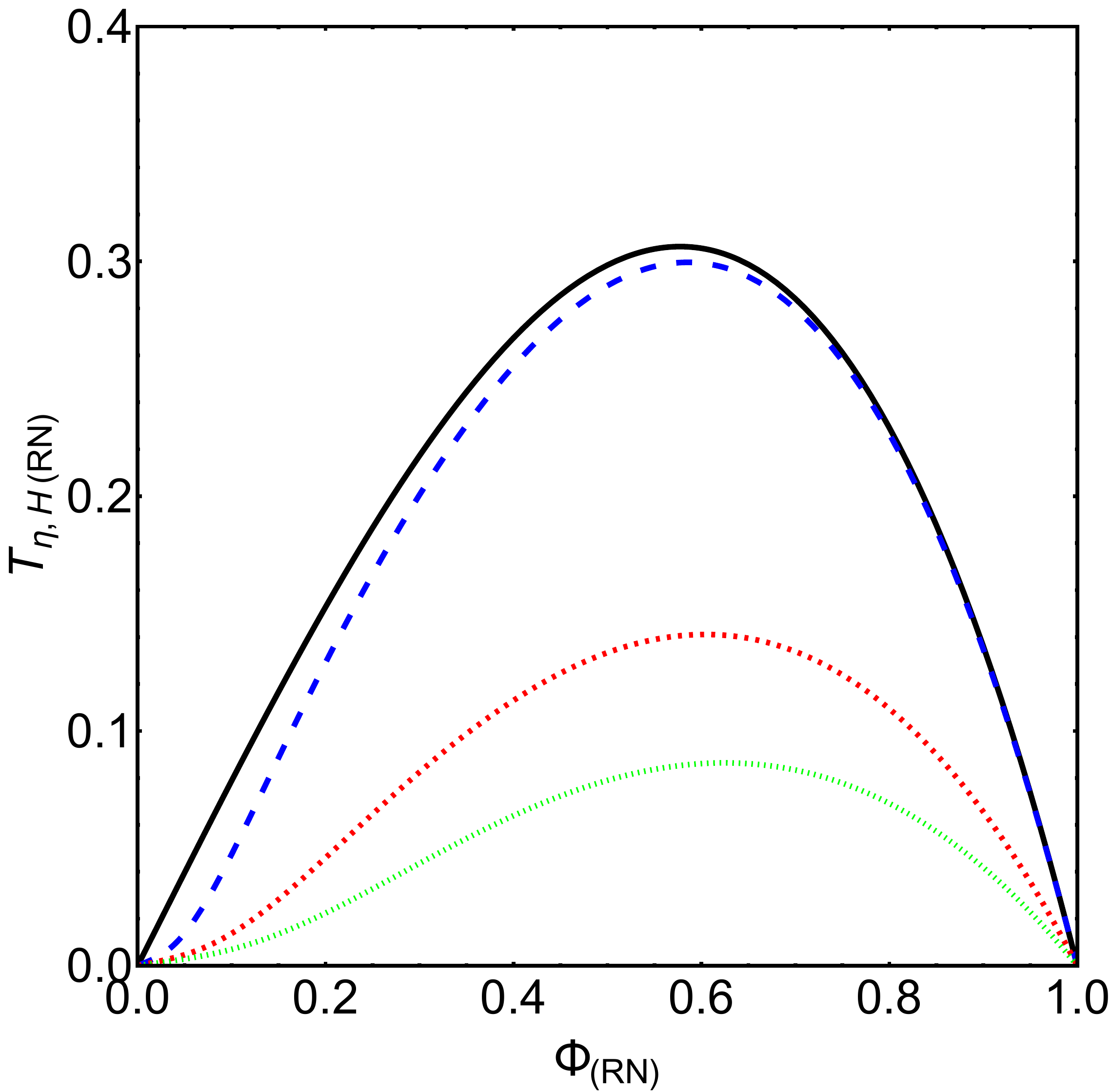} 
\qquad
\includegraphics[width=6cm]{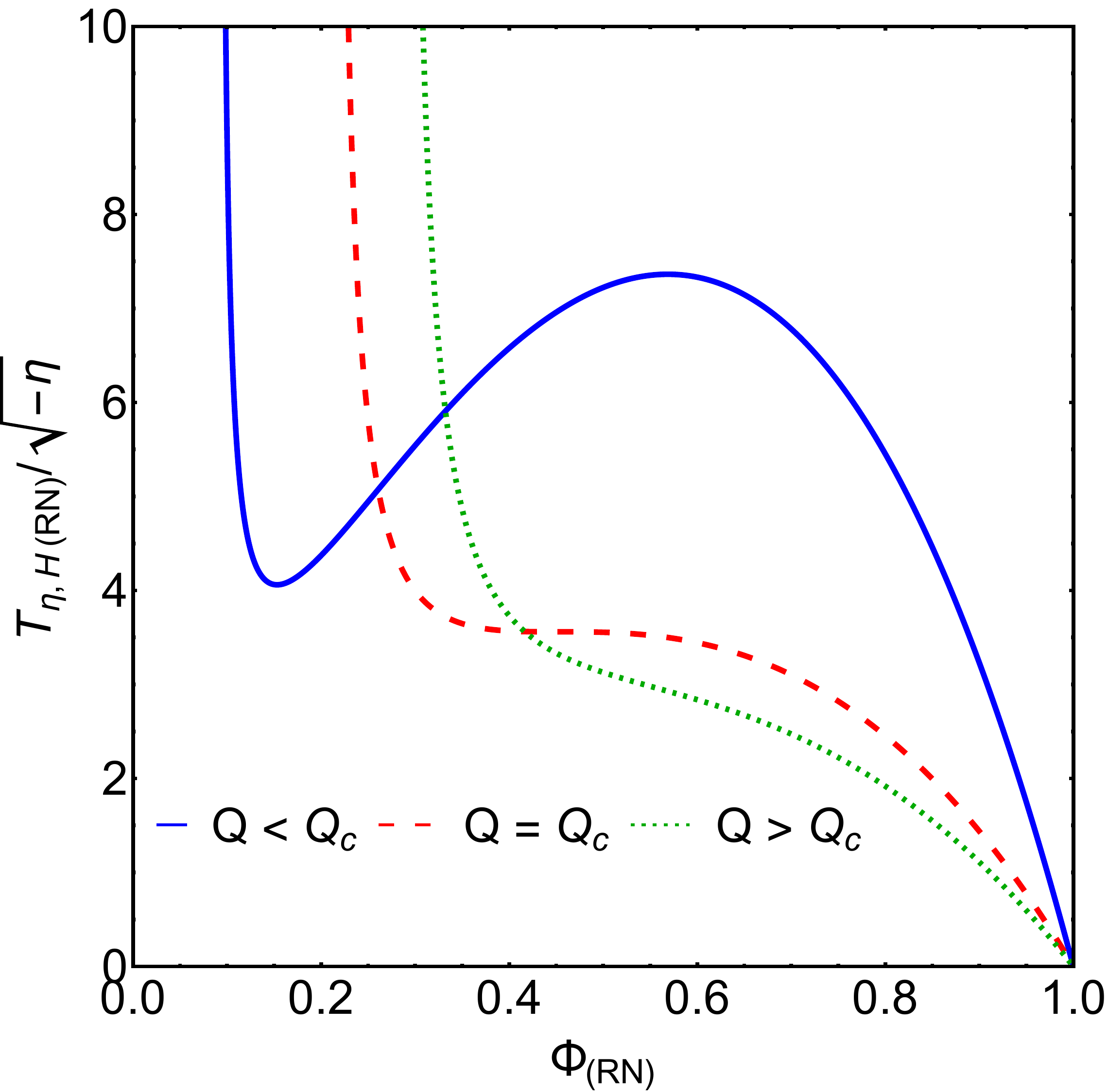}
\caption{Isobaric profiles in the \( T_{\eta, \mathrm{H(RN)}}\text{--}\Phi_{(\mathrm{RN})} \) plane at fixed charge \( Q \), illustrating the thermodynamic behavior of the Tsallis-deformed black hole fluid. Left panel: undeformed case (\( \eta = 0 \)) for \( Q = 0.1 \) (black solid), alongside deformed isobars with \( \eta = 0.002 \) for \( Q = 0.1 \) (blue dashed), \( 0.2 \) (red dotted), and \( 0.3 \) (green tiny-dotted). Right panel: profiles for \( \eta = -0.002 \), showing \( Q = 0.095 < Q_{c} \) (blue solid), \( Q = Q_{c} \) (red dashed), and \( Q = 0.3 > Q_{c} \) (green dotted), highlighting critical behavior in the sub-extensive regime.}
\label{eos1}
\end{figure}
\par
In the fluid reinterpretation of the black hole, the equation of state is given by 
\(T_{\eta,\rm BH} = T_{\eta,\rm BH}(\Phi, Q)\). 
Using Eq.~\eqref{charge Phi 1}, the generalized Hawking temperature can be rewritten in terms of \(Q\), \(\Phi_{(\mathrm{RN})}\), and the deformation parameter \(\eta\) as
\begin{equation}
\displaystyle
T_{\eta, \mathrm{H}(\mathrm{RN})} = 
\frac{
\Phi_{(\mathrm{RN})} (1 - \Phi_{(\mathrm{RN})}^{2}) \left(12 \pi \eta Q^{2} + \Phi_{(\mathrm{RN})}^{2}\right) }
{4 \pi Q \left(144 \pi \eta Q^{2} + \Phi_{(\mathrm{RN})}^{2}\right)} \exp \left[\frac{11 \pi \eta Q^{2}}{12 \pi \eta Q^{2} + \Phi_{(\mathrm{RN})}^{2}} \right].
\label{eos}
\end{equation}
Due to its transcendental form, Eq.~\eqref{eos} cannot be explicitly inverted to yield,
\begin{equation}
\displaystyle
Q = Q\!\left(T_{\eta, \mathrm{H}(\mathrm{RN})}, \, \Phi_{(\mathrm{RN})}\right),
\nonumber
\end{equation}
thereby precluding a conventional analytic expression for the equation of state. Instead, we analyze the behavior of the temperature as a function of the electric potential at fixed charge. In analogy with ordinary fluids, curves at constant charge \(Q\) may be regarded as isobaric curves. These isobars are displayed in the \( T_{\eta, \mathrm{H}(\mathrm{RN})}\text{--}\Phi_{(\mathrm{RN})} \) plane in Fig.~\ref{eos1}, providing a visual representation of the thermodynamic response of the black hole system under variations of the charge and deformation parameter.
\par
For $\eta > 0$, as shown in the left panel of Fig.~\ref{eos1}, the temperature profile 
$T_{\eta,\mathrm{H(RN)}}(\Phi_{(\mathrm{RN})})$ 
exhibits only a single extremum for all values of the charge. 
This isolated maximum does not give rise to any non-trivial phase structure, and we therefore do not consider this case further. In contrast, for $\eta < 0$, illustrated in the right panel of Fig.~\ref{eos1}, 
the temperature curve can develop two distinct extrema when $Q < Q_c$, 
leading to a multi-branch structure analogous to that of a VdW fluid. This behavior signals the presence of non-trivial critical phenomena in the sub-extensive regime. For representative values, when $Q = 0.095 < Q_c$, the temperature profile exhibits two extrema, 
separating stable and metastable (spinodal-like) branches. 
At the critical charge $Q = Q_c$, these extrema merge into a single inflection point, terminating the coexistence curve. 
For $Q = 0.3 > Q_c$, the system enters a supercritical regime characterized by smooth thermodynamic behavior 
without phase separation.
\par
As the charge increases toward the critical value, the two extrema merge into a single inflection point. 
The critical point is therefore defined by the simultaneous vanishing of the first and second derivatives of the generalized Hawking temperature with respect to the electric potential at fixed $Q$:
\begin{equation}
\displaystyle
\left(\frac{\partial T_{\eta,\mathrm{H(RN)}}}{\partial \Phi_{(\mathrm{RN})}}\right)_Q = 0,
\qquad
\left(\frac{\partial^{2} T_{\eta,\mathrm{H(RN)}}}{\partial \Phi_{(\mathrm{RN})}^{2}}\right)_Q = 0.
\label{crit}
\end{equation}
\par
The value $Q = Q_c$ satisfying these conditions marks the onset of a continuous (second-order) phase transition, analogous to the critical point in classical fluid systems. At this point, the two extrema of the temperature profile coalesce into a single inflection point, and the associated thermodynamic response functions diverge, signaling the emergence of genuine critical phenomena in the Tsallis-deformed charged black hole fluid.
\par
A direct computation yields the first derivative of the generalized Hawking temperature with respect to the electric potential:
\begin{equation}
\displaystyle
\left(\frac{\partial T_{\eta, \mathrm{H(RN)}}}{\partial \Phi_{(\mathrm{RN})}}\right)_{Q} 
= \frac{\mathcal{P}_{\eta}^{(1)}(\Phi_{(\mathrm{RN})})}{4 \pi Q \,\bigl(12 \pi \eta Q^{2} + \Phi_{(\mathrm{RN})}^{2}\bigr)\, \bigl(144 \pi \eta Q^{2} + \Phi_{(\mathrm{RN})}^{2}\bigr)^{2}} 
\exp\!\left[\frac{11 \pi \eta Q^{2}}{12 \pi \eta Q^{2} + \Phi_{(\mathrm{RN})}^{2}}\right],
\label{crit 1}
\end{equation}
where $\mathcal{P}_{\eta}^{(1)}(\Phi_{(\mathrm{RN})})$ is an even polynomial in $\Phi_{(\mathrm{RN})}$ of degree~8, given by
\begin{equation}
\displaystyle
\mathcal{P}_{\eta}^{(1)}(\Phi_{(\mathrm{RN})}) = \sum_{k=0}^{4} \mathcal{B}_{k+1}\, \Phi_{(\mathrm{RN})}^{2k},
\end{equation}
with coefficients
\[
\begin{aligned}
\mathcal{B}_{1} &= 20736 \pi^{3} \eta^{3} Q^{6}, 
& \mathcal{B}_{2} &= 144 \pi^{2} \eta^{2} Q^{4} \left(25 - 432 \pi \eta Q^{2}\right), \\
\mathcal{B}_{3} &= 10 \pi \eta Q^{2} \left(41 - 1080 \pi \eta Q^{2}\right), 
& \mathcal{B}_{4} &= 1 - 746 \pi \eta Q^{2}, \\
\mathcal{B}_{5} &= -3.
\end{aligned}
\]
\par
Similarly, a further differentiation yields the second derivative of the generalized Hawking temperature with respect to the electric potential:
\begin{equation}
\displaystyle
\left(\frac{\partial^{2} T_{\eta, \mathrm{H(RN)}}}{\partial \Phi_{(\mathrm{RN})}^{2}}\right)_{Q} 
= - \frac{\Phi_{(\mathrm{RN})} \, \mathcal{P}_{\eta}^{(2)}(\Phi_{(\mathrm{RN})})}{2 \pi Q \,\bigl(12 \pi \eta Q^{2} + \Phi_{(\mathrm{RN})}^{2}\bigr)^{3}\, \bigl(144 \pi \eta Q^{2} + \Phi_{(\mathrm{RN})}^{2}\bigr)^{3}} 
\exp\!\left[\frac{11 \pi \eta Q^{2}}{12 \pi \eta Q^{2} + \Phi_{(\mathrm{RN})}^{2}}\right],
\label{crit 2}
\end{equation}
where $\mathcal{P}_{\eta}^{(2)}(\Phi_{(\mathrm{RN})})$ is an even polynomial in $\Phi_{(\mathrm{RN})}$ of degree~12, given by
\begin{equation}
\displaystyle
\mathcal{P}_{\eta}^{(2)}(\Phi_{(\mathrm{RN})}) = \sum_{k=0}^{6} \mathcal{C}_{k+1}\, \Phi_{(\mathrm{RN})}^{2k},
\end{equation}
with coefficients
\[
\begin{aligned}
\mathcal{C}_{1} &= 1289945088\, \pi^{6} \eta^{6} Q^{12}, 
& \mathcal{C}_{2} &= -103680\, \pi^{4} \eta^{4} Q^{8} \left(121 - 4320\, \pi \eta Q^{2}\right), \\
\mathcal{C}_{3} &= -432\, \pi^{3} \eta^{3} Q^{6} \left(3025 - 190656\, \pi \eta Q^{2}\right), 
& \mathcal{C}_{4} &= -2\, \pi^{2} \eta^{2} Q^{4} \left(24805 - 3401208\, \pi \eta Q^{2}\right), \\
\mathcal{C}_{5} &= \pi \eta Q^{2} \left(121 + 237194\, \pi \eta Q^{2}\right), 
& \mathcal{C}_{6} &= 1371\, \pi \eta Q^{2}, \\
\mathcal{C}_{7} &= 3.
\end{aligned}
\]
\par
Accordingly, the criticality condition in Eq.~\eqref{crit} is equivalent to solving the pair of polynomial equations
\begin{equation}
\displaystyle
\mathcal{P}_{\eta}^{(1)} (\Phi_{(\mathrm{RN})}) = 0, 
\qquad 
\mathcal{P}_{\eta}^{(2)} (\Phi_{(\mathrm{RN})}) = 0,
\label{crit 3}
\end{equation}
which encode the simultaneous vanishing of the first and second derivatives of the generalized Hawking temperature with respect to the electric potential, thereby identifying the precise locations of the critical points in the sub-extensive regime.
\par
Although closed-form analytic solutions are generally inaccessible due to the high-degree polynomials involved, high-precision numerical evaluation allows for the accurate determination of the critical parameters. For instance, for a deformation parameter \( \eta = -0.002 \), the critical point is found to be
\[
T_{c} \approx 0.159206,
\qquad
Q_{c} \approx 0.221952, 
\qquad 
\Phi_{c} \approx 0.451425,
\]
which marks the onset of VdW–like criticality in the sub-extensive Tsallis-deformed black hole system. This critical point corresponds to the merging of two second-order extrema in the temperature profile, delineating the transition between SBH and LBH phases.
\par
Having established the Tsallis-deformed equation of state and its critical structure, we now proceed to a deeper examination of the system’s mechanical stability. In classical thermodynamics, such stability is quantified by the isothermal compressibility, which measures the system's response to infinitesimal pressure fluctuations at fixed temperature. As a thermodynamic susceptibility, the isothermal compressibility provides a natural diagnostic for the onset of phase transitions and the breakdown of stability within the black hole ensemble.
\par
In standard thermodynamics, mechanical stability requires a positive isothermal compressibility, 
\begin{equation}
\displaystyle
k_{T} = - \frac{1}{V} \left(\frac{\partial V}{\partial P}\right)_{T},
\end{equation}
where the negative sign ensures that an increase in pressure leads to a decrease in volume. A negative compressibility indicates anomalous mechanical behavior, signaling instability. By analogy, black hole thermodynamics admits a similar criterion: under the thermodynamic reinterpretation \( Q \leftrightarrow P \) and \( \Phi_{(\mathrm{RN})} \leftrightarrow V \), the isothermal compressibility of the black hole “fluid” is defined as
\begin{equation}
\displaystyle
\kappa_{T} = - \frac{1}{\Phi_{(\mathrm{RN})}} \left( \frac{\partial \Phi_{(\mathrm{RN})}}{\partial Q} \right)_{T_{\eta, \mathrm{H}(\mathrm{RN})}},
\end{equation}
which characterizes the response of the system to infinitesimal charge fluctuations at fixed generalized Hawking temperature. Positive values of \(\kappa_{T}\) correspond to mechanically stable configurations, whereas negative values indicate instability, in which small perturbations in the effective pressure \(Q\) lead to runaway changes in the conjugate volume \(\Phi_{(\mathrm{RN})}\). Divergences of \(\kappa_{T}\) coincide with extrema of the temperature profile, marking the spinodal boundaries of the black hole phase diagram and establishing a direct link between mechanical response and the temperature landscape, analogous to classical fluids.
\par
A practical evaluation of the isothermal compressibility can be performed using the cyclic relation
\begin{equation}
\displaystyle
\left(\frac{\partial \Phi_{(\mathrm{RN})}}{\partial Q}\right)_{T_{\eta, \mathrm{H}(\mathrm{RN})}} 
\left(\frac{\partial Q}{\partial T_{\eta, \mathrm{H}(\mathrm{RN})}}\right)_{\Phi_{(\mathrm{RN})}} 
\left(\frac{\partial T_{\eta, \mathrm{H}(\mathrm{RN})}}{\partial \Phi_{(\mathrm{RN})}}\right)_{Q} 
= -1,
\end{equation}
which allows the compressibility to be expressed in terms of temperature derivatives:
\begin{equation}
\displaystyle
\kappa_{T_{\eta, \mathrm{H}(\mathrm{RN})}} 
= \frac{1}{\Phi_{(\mathrm{RN})}} 
\left(\frac{\partial T_{\eta, \mathrm{H}(\mathrm{RN})}}{\partial Q}\right)_{\Phi_{(\mathrm{RN})}} 
\left(\frac{\partial T_{\eta, \mathrm{H}(\mathrm{RN})}}{\partial \Phi_{(\mathrm{RN})}}\right)^{-1}_{Q} 
= \frac{(1 - \Phi_{(\mathrm{RN})}^{2}) \, \mathcal{Q}_{\eta(1)} (\Phi_{(\mathrm{RN})})}
{\mathcal{Q}_{\eta(2)} (\Phi_{(\mathrm{RN})})}.
\label{q compress 1}
\end{equation}
This formulation ties the isothermal compressibility directly to the previously analyzed temperature profile, thereby integrating mechanical stability with thermal behavior. 
It enables a clear and quantitative identification of instability onsets via the behavior of $\kappa_{T}$, consistently reproducing the phase structure inferred from temperature extrema and heat capacity. In particular, divergences of $\kappa_{T}$ delineate spinodal boundaries separating metastable and unstable branches, providing a direct analogue of the classical fluid picture within the Tsallis-deformed black hole ensemble. 
By linking compressibility, temperature extrema, and phase stability, this framework establishes a unified thermodynamic description of black hole phase structure.
\par
Here, the numerator polynomial is defined as
\begin{equation}
\displaystyle
\mathcal{Q}_{\eta(1)}(\Phi_{(\mathrm{RN})}) :=
\sum_{k=0}^{4} \mathcal{D}_{k+1}\, \Phi_{(\mathrm{RN})}^{2k},
\end{equation}
with coefficients
\[
\begin{aligned}
\mathcal{D}_{1} &= -20736\, \pi^{3} \eta^{3} Q^{7}, &
\mathcal{D}_{2} &= -144\, \pi^{2} \eta^{2} Q^{5} \left( 25 - 432\, \pi \eta Q^{2} \right), \\
\mathcal{D}_{3} &= -10\, \pi \eta Q^{3} \left( 41 - 1080\, \pi \eta Q^{2} \right), &
\mathcal{D}_{4} &= -Q \left( 1 - 746\, \pi \eta Q^{2} \right), &
\mathcal{D}_{5} &= 3 Q .
\end{aligned}
\]
\par
The denominator polynomial is defined as
\begin{equation}
\displaystyle
\mathcal{Q}_{\eta(2)}(\Phi_{(\mathrm{RN})}) :=
\sum_{k=0}^{4} \mathcal{E}_{k+1}\, \Phi_{(\mathrm{RN})}^{2k},
\end{equation}
with coefficients
\[
\begin{aligned}
\mathcal{E}_{1} &= 20736\, \pi^{3} \eta^{3} Q^{6}, &
\mathcal{E}_{2} &= \pi^{2} \eta^{2} Q^{4} \left(3600 - 20736\, \pi \eta Q^{2}\right), \\
\mathcal{E}_{3} &= \pi \eta Q^{2} \left(410 - 3600\, \pi \eta Q^{2}\right), &
\mathcal{E}_{4} &= 1 - 410\, \pi \eta Q^{2}, &
\mathcal{E}_{5} &= -1 .
\end{aligned}
\]
\begin{figure}[ht]
\centering
\includegraphics[width=6cm]{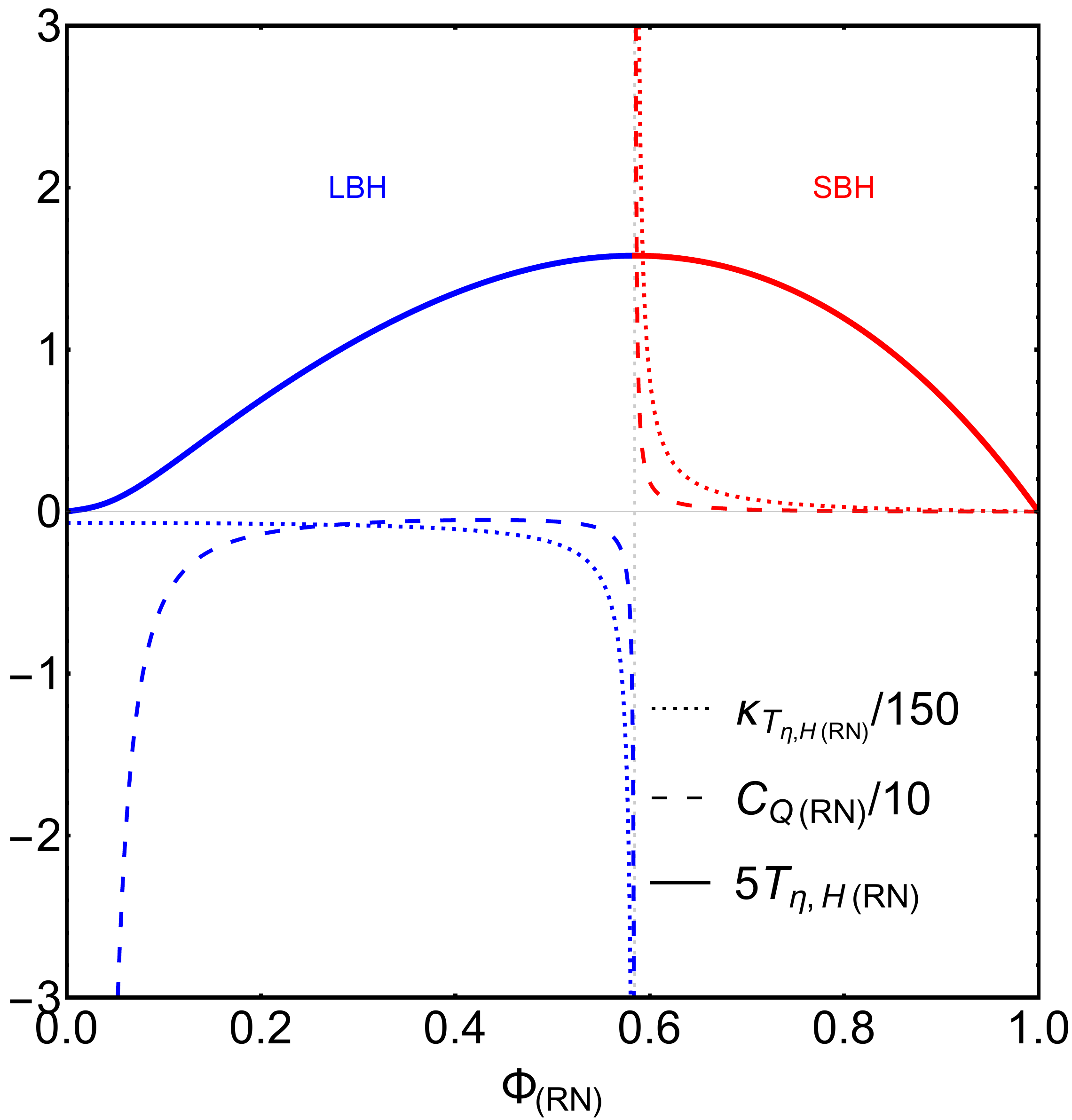}
\caption{Profiles of the isothermal compressibility, heat capacity, and generalized Hawking temperature as functions of the electric potential \( \Phi_{(\mathrm{RN})} \) at fixed electric charge \( Q = 0.095 \). The Tsallis deformation parameter is set to \( \eta = 0.002 \), corresponding to the super-extensive regime.}
\label{mech stable 1}
\end{figure}
\par
Fig.~\ref{mech stable 1} presents the isothermal compressibility across the Tsallis parameter space, plotted alongside the corresponding heat capacity profiles at fixed electric charge. In the super-extensive regime (\( \eta > 0 \)), the SBH branch exhibits both positive compressibility and positive heat capacity, indicating thermodynamic stability. By contrast, the LBH branch develops negative compressibility together with negative heat capacity, signaling the onset of both mechanical and thermal instabilities. This behavior can be understood from Eq.~\eqref{q compress 1}, where the isothermal compressibility scales as 
\(\kappa_{T_{\eta, \rm H}} \sim \left(\partial T_{\eta, \rm H}/\partial \Phi_{(\rm RN)}\right)^{-1}\). Similarly, the heat capacity is governed by the same underlying derivative structure. 
Consequently, divergences in the heat capacity coincide with those of the isothermal compressibility, consistently marking the spinodal boundaries of the system.
\par
\begin{figure}[ht]
\centering
\includegraphics[width=6cm]{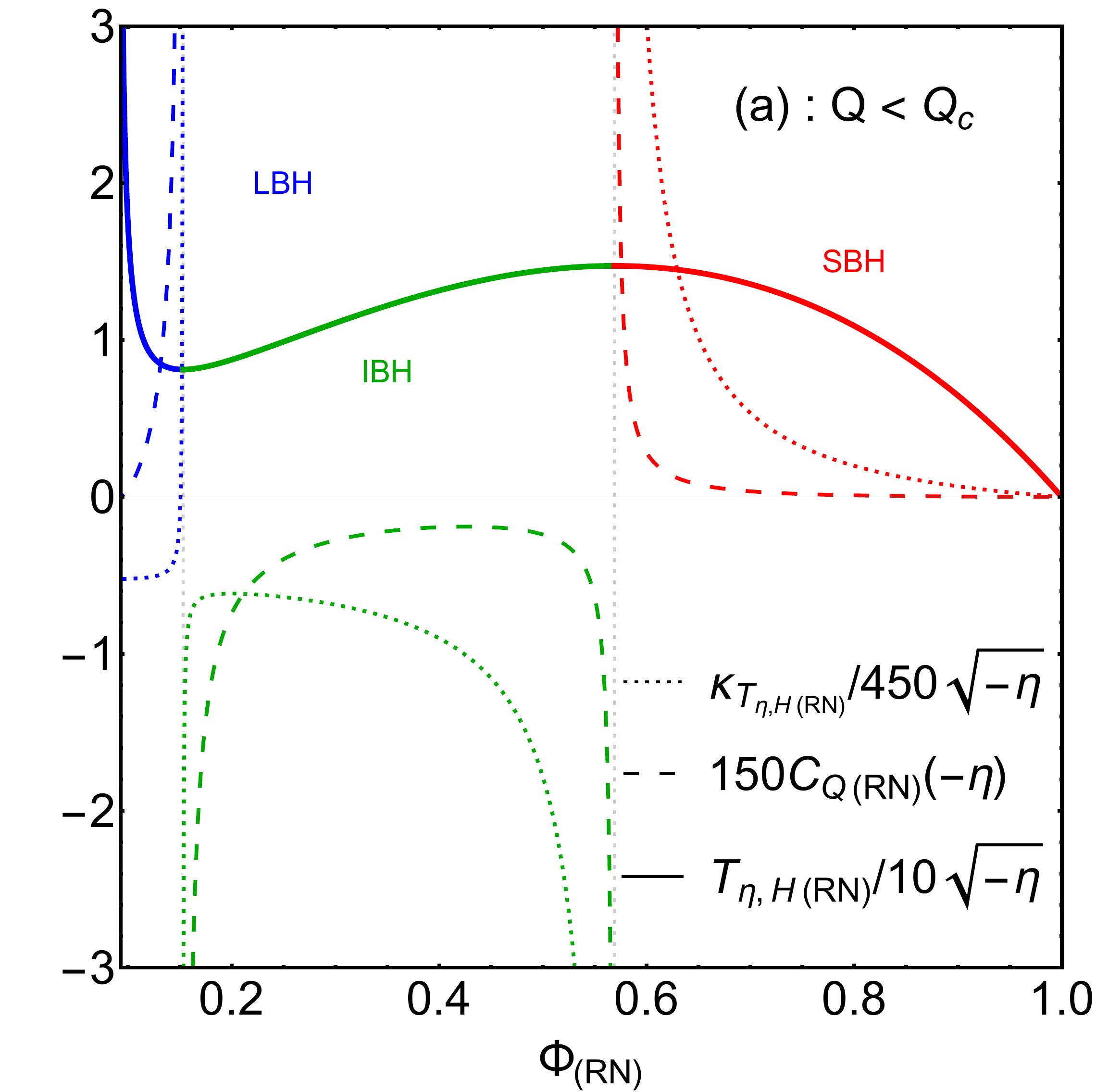}
\quad
\includegraphics[width=6cm]{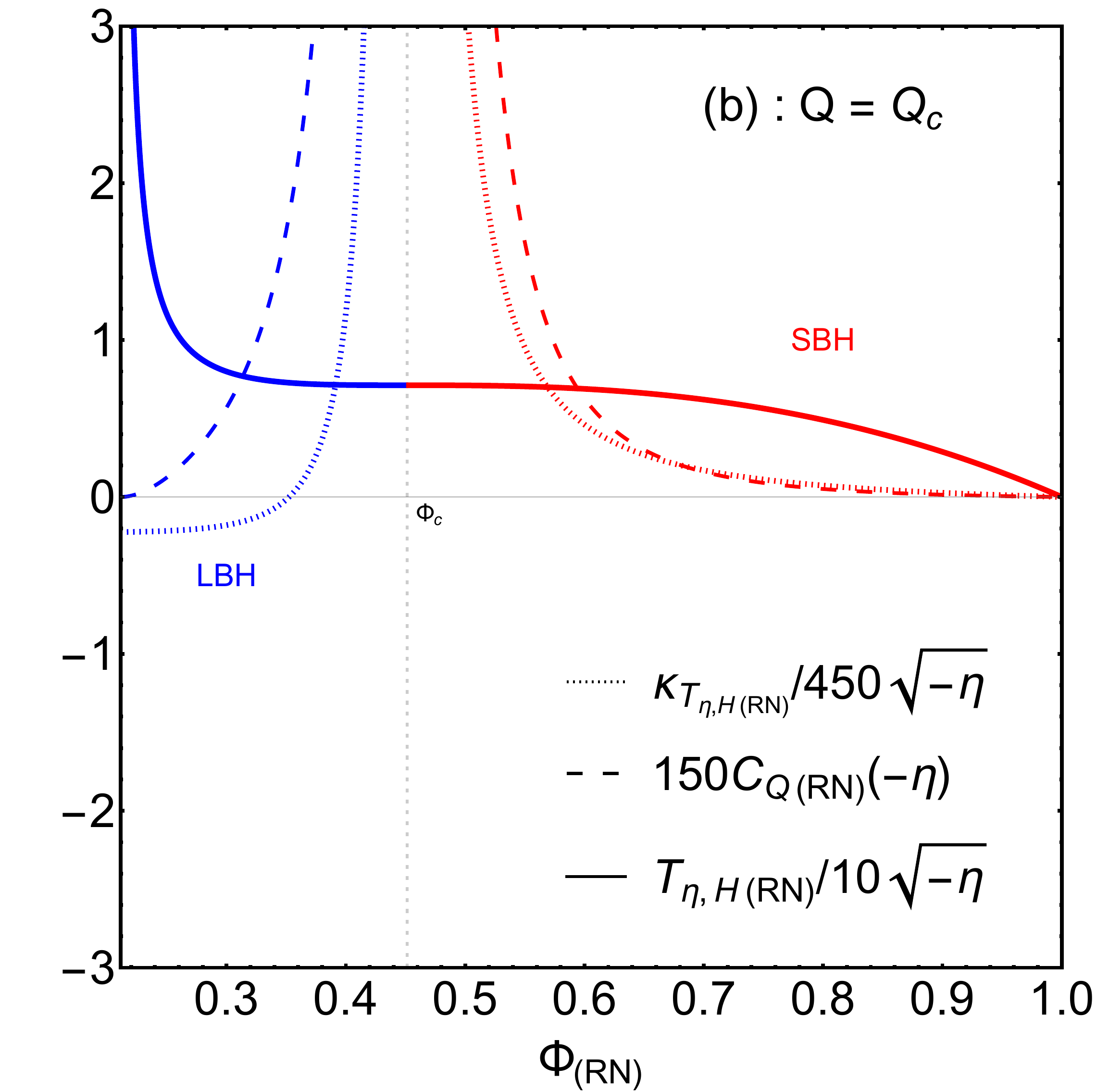}
\quad
\includegraphics[width=6cm]{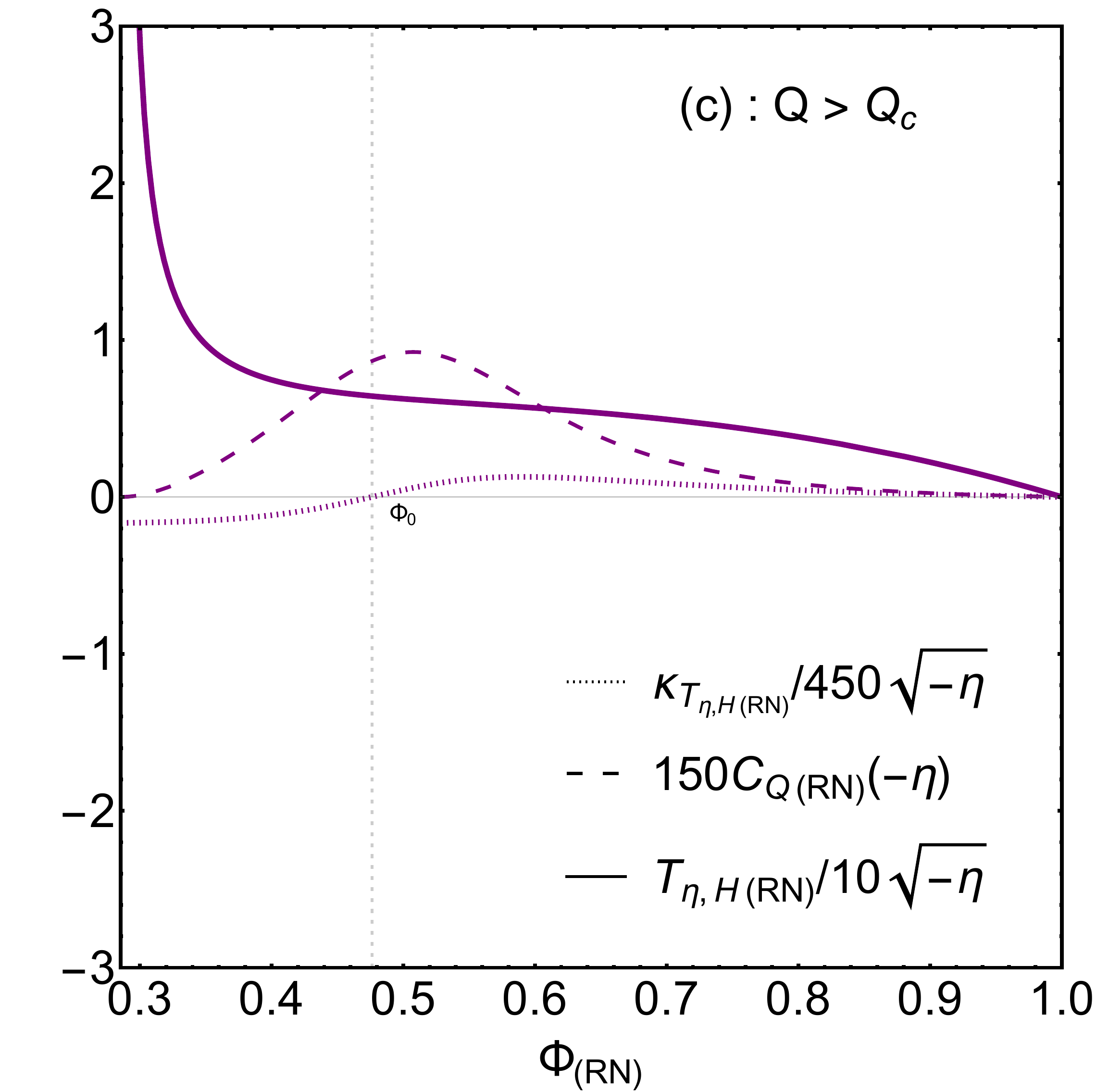}
\caption{Profiles of isothermal compressibility, heat capacity, and Hawking temperature as functions of electric potential \( \Phi \), evaluated at fixed charge \( Q = 0.095 < Q_c \) (a), \( Q = Q_c \) (b), and \( Q = 0.3 > Q_c \) (c), for Tsallis deformation parameter \( \eta = -0.002 \). The plots highlight critical scaling and divergent response near \( Q_c \).}
\label{mech stable 2}
\end{figure}

The sub-extensive regime (\( \eta < 0 \)) is shown in Fig.~\ref{mech stable 2}. For \( Q = 0.095 < Q_{c} \) (panel~(a)), the temperature profile has two extrema—a local minimum and a local maximum. In this case, the LBH branch has positive heat capacity and predominantly negative isothermal compressibility, except near the boundary separating the SBH and LBH branches. This indicates that the LBH is thermally stable but mechanically unstable. The IBH branch is unstable in both respects, while the SBH branch remains stable. For \( Q = Q_{c} \) and \( Q = 0.3 > Q_{c} \) (panels~(b) and (c)), the black hole is locally stable, both thermally and mechanically, for \(\Phi > \Phi_{c}\) and \(\Phi > \Phi_{0}\), respectively, where \(\Phi_{0}\) is determined from the condition \(\kappa_{T_{\eta, \mathrm{H}}} = 0\).
\par
At subcritical charges, the region with negative compressibility indicates mechanical instability and corresponds to the phase transition region. 
This behavior is similar to that of ordinary fluids undergoing a first-order phase transition, showing a clear connection between compressibility and phase structure.
\par
The emergence of a region with negative isothermal compressibility 
(\(\kappa_{T_{\eta, \rm H}} < 0\)) signals mechanical instability, indicating that the corresponding branch does not represent an equilibrium state. In this regime, the system cannot remain in a single homogeneous phase and instead separates into two distinct equilibrium phases. The mechanically unstable segment of the temperature curve is therefore replaced by a coexistence line connecting the two stable endpoints, representing the transition between these phases. 
This replacement is implemented via the Maxwell equal-area construction. At the onset of phase coexistence, the temperature remains constant at \(T^{(*)}_{\mathrm{HP}}\), with the corresponding charge \(Q^{(*)}\), forming a horizontal plateau in the \(T_{\eta, \mathrm{H}}-\Phi\) diagram. 
This behavior is directly analogous to the flat segment of the VdW isotherm describing liquid--gas equilibrium in conventional fluids.
\par
To determine the precise equilibrium points along the coexistence line, we impose the condition of equal Gibbs free energy between the coexisting phases. We begin with the thermodynamic relation for the charged canonical-like ensemble
\begin{equation}
\displaystyle
d\mathcal{G}_{\eta}
= - S_{\eta,\mathrm{BH}}\, dT_{\eta,\mathrm H}
+ \Phi\, dQ.
\end{equation}
At fixed charge (\(dQ=0\)), this reduces to
\begin{equation}
\displaystyle
d\mathcal{G}_{\eta}
= - S_{\eta,\mathrm{BH}}\, dT_{\eta,\mathrm H}.
\end{equation}
Along the coexistence curve, the temperature can be expressed as \(T_{\eta,\mathrm H}=T_{\eta,\mathrm H}(\Phi,Q)\), so that
\begin{equation}
\displaystyle
\left(\frac{\partial \mathcal{G}_{\eta}}{\partial \Phi}\right)_{Q}
=
- S_{\eta,\mathrm{BH}}
\left(\frac{\partial T_{\eta,\mathrm H}}{\partial \Phi}\right)_{Q}.
\end{equation}
Integrating between the two coexistence points \(\Phi_{\rm LBH}\) and 
\(\Phi_{\rm SBH}\), we obtain
\begin{equation}
\displaystyle
\Delta \mathcal{G}_{\eta}
=
- \int_{\Phi_{\rm LBH}}^{\Phi_{\rm SBH}}
d\Phi \,
S_{\eta,\mathrm{BH}}(\Phi)
\left(\frac{\partial T_{\eta,\mathrm H}}{\partial \Phi}\right)_{Q}.
\end{equation}
At phase equilibrium, the Gibbs free energies of the SBH and LBH branches are equal, implying \(\Delta \mathcal{G}_{\eta}=0\). 
This condition leads to
\begin{equation}
\displaystyle
\int_{\Phi_{\rm LBH}}^{\Phi_{\rm SBH}}
d\Phi \,
S_{\eta,\mathrm{BH}}(\Phi)
\left(\frac{\partial T_{\eta,\mathrm H}}{\partial \Phi}\right)_{Q}
= 0,
\label{Maxwell-equal-area-1}
\end{equation}
which represents the Maxwell equal-area construction in the present variables, ensuring phase coexistence between the SBH and LBH states. This condition replaces the mechanically unstable region with a thermodynamically consistent coexistence configuration.
\par
We may equivalently reformulate the Maxwell condition by changing variables from the electric potential \(\Phi\) to the generalized black hole entropy \(S_{\eta,\rm BH}\) along the coexistence branches, where \(S_{\eta,\rm BH}(\Phi)\) is single-valued at fixed charge. Using the relation
\begin{equation}
\displaystyle
\left(\frac{\partial T_{\eta,\rm H}}{\partial \Phi}\right)_Q d\Phi
= dT_{\eta,\rm H},
\end{equation}
the integral in Eq.~\eqref{Maxwell-equal-area-1} can be rewritten as
\begin{equation}
\displaystyle
\int_{\Phi_{\rm LBH}}^{\Phi_{\rm SBH}} 
d\Phi \,
S_{\eta,\rm BH}(\Phi)
\left(\frac{\partial T_{\eta,\rm H}}{\partial \Phi}\right)_Q
=
\int S_{\eta,\rm BH}\, dT_{\eta,\rm H}.
\end{equation}
Applying integration by parts yields
\begin{align}
\int S_{\eta,\rm BH}\, dT_{\eta,\rm H}
&=
S_{\eta,\rm BH} T_{\eta,\rm H}
\Big|_{\Phi_{\rm LBH}}^{\Phi_{\rm SBH}}
-
\int_{S_{\rm LBH}}^{S_{\rm SBH}}
T_{\eta,\rm H}(S_{\eta,\rm BH})\, dS_{\eta,\rm BH}
\nonumber\\
&=
T^{(*)}_{\rm HP}
\left(S_{\rm SBH}-S_{\rm LBH}\right)
-
\int_{S_{\rm LBH}}^{S_{\rm SBH}}
T_{\eta,\rm H}(S_{\eta,\rm BH})\, dS_{\eta,\rm BH},
\end{align}
where we used 
\(T_{\eta,\rm H}(\Phi_{\rm SBH})
=
T_{\eta,\rm H}(\Phi_{\rm LBH})
=
T^{(*)}_{\rm HP}\). This leads to the standard Maxwell equal-area condition in the \(T_{\eta, \rm H}\text{--}S_{\eta, \rm BH}\) plane:
\begin{equation}
\displaystyle
\int_{S_{\rm LBH}}^{S_{\rm SBH}} 
dS_{\eta, \rm BH} 
\left[ T_{\eta, \rm H}(S_{\eta, \rm BH}) - T^{(*)}_{\rm HP} \right] = 0,
\label{Maxwell equal-area 3}
\end{equation}
which expresses the equal-area construction in terms of entropy, providing a more direct thermodynamic interpretation of the coexistence condition.
\par
The coexistence potentials \(\Phi_{\rm LBH}\) and \(\Phi_{\rm SBH}\) 
are obtained by evaluating the equation of state on the respective branches:
\begin{eqnarray}
\displaystyle
T^{(*)}_{\rm HP} &=& 
\frac{\Phi_{\rm LBH} (1 - \Phi_{\rm LBH}^2) 
(12 \pi \eta Q^2 + \Phi_{\rm LBH}^2)}
{4 \pi Q (144 \pi \eta Q^2 + \Phi_{\rm LBH}^2)} \exp\!\left[\frac{11 \pi \eta Q^2}
{12 \pi \eta Q^2 + \Phi_{\rm LBH}^2} \right], \\
\displaystyle
T^{(*)}_{\rm HP} &=& 
\frac{\Phi_{\rm SBH} (1 - \Phi_{\rm SBH}^2) 
(12 \pi \eta Q^2 + \Phi_{\rm SBH}^2)}
{4 \pi Q (144 \pi \eta Q^2 + \Phi_{\rm SBH}^2)} \exp\!\left[\frac{11 \pi \eta Q^2}
{12 \pi \eta Q^2 + \Phi_{\rm SBH}^2} \right].
\end{eqnarray}
Together with Eq.~\eqref{Maxwell equal-area 3}, these relations form a coupled transcendental system for 
\(T^{(*)}_{\rm HP}\), \(\Phi_{\rm LBH}\), and \(\Phi_{\rm SBH}\), which cannot be solved analytically. For a representative choice of parameters in the sub-extensive regime, 
\(\eta = -0.002\) and \(Q = 0.095 < Q_c\), numerical evaluation yields
\[
T^{(*)}_{\rm HP} \approx 0.213434, \quad
\Phi_{\rm LBH} \approx 0.11814, \quad
\Phi_{\rm SBH} \approx 0.83548.
\]
These values provide an explicit illustration of the coexistence point determined by the Maxwell construction.
\begin{figure}[ht]
\centering
\includegraphics[width=6cm]{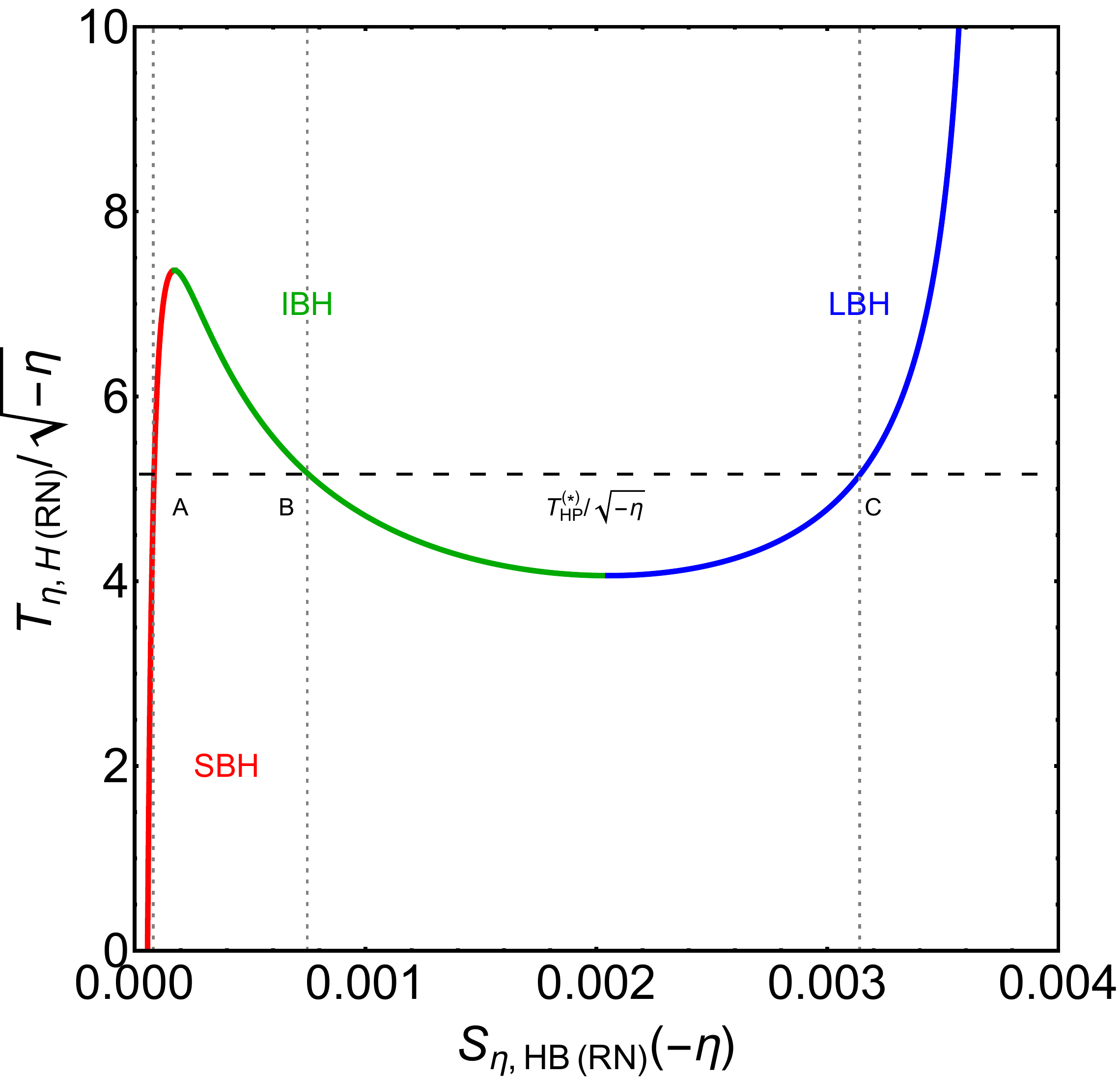}
\caption{Profiles of the isobaric curves on the \( T_{\eta, \mathrm{H}}{-}S_{\eta, \rm BH} \) plane at fixed charge \( Q = 0.095 < Q_{c} \) and Tsallis deformation \( \eta = -0.002 \).}
\label{Maxwell area 1}
\end{figure}
\par
In the $T_{\eta,\rm H}-S_{\eta,\rm BH}$ plane (shown in Fig.~\ref{Maxwell area 1}), 
the equation-of-state curve intersects the coexistence isotherm 
$T^{(*)}_{\rm HP}$ at three points satisfying
\(
S_{A} < S_{B} < S_{C}.
\)
Point $A$ lies on the lower locally stable branch (SBH), point $C$ on the upper stable branch (LBH), while $B$ belongs to the locally unstable branch (IBH). The segment between $A$ and $C$ therefore contains an intermediate unstable region around $B$, which does not correspond to an equilibrium state. The Maxwell construction replaces this entire segment by a horizontal coexistence line at $T^{(*)}_{\rm HP}$, connecting the two stable endpoints $A$ and $C$. At this temperature, the SBH and LBH phases coexist in thermodynamic equilibrium, with equal Gibbs free energies,
\(
\mathcal{G}_{\eta (\rm SBH)} 
=
\mathcal{G}_{\eta (\rm LBH)}.
\)
Numerically, the areas above ($A \to B$) and below ($B \to C$) the isotherm cancel to high precision ($\lesssim 10^{-16}$), confirming the equal-area law. This geometric condition is equivalent to the equality of the Gibbs free energies of the two coexisting phases.
\par
Motivated by Eq.~\eqref{Maxwell equal-area 3}, we define the 
reversible heat exchanged in a general $i \to j$ transition as
\begin{equation}
\displaystyle
Q_{ij} = \int_{S_{i}}^{S_{j}} T_{\eta,\rm H}(S_{\eta, \rm BH}) \, dS_{\eta, \rm BH},
\end{equation}
and introduce the deviation from the coexistence isotherm 
in terms of an \emph{excess heat},
\begin{equation}
\displaystyle
\Delta Q_{ij} = \int_{S_{i}}^{S_{j}} 
\Big[T_{\eta,\rm H}(S_{\eta, \rm BH}) - T^{(*)}_{\rm HP} \Big] \, dS_{\eta, \rm BH}.
\end{equation}
The Maxwell construction can then be expressed compactly as
\begin{equation}
\displaystyle
0 = \int_{S_{A}}^{S_{C}} 
\Big[ T_{\eta,\rm H}(S_{\eta, \rm BH}) - T^{(*)}_{\rm HP} \Big] \, dS_{\eta, \rm BH},
\end{equation}
which guarantees that the net excess heat over the coexistence region vanishes. 
This condition enforces thermodynamic equilibrium between the SBH and LBH branches at $T^{(*)}_{\rm HP}$.
\par
Since $S_{A} < S_{B} < S_{C}$, both entropy increments 
$S_{B}-S_{A}$ and $S_{C}-S_{B}$ are strictly positive. 
The sign of $\Delta Q_{ij}$ is determined by whether 
$T_{\eta,\rm H}(S_{\eta, \rm BH})$ lies above or below the coexistence isotherm.
\par
\begin{itemize}
\item For $A \to B$, the equation-of-state curve lies above the 
coexistence isotherm, and therefore
\begin{equation}
\displaystyle
\Delta Q_{AB} > 0,
\end{equation}
signifying a net heat absorption as the SBH approaches 
the intermediate unstable configuration.
\par
\item For $B \to C$, the equation-of-state curve lies below the 
coexistence isotherm, hence
\begin{equation}
\displaystyle
\Delta Q_{BC} < 0,
\end{equation}
corresponding to a net heat release as the system moves 
toward the LBH branch.
\end{itemize}
\par
By construction,
\begin{equation}
\displaystyle
\Delta Q_{AB} + \Delta Q_{BC} = 0,
\end{equation}
which encapsulates the balance of excess heat and establishes 
exact phase equilibrium. In physical terms, the heat absorbed 
along $A \to B$ is precisely compensated by the heat released 
along $B \to C$. Thermodynamically, the SBH absorbs heat from an external reservoir at fixed $(T_{\rm HP}^{(*)}, Q)$ in order to lower its Gibbs free energy, while the LBH releases heat to the same reservoir until its Gibbs free energy decreases to the same equilibrium value. This interplay guarantees coexistence at $T_{\rm HP}^{(*)}$.
\par
Altogether, this analysis demonstrates that the introduction of the Tsallis non-extensive parameter \(\eta\) leads to a non-trivial modification of the thermodynamic structure of charged black holes. In particular, the generalized entropy induces the emergence of a novel phase structure absent in the standard RN case, characterized by phase coexistence and first-order phase transitions. This modified thermodynamic behavior exhibits a close analogy with that of VdW fluids, allowing for a consistent reinterpretation of black hole thermodynamics in a fluid-like framework. 
Within this picture, quantities such as temperature, electric potential, and charge play roles analogous to their fluid counterparts, and the Maxwell equal-area construction naturally encodes the coexistence of distinct phases.
\par
In the next subsection, we turn to the critical behavior of this system and investigate the scaling properties and critical exponents associated with the phase transition in the presence of non-extensive effects.

\subsection{Scaling Behavior and Universal Properties Near the Critical Point}
\par
Having reformulated black hole thermodynamics within the Tsallis non-extensive framework, we now investigate its critical behavior. 
By fixing the deformation parameter \(\eta\), we consider a specific thermodynamic setting in which the Hawking temperature \(T_{\eta,\mathrm{H}}\), entropy \(S_{\eta,\mathrm{BH}}\), electric potential \(\Phi\), and Gibbs free energy \(\mathcal{G}_{\eta}\) can be consistently analyzed, providing a clear basis for identifying phase transitions.
\par
The resulting equation of state, expressed in either $(Q,\Phi)$ or $(T_{\eta,\mathrm{H}},Q)$ variables, reveals multiple thermodynamic branches analogous to those of VdW fluids. This structure motivates us to study of phase transitions and critical phenomena in the black hole system.
\par
In classical thermodynamics, second-order phase transitions are 
universally characterized by scaling laws of thermodynamic response 
functions~\cite{Fisher1971,Stanley1971,Goldenfeld1992}. A paradigmatic example is the VdW fluid, 
for which the relations near the critical point take the form
\begin{eqnarray}
\displaystyle
P - P_{c} &\sim& |V - V_{c}|^{\delta}, \\
V_{g} - V_{l} &\sim& |T - T_{c}|^{\beta}, \\
C_{V} &\sim& |T - T_{c}|^{-\alpha}, \\
k_{T} &\sim& |T - T_{c}|^{-\gamma},
\end{eqnarray}
where $P$, $V$, and $T$ denote the pressure, volume, and temperature, respectively; $C_V$ is the heat capacity at constant volume; $k_T$ the isothermal compressibility; and $V_g$ and $V_l$ the gas and liquid volumes. The associated mean-field critical exponents are universal \cite{Landau1980,Callen1985,Chaikin1995}
\begin{equation}
\displaystyle
\alpha = 0, \qquad \beta = \frac{1}{2}, \qquad \gamma = 1, \qquad \delta = 3.
\end{equation}
They depend solely on global symmetries and dimensionality, rather than on microscopic details~\cite{Wilson:1974mb,Binney1992}. This universality strongly motivates the search for analogous scaling behavior in black hole thermodynamics, where the precise microscopic degrees of freedom remain unknown but the macroscopic behavior suggests deep structural parallels with fluid systems.
\par
Motivated by the universality of critical exponents in conventional fluids, it is natural to cast the Tsallis-deformed black hole thermodynamic variables into a dimensionless form that makes the emergence of critical behavior transparent. To this end, we introduce the \textit{dimensionless reduced variables}:
\begin{equation}
\displaystyle
\bar{T} := \frac{T_{\eta, \rm H}}{T_{c}} - 1, \qquad 
\bar{Q} := \frac{Q}{Q_{c}} - 1, \qquad 
\bar{\Phi} := \frac{\Phi}{\Phi_{c}} - 1,
\label{reduced variable}
\end{equation}
where $(T_{c}, Q_{c}, \Phi_{c})$ denote the corresponding critical values. By construction, the critical point is located at $(\bar{T}_{c}, \bar{Q}_{c}, \bar{\Phi}_{c}) = (0,0,0)$.
\par
The onset of criticality is identified by the inflection point of the equation of state with respect to the order parameter—taken here to be the electric potential \(\Phi\)—at fixed charge \(Q\):
\begin{equation}
\displaystyle
\left(\frac{\partial \bar{T}}{\partial \bar{\Phi}}\right)_{\bar{Q}} = 0, \qquad
\left(\frac{\partial^{2} \bar{T}}{\partial \bar{\Phi}^{2}}\right)_{\bar{Q}} = 0.
\label{eq:crit_condition}
\end{equation}
These conditions define the critical point, corresponding to the merging of the two phases into a single phase. 
They also provide the starting point for a systematic expansion in the vicinity of criticality.
\par
Tsallis-deformed Hawking temperature can be written in terms of the dimensionless variables as
\begin{align}
\displaystyle
\bar{T}(\bar{Q},\bar{\Phi})
&= \frac{1}{T_{c}} \Bigg[\frac{\Phi_{c} (1+\bar{\Phi}) \Big[1 - \Phi_{c}^{2} (1+\bar{\Phi})^{2} \Big] \Big[12 \pi \eta Q_{c}^{2} (1+\bar{Q})^{2} + \Phi_{c}^{2} (1+\bar{\Phi})^{2}\Big]}{4 \pi Q_{c} (1+\bar{Q}) \Big[144 \pi \eta Q_{c}^{2} (1+\bar{Q})^{2} + \Phi_{c}^{2} (1+\bar{\Phi})^{2}\Big]} \nonumber\\
\displaystyle
&\quad \times \exp\left[ \frac{11 \pi \eta Q_{c}^{2} (1+\bar{Q})^{2}}{12 \pi \eta Q_{c}^{2} (1+\bar{Q})^{2} + \Phi_{c}^{2} (1+\bar{\Phi})^{2}} \right] \Bigg] - 1.
\label{eos dimensionless}
\end{align}
Expanding $\bar{T}(\bar{Q}, \bar{\Phi})$ as a bivariate Taylor series about $(\bar{Q}_{c}, \bar{\Phi}_{c}) = (0,0)$ yields
\begin{align}
\displaystyle
\bar{T}(\bar{Q}, \bar{\Phi}) &= a_{00} + a_{10} \bar{Q} + a_{01} \bar{\Phi} 
+ \frac{1}{2} a_{20} \bar{Q}^{2} + a_{11} \bar{Q} \bar{\Phi} + \frac{1}{2} a_{02} \bar{\Phi}^{2} \nonumber \\
\displaystyle
&\quad + \frac{1}{6} a_{30} \bar{Q}^{3} + \frac{1}{2} a_{21} \bar{Q}^{2} \bar{\Phi} 
+ \frac{1}{2} a_{12} \bar{Q} \bar{\Phi}^{2} + \frac{1}{6} a_{03} \bar{\Phi}^{3} + \cdots,
\label{eq:expand_T}
\end{align}
where
\begin{equation}
\displaystyle
a_{mn} := \frac{1}{m! n!} \left. \frac{\partial^{m+n} \bar{T}}{\partial \bar{Q}^{m} \partial \bar{\Phi}^{n}} \right|_{(0,0)},
\end{equation}
and the coefficients depend explicitly on the deformation parameter $\eta$.
\par
Applying the criticality conditions~\eqref{eq:crit_condition} immediately imposes
\begin{equation}
\displaystyle
a_{01} = 0, \qquad a_{02} = 0,
\end{equation}
so that the expansion reduces to
\begin{align}
\displaystyle
\bar{T}(\bar{Q}, \bar{\Phi}) &= a_{00} + a_{10} \bar{Q} + \frac{1}{2} a_{20} \bar{Q}^{2} + a_{11} \bar{Q} \bar{\Phi} 
+ \frac{1}{6} a_{30} \bar{Q}^{3} + \frac{1}{2} a_{21} \bar{Q}^{2} \bar{\Phi} \nonumber\\
\displaystyle
&\quad + \frac{1}{2} a_{12} \bar{Q} \bar{\Phi}^{2} + \frac{1}{6} a_{03} \bar{\Phi}^{3} + \cdots.
\label{eq:expand_T2}
\end{align}
\par
At the critical temperature \(\bar{T}_{c}\), the critical exponent \(\delta\) characterizes how the order parameter (the electric potential \(\Phi\)) responds to variations of the control parameter (the charge \(Q\)). 
In the spirit of Landau theory, we expand the equation of state around criticality under the isothermal condition \(\bar{T} = \bar{T}_{c}\). 
Under this constraint, the equation of state implicitly determines \(\bar{Q}\) as a function of \(\bar{\Phi}\), which can be expanded as
\begin{equation}
\displaystyle
\bar{Q} = \bar{Q}(\bar{\Phi}) = b_{1} \bar{\Phi} + b_{2} \bar{\Phi}^{2} + b_{3} \bar{\Phi}^{3} + \cdots.
\end{equation}
The coefficients \(b_n\) are determined by differentiating the constraint 
\(\bar{T}(\bar{Q}(\bar{\Phi}), \bar{\Phi}) = \bar{T}_c\) with respect to \(\bar{\Phi}\). 
The first derivative gives
\begin{equation}
\displaystyle
\frac{d\bar{T}}{d\bar{\Phi}} 
= \left(\frac{\partial \bar{T}}{\partial \bar{Q}}\right) 
\frac{d \bar{Q}}{d \bar{\Phi}} 
+ \left(\frac{\partial \bar{T}}{\partial \bar{\Phi}}\right) = 0.
\end{equation}
At criticality, the condition 
\(\left(\partial \bar{T}/\partial \bar{\Phi}\right)_{\bar{Q}} = 0\) 
implies \(b_{1} = 0\), eliminating the linear term. 
Taking the second derivative and using 
\(\left(\partial^{2} \bar{T}/\partial \bar{\Phi}^{2}\right)_{\bar{Q}} = 0\) 
yields $b_{2} = 0$, so that the quadratic term also vanishes. 
Thus, the leading non-trivial contribution arises at cubic order, with
\begin{equation}
\displaystyle
b_{3} = -\frac{1}{6}
\left.
\frac{\partial^{3} \bar{T}/\partial \bar{\Phi}^{3}}
{\partial \bar{T}/\partial \bar{Q}}
\right|_{\rm crit}.
\end{equation}
\par
This hierarchy demonstrates that all lower-order contributions vanish at criticality, leaving the cubic non-linearity as the leading term. 
Accordingly, the scaling behavior of \(\bar{Q}(\bar{\Phi})\) near the critical point is
\begin{equation}
\displaystyle
\bar{Q}(\bar{\Phi}) \simeq 
-\frac{1}{6}
\left.
\frac{\partial^{3} \bar{T}/\partial \bar{\Phi}^{3}}
{\partial \bar{T}/\partial \bar{Q}}
\right|_{\rm crit}
\bar{\Phi}^{3},
\quad \Rightarrow \quad 
(Q-Q_{c}) \sim (\Phi-\Phi_{c})^{3}.
\end{equation}
By analogy with a VdW fluid system, where $P-P_{c} \sim |V-V_{c}|^{\delta}$, 
we identify the electric potential \(\Phi\) as the effective volume and the charge \(Q\) as the pressure-like variable. Consequently, the critical exponent is obtained as
\begin{equation}
\displaystyle
\delta = 3.
\end{equation}
This result shows that the black hole system reproduces the mean-field critical exponent \(\delta\), in agreement with that of the VdW fluid.
\par
To proceed further, we exploit the left--right symmetry between the two coexistence branches in the vicinity of the critical point $(\bar{T}_{c}, \bar{Q}_{c}, \bar{\Phi}_{c})$. This symmetry allows the order parameter to be expanded as
\begin{equation}
\displaystyle
\bar{\Phi}_{\rm SBH} = \frac{\epsilon}{\Phi_{c}}, 
\qquad 
\bar{\Phi}_{\rm LBH} = -\frac{\epsilon}{\Phi_{c}},
\label{different Phi 1}
\end{equation}
or equivalently, in the original variables,
\begin{equation}
\displaystyle
\Phi_{\rm SBH} = \Phi_{c} + \epsilon, 
\qquad 
\Phi_{\rm LBH} = \Phi_{c} - \epsilon,
\label{different Phi 2}
\end{equation}
where \(\epsilon\) measures the deviation from the critical point. Near criticality, the two coexisting phases are $\bar{\Phi}_{\rm LBH}$ and $\bar{\Phi}_{\rm SBH}$, with $\bar{\Phi}_{\rm SBH} > \bar{\Phi}_{\rm LBH}$ at fixed $\bar{Q}$ and $\bar{T} = \bar{T}_{c}$. To analyze their coexistence quantitatively, we employ the Maxwell construction near the critical point, which enforces thermodynamic consistency through the equal-area condition,
\begin{equation} 
\displaystyle 
\int_{\bar{\Phi}_{\rm LBH}}^{\bar{\Phi}_{\rm SBH}} d\bar{\Phi} \, S(\bar{\Phi}) \left(\frac{\partial \bar{T}}{\partial \bar{\Phi}}\right)_{\bar{Q}} = 0. 
\label{eq:MaxwellPhi} 
\end{equation} 
We then expand the relevant thermodynamic quantities around the critical point, retaining only the leading-order contributions in $\bar{Q}$:
\begin{eqnarray} 
\displaystyle 
S(\bar{Q}, \bar{\Phi}) &=& S_{c} + \frac{1}{2} \left. \frac{\partial^{3} S}{\partial \bar{Q} \, \partial \bar{\Phi}^{2}} \right|_{(0, 0)} \bar{Q} \bar{\Phi}^{2} + \cdots, \\ 
\displaystyle 
\left(\frac{\partial \bar{T}}{\partial \bar{\Phi}}\right)_{\bar{Q}} &=& \left. \frac{\partial \dot{\bar{T}}}{\partial \bar{Q}} \right|_{(0, 0)} \bar{Q} + \frac{1}{2} \dddot{\bar{T}}(0, 0) \, \bar{\Phi}^{2} + \cdots,
\end{eqnarray} 
where the dot denotes differentiation with respect to $\bar{\Phi}$, and $S_{c} = S_{\eta, \rm BH}(0, 0)$. Substituting the symmetric endpoints \( \bar{\Phi}_{\rm SBH} = \epsilon/\Phi_{c} \) and \( \bar{\Phi}_{\rm LBH} = -\epsilon/\Phi_{c} \) into Eq.~\eqref{eq:MaxwellPhi}, and keeping only the leading-order terms, the integral reduces to
\begin{eqnarray} 
\displaystyle 
0 &\simeq& \frac{1}{6} S_{c} \, \dddot{\bar{T}}(0, 0) \, \bar{\Phi}^{3} \Big|_{-\epsilon/\Phi_{c}}^{\epsilon/\Phi_{c}} 
+ S_{c} \left. \frac{\partial \dot{\bar{T}}}{\partial \bar{Q}} \right|_{(0, 0)} \bar{Q} \, \bar{\Phi} \Big|_{-\epsilon/\Phi_{c}}^{\epsilon/\Phi_{c}}.
\end{eqnarray} 
Due to the symmetric integration limits, contributions from odd functions of \(\bar{\Phi}\) vanish, while even terms survive. To leading order, this yields
\begin{equation} 
\displaystyle 
\bar{\Phi} \simeq \left[-\frac{6 \, \partial_{\bar{Q}} \dot{\bar{T}} (0, 0)}{\dddot{\bar{T}}(0, 0)}\right]^{1/2} \bar{Q}^{1/2}, 
\quad \Rightarrow \quad \bar{\Phi} \sim \bar{Q}^{1/2}, 
\label{Qphi1} 
\end{equation} 
with the requirement that $\dddot{\bar{T}}(0, 0)$ and $\partial_{\bar{Q}} \dot{\bar{T}}(0, 0)$ have opposite signs to ensure a positive real value. Expanding the temperature near $(0,0)$ gives
\begin{equation} 
\displaystyle 
\bar{T} \simeq (\partial_{\bar{Q}} \bar{T}(0, 0)) \, \bar{Q}, 
\end{equation} 
so that
\begin{equation} 
\displaystyle 
\bar{\Phi} \simeq \left[-\frac{6 \, \partial_{\bar{Q}} \dot{\bar{T}} (0, 0)}{\dddot{\bar{T}}(0, 0) \, (\partial_{\bar{Q}} \bar{T}(0, 0))}\right]^{1/2} \bar{T}^{1/2}.
\label{Qphi2} 
\end{equation} 
Using Eqs.~\eqref{different Phi 1} and~\eqref{different Phi 2}, we obtain
\begin{equation} 
\displaystyle 
2\epsilon = \Phi_{\rm SBH} - \Phi_{\rm LBH}, 
\qquad \bar{\Phi} \sim \epsilon 
\quad \Rightarrow \quad \Phi_{\rm SBH} - \Phi_{\rm LBH} \sim |T-T_{c}|^{1/2}.
\end{equation}
Thus, the order parameter vanishes as $|T-T_{c}|^{1/2}$, identifying the critical exponent as
\begin{equation}
\displaystyle
\beta = \frac{1}{2}.
\end{equation}
\par
Physically, $\beta$ characterizes the onset of SBH--LBH separation, in close analogy with the volume contrast between liquid and gas phases near criticality. For $T<T_{c}$, the two branches coexist as distinct metastable states; at $T=T_{c}$, they merge continuously; and for $T>T_{c}$, no separation persists, leaving a single homogeneous phase. 
In this way, the black hole system faithfully reproduces the mean-field universality class.
\par
Next, we examine the heat capacity at fixed charge,
\begin{equation}
\displaystyle
C_{Q} = T_{\eta,\mathrm H}\left(\frac{\partial S_{\eta,\mathrm{BH}}}{\partial T_{\eta,\mathrm H}}\right)_{Q},
\end{equation}
which characterizes the response of the entropy to temperature variations along an isocharged process. Using the reduced temperature $\bar{T}$ defined in Eq.~\eqref{reduced variable}, we expand the Tsallis-deformed entropy near the critical point at fixed $Q$ as
\begin{equation}
\displaystyle
S_{\eta,\mathrm{BH}}(\bar{T}) = S_{c} + s_{1} \bar{T} + s_{2} \bar{T}^{2} + \mathcal{O}(\bar{T}^{3}),
\end{equation}
where $s_{1}$ and $s_{2}$ encode the leading-order response of the entropy to temperature fluctuations. Differentiating with respect to $T_{\eta,\mathrm H}$ gives
\begin{equation}
\displaystyle
\left(\frac{\partial S_{\eta,\mathrm{BH}}}{\partial T_{\eta,\mathrm H}}\right)_{Q} 
= \frac{1}{T_{c}}\left(s_{1} + 2 s_{2} \bar{T} + \mathcal{O}(\bar{T}^{2})\right),
\end{equation}
which remains finite at $\bar{T} = 0$. Consequently, the heat capacity behaves as
\begin{equation}
\displaystyle
C_{Q} = T_{\eta,\mathrm H}\left(\frac{\partial S_{\eta,\mathrm{BH}}}{\partial T_{\eta,\mathrm H}}\right)_{Q} 
= s_{1} + \mathcal{O}(\bar{T}),
\end{equation}
and therefore remains finite at the critical point. In the special case $s_{1}=0$, the leading contribution arises from the quadratic term,
\begin{equation}
\displaystyle
C_{Q} \sim s_{2} \bar{T},
\end{equation}
which vanishes smoothly as $\bar{T} \to 0$, without divergence. Thus, for the Tsallis-deformed RN black hole, the critical exponent $\alpha$, defined through
\begin{equation}
\displaystyle
C_{Q} \sim |T-T_{c}|^{-\alpha},
\end{equation}
is found to be
\begin{equation}
\displaystyle
\alpha = 0,
\end{equation}
indicating that the heat capacity remains finite at the critical point, in agreement with the mean-field universality class.
\par
We now consider the isothermal compressibility at fixed charge, expressed in terms of the reduced variables,
\begin{equation}
\displaystyle
\kappa_{T_{\eta,\mathrm{H}}} := \frac{1}{Q_{c} (\bar{\Phi}+1)} \left(\frac{\partial \bar{\Phi}}{\partial \bar{Q}}\right)_{\bar{T}},
\end{equation}
which quantifies the response of the black hole to small variations in $Q$. Near the critical point, $\kappa_{T_{\eta,\mathrm{H}}}$ diverges, indicating susceptibility-like behavior analogous to that of VdW fluids. Using the equation of state $\bar{T}(\bar{Q}, \bar{\Phi})$, the derivative $(\partial \bar{\Phi}/\partial \bar{Q})_{\bar{T}}$ can be obtained via implicit differentiation at fixed $\bar{T}$:
\begin{equation}
\displaystyle
\left(\frac{\partial \bar{\Phi}}{\partial \bar{Q}}\right)_{\bar{T}} 
= - \frac{\left(\partial \bar{T}/\partial \bar{Q}\right)}{\left(\partial \bar{T}/\partial \bar{\Phi}\right)}.
\end{equation}
Near the critical point, the expansion of the equation of state implies
\begin{equation}
\displaystyle
\left(\frac{\partial \bar{T}}{\partial \bar{\Phi}}\right) \sim \bar{\Phi}^{2},
\end{equation}
while $\left(\partial \bar{T}/\partial \bar{Q}\right)$ remains finite. Consequently,
\begin{equation}
\displaystyle
\left(\frac{\partial \bar{\Phi}}{\partial \bar{Q}}\right)_{\bar{T}} \sim \bar{\Phi}^{-2},
\end{equation}
which diverges as $\bar{\Phi} \to 0$. Along the coexistence line, the order parameter scales as
\begin{equation}
\displaystyle
\Phi - \Phi_{c} \sim |T-T_{c}|^{1/2}.
\end{equation}
Substituting this relation into the definition of $\kappa_{T_{\eta,\mathrm{H}}}$ yields
\begin{equation}
\displaystyle
\kappa_{T_{\eta,\mathrm{H}}} \sim |T-T_c|^{-1},
\end{equation}
so that the corresponding critical exponent is
\begin{equation}
\displaystyle
\gamma = 1.
\end{equation}
\par
Collecting all preceding results, the critical exponents of the Tsallis-deformed RN black hole at fixed $\eta$ are
\begin{equation}
\displaystyle
\alpha = 0, \qquad \beta = \frac{1}{2}, \qquad \gamma = 1, \qquad \delta = 3,
\label{BH critical}
\end{equation}
which exactly coincide with those of the classical VdW universality class. 
This correspondence arises naturally from the VdW fluid reinterpretation of charged black holes, in which the thermodynamic variables $(\Phi, Q)$ play roles analogous to $(V, P)$ in ordinary fluids. 
In this framework, a coexistence line between the SBH and LBH branches emerges near the critical temperature $T_c$, directly analogous to the liquid--gas transition in VdW systems. 
This fluid-like description provides a consistent and systematic framework for applying standard thermodynamic and critical-scaling methods to analyze phase transitions in black hole systems.
\par
Having established the thermodynamic universality of the Tsallis-deformed RN black hole, we now turn to potential observational implications. 
In particular, photon-related quantities—such as the photon sphere radius, orbital periods, and Lyapunov exponents—provide complementary geometric probes of the black hole spacetime. A detailed analysis of these optical observables and their possible connection to the underlying thermodynamic structure will be presented in the next section.

\section{Photon Trajectories and Optical Features of Black Holes}\label{optics}
\par
After establishing the thermodynamic properties, fluid-like reinterpretation, and critical behavior of charged black holes, we now extend the analysis to an observational perspective. In particular, we focus on optical observables arising from photon motion in the strong-gravity regime, which can serve as effective probes of black-hole phase transitions. This approach provides a concrete bridge between theoretical predictions and measurable quantities in high-resolution astrophysical imaging.  
\par
Photons emitted from distant sources follow null geodesics that are strongly deflected by the spacetime curvature, giving rise to distinctive \emph{photon rings} on the observer’s image plane. These sharply defined structures encode information about the near-horizon geometry, as recently highlighted by the EHT observations~\cite{EHTC:2019a, EHTC:2019f, EHTC:2022a, EHTC:2022f}. Beyond their geometric role, photon rings also carry dynamical imprints that reflect the stability properties of the underlying photon trajectories.  
\par
The dynamical characterization of photon rings can be formalized in terms of three fundamental quantities~\cite{Guo:2022, Yang:2023, Lyu:2024, Kumara:2024, Hale:2024lzh, Promsiri:2024hrl}:  
\begin{itemize}
\item The \emph{orbital half-period} $\tau$, which represents the coordinate time required for a photon to complete half of a circular orbit and determines the spacing of sub-rings.  
\item The \emph{angular Lyapunov exponent} $\lambda_{\rm L}$, which quantifies the exponential divergence of nearby trajectories in the angular direction; larger values correspond to stronger instability and sharper ring features.  
\item The \emph{temporal Lyapunov exponent} $\gamma_{\rm L}$, which governs divergence along the temporal direction and controls the escape rate of photons, influencing the temporal broadening of observed signals.  
\end{itemize}
\par
Taken together, the triplet $(\tau, \lambda_{\rm L}, \gamma_{\rm L})$ provides a unified optical framework for characterizing the stability properties of photon dynamics in the vicinity of a black hole. In particular, these parameters govern how photon trajectories behave near the photon sphere and therefore control the dynamical stability of the associated null geodesics. As a result, observable features of black-hole images—such as the brightness, thickness, and fine substructure of photon rings—can be understood as direct manifestations of these stability properties. The morphology of the photon ring thus encodes information about the underlying photon dynamics and, indirectly, about the thermodynamic state of the black hole.
\par
In what follows, we consider analytic expressions for $\tau$, $\lambda_{\rm L}$, and $\gamma_{\rm L}$ to clarify their observational signatures and thermodynamic significance. This framework establishes a quantitative connection between orbital instabilities and variations in thermodynamic quantities, including the Hawking temperature and response functions.

\subsection{Analysis of Critical Parameters near the Photon Sphere}
\par
Building on the preceding discussion of optical probes of black hole thermodynamics, we now establish the theoretical foundations for photon-ring orbits. To this end, we begin by formulating the Lagrangian for photon motion in a general static, spherically symmetric spacetime, thereby providing a framework that links geodesic dynamics to observable image features. The line element introduced in Eq.~\eqref{ds} reads
\begin{equation}
\displaystyle
ds^{2} = -g(r)\, dt^{2} + \frac{dr^{2}}{g(r)} + r^{2}\left(d\theta^{2} + \sin^{2}\theta\, d\phi^{2}\right). 
\end{equation}
The metric possesses static symmetry, which guarantees invariance under time translations, while spherical symmetry ensures invariance under spatial rotations. These symmetries correspond to Killing vector fields, which in turn generate conserved quantities along geodesics, simplifying the analysis of photon trajectories.
\par
For a photon following a null geodesic, the associated Lagrangian is
\begin{equation}
\displaystyle
\mathcal{L} = \frac{1}{2} g_{\mu\nu} \frac{dx^{\mu}}{d \tilde{k}} \frac{dx^{\nu}}{d \tilde{k}} 
= -\frac{1}{2} g(r) \left(\frac{dt}{d \tilde{k}}\right)^{2} + \frac{1}{2} \frac{1}{g(r)} \left(\frac{dr}{d \tilde{k}}\right)^{2} + \frac{1}{2} r^{2} \left[ \left(\frac{d \theta}{d \tilde{k}}\right)^{2} + \sin^{2}\theta \left(\frac{d\phi}{d \tilde{k}}\right)^{2} \right],
\end{equation}
with $\tilde{k}$ an affine parameter providing a coordinate-invariant parametrization of motion. The canonical momenta $p^{\mu} = g^{\mu\nu}\, (dx_{\nu}/d\tilde{k})$ are conserved along directions generated by Killing vectors. In particular, the vectors $K^{\mu}_{(t)}$ and $K^{\mu}_{(\phi)}$, associated with time translations and azimuthal rotations, give rise to the conserved energy $\bar{E}$ and angular momentum $\bar{L}$:
\begin{equation}
\displaystyle
\bar{E} \equiv -p_{\mu} K^{\mu}_{(t)} = -p_{t} = g(r)\frac{dt}{d\tilde{k}}, 
\qquad 
\bar{L} \equiv p_{\mu} K^{\mu}_{(\phi)} = p_{\phi} = r^{2}\sin^{2}\theta\,\frac{d\phi}{d\tilde{k}} .
\end{equation}
\par
By spherical symmetry, the motion can be confined to the equatorial plane $\theta = \pi/2$, giving
\begin{equation}
\displaystyle
\bar{E} = g(r)\frac{dt}{d \tilde{k}}, \qquad \bar{L} = r^{2}\frac{d\phi}{d \tilde{k}},
\label{conserved_equatorial}
\end{equation}
where $\bar{E}$ corresponds to the photon frequency measured at infinity, while $\bar{L}$ encodes the angular momentum responsible for trajectory bending.
\par
We define the \emph{impact parameter} $b \equiv \bar{L}/\bar{E}$, representing the asymptotic perpendicular distance between the photon trajectory and the black-hole center for a distant observer. This parameter determines the qualitative behavior of the geodesic: sufficiently large $b$ leads to scattering to infinity, smaller $b$ results in capture by the black hole, while the critical value corresponds to trajectories asymptotically approaching the photon sphere. Consequently, $b$ fixes the apparent angular displacement of the photon on the observer’s sky and directly sets the shadow radius. Hence, the conserved quantities and the impact parameter establish a direct link between near-horizon photon dynamics and observable optical features.
\par
Imposing the null condition $ds^{2} = 0$ and using the conserved quantities, the radial motion can be recast in an energy-like form:
\begin{equation}
\displaystyle
\left(\frac{dr}{d\bar{k}}\right)^{2} + V_{\mathrm{eff}}(r) = \frac{1}{b^{2}}, 
\label{eom}
\end{equation}
where $\bar{k} \equiv \bar{L} \tilde{k}$ is a rescaled affine parameter, and the \emph{effective potential} is
\begin{equation}
\displaystyle
V_{\mathrm{eff}}(r) \equiv \frac{g(r)}{r^{2}}.
\end{equation}
Circular null orbits, or photon spheres, occur at radii $r_{0}$ satisfying $V_{\mathrm{eff}}(r_0) = 1/b_{0}^{2}$ and $V_{\mathrm{eff}}'(r_{0}) = 0$, with stability determined by the sign of $V_{\mathrm{eff}}''(r_{0})$. A negative second derivative signals instability, implying that small perturbations drive photons either inward to the horizon or outward to infinity. This formalism thus provides a rigorous framework for connecting local spacetime geometry to observable ring structures.
\par
For a RN black hole of mass $M$ and charge $Q$, recalling Eq.~\eqref{horizon function Q}, the horizon function is
\begin{equation}
\displaystyle
g(r) = 1 - \frac{2M}{r} + \frac{Q^{2}}{r^{2}}, \nonumber
\end{equation}
leading to the effective potential
\begin{equation}
\displaystyle
V_{\mathrm{eff(RN)}}(r) = \frac{1}{r^{2}} \left(1 - \frac{2M}{r} + \frac{Q^{2}}{r^{2}} \right).
\label{Veff}
\end{equation}
The potential exhibits a local maximum at
\begin{equation}
\displaystyle
r_0 = \frac{3M + \sqrt{9 M^{2} - 8 Q^{2}}}{2},
\end{equation}
which always lies outside the event horizon $r_{bh} = M + \sqrt{M^{2} - Q^{2}}$. Evaluating the second derivative,
\begin{equation}
\displaystyle
\frac{d^{2} V_{\mathrm{eff(RN)}}}{dr^{2}}\Big|_{r_{0}} = \frac{6}{r_{0}^{4}} - \frac{24 M}{r_{0}^{5}} + \frac{20 Q^{2}}{r_{0}^{6}} < 0,
\end{equation}
confirms that the photon sphere is intrinsically unstable for $0 \le Q \le M$.
\par
The corresponding \emph{critical impact parameter} is
\begin{equation}
\displaystyle
b_{0} = \frac{1}{\sqrt{V_{\mathrm{eff(RN)}}(r_{0})}} = \frac{(3M + \sqrt{9 M^{2} - 8 Q^{2}})^{2}}{2 \sqrt{2 M (3M + \sqrt{9 M^{2} - 8 Q^{2}}) - 4 Q^{2}}},
\end{equation}
which determines the observable shadow radius. The quantities $r_{0}$ and $b_{0}$ are related but distinct: $r_{0}$ denotes the radius of the photon sphere, corresponding to the circular photon orbit near the black hole, whereas $b_{0}$ sets the apparent shadow size measured by a distant observer. Photons with $b < b_{0}$ fall into the black hole, those with $b > b_{0}$ escape to infinity, and photons with $b = b_{0}$ asymptotically orbit at $r_{0}$ before escaping or plunging.
\par
In the Sch limit $(Q \to 0)$, one finds $r_{0} = 3M$ and $b_{0} = 3 \sqrt{3} M$, while in the extremal RN limit $(Q \to M)$, $r_{0} = 2M$ and $b_{0} = 4M$. This behavior shows that increasing charge shrinks the shadow, while the photon sphere remains unstable. The instability is characterized by Lyapunov exponents that govern orbital decay, linking the local geometry at $r_{0}$ to observable photon-ring features at infinity~\cite{EHTC:2022a}.

\subsubsection{Orbital Half-Period}
\par
Continuing from the discussion of photon spheres, we now introduce the \emph{orbital half-period} $\tau$, which characterizes the coordinate time required for a photon to complete half of its unstable circular orbit around the black hole, as measured by a static observer at infinity. This timescale captures the dynamics of light propagation near the photon sphere, shaping the optical appearance of the black hole and providing a natural bridge between local geodesic structure and observable shadow features.
\par
From an observational perspective, the orbital timescale $\tau$ governs the dominant time delay between successive photon arrivals on the observer’s screen for trajectories that execute partial loops before escaping to infinity~\cite{Gralla:2019xty,Guo:2022}. This delay arises from gravitational time dilation and the tendency of photons to linger near the horizon. As a consequence, $\tau$ leaves a clear imprint on the temporal structure of black-hole images, most notably in the spacing of ``photon echoes'' that compose the photon ring~\cite{Johnson:2020}. Thus, $\tau$ provides a direct link between local photon dynamics and observable timing signatures, underscoring its importance for high-resolution interferometric or strong-lensing measurements~\cite{EHTC:2022a}.
\par
To characterize the relevant timescale $\tau$, we consider null geodesics in a static, spherically symmetric spacetime with line element given in Eq.~\eqref{ds}. Owing to the time-translation and rotational symmetries of the spacetime, the motion admits two conserved quantities, as defined in Eq.~\eqref{conserved_equatorial}.
\par  
The photon angular velocity measured with respect to the coordinate time at infinity is defined as
\begin{equation}
\displaystyle
\Omega \equiv \frac{d\phi}{dt}.
\end{equation}
Using the relations above, this can be written as
\begin{equation}
\displaystyle
\Omega = \frac{d\phi/d\tilde{k}}{dt/d\tilde{k}} = \frac{\bar{L}}{\bar{E}}\,\frac{g(r)}{r^{2}}.
\end{equation}
Using the definition of the impact parameter $b = \bar{L}/\bar{E}$, the angular velocity becomes
\begin{equation}
\displaystyle
\Omega = b\,\frac{g(r)}{r^{2}}.
\end{equation}
\par
The corresponding orbital period,
\begin{equation}
\displaystyle
\mathcal{T} = \frac{2\pi}{\Omega},
\end{equation}
represents the coordinate time required for a photon to complete one full revolution around the black hole. In photon-ring observations, however, the physically relevant quantity is not the full orbital period but the temporal separation between successive photon echoes on the observer's screen. Photons contributing to higher-order sub-rings execute additional partial loops around the photon sphere before escaping to infinity. Successive image orders correspond approximately to photons completing an additional half-orbit ($\Delta\phi \simeq \pi$), which produces a characteristic time delay
\begin{equation}
\displaystyle
\tau = \frac{\pi}{\Omega} = \frac{\mathcal{T}}{2}.
\end{equation}
\par
For photons on unstable circular null orbits at radius $r_{0}$, the angular velocity becomes
\begin{equation}
\displaystyle
\Omega_{0} = b_{0}\frac{g(r_{0})}{r_{0}^{2}}.
\end{equation}
At the photon sphere, the critical impact parameter is fixed by the condition for a circular null orbit. The radial motion must vanish, $dr/d\bar{k} = 0$, which, using the radial equation of motion found in Eq.~\eqref{eom}, implies
\begin{equation}
\displaystyle
b_{0} = \frac{r_{0}}{\sqrt{g(r_{0})}}.
\end{equation}
Substituting this relation gives the orbital-half period
\begin{equation}
\displaystyle
\tau = \frac{\pi r_{0}}{\sqrt{g(r_{0})}}.
\end{equation}
\par
This relation shows that the delay scale $\tau$ is directly determined by the photon-sphere radius and the local spacetime curvature, linking the near-horizon geometry to observable timing signatures of photon-ring structures.
\par
For a RN black hole, the half-period evaluates to
\begin{equation}
\displaystyle
\tau_{\mathrm{(RN)}} = \frac{\pi (3M + \sqrt{9M^{2} - 8Q^{2}})^{2}}{2 \sqrt{2M(3M + \sqrt{9M^{2} - 8Q^{2}}) - 4Q^{2}}}, 
\label{tau1}
\end{equation}
valid for $0 \le Q \le M$, making explicit the dependence of $\tau$ on the black hole mass $M$ and charge $Q$. In the Sch limit $(Q \to 0)$, one recovers $r_{0} = 3M$ and $g(r_{0}) = 1/3$, giving~\cite{Gralla:2019xty}
\begin{equation}
\displaystyle
\lim_{Q \to 0} \tau_{\mathrm{(RN)}} = 3\pi\sqrt{3}\,M.
\end{equation}
For small charge $Q \ll M$, the half-period with leading-order correction reads
\begin{equation}
\displaystyle
\tau_{\mathrm{(RN)}} \simeq 3\pi\sqrt{3}\,M \left[ 1 - \left(\frac{Q}{\sqrt{6}M}\right)^2 \right],
\end{equation}
Note that the correction contributes from the fact that the inward shift of $r_{0}$ together with the modification of $g(r_{0})$, leads to an enhanced $\Omega_{0}$. In the extremal limit $(Q \to M)$, with $r_{0} = 2M$ and $g(r_{0}) = 1/4$, one obtains~\cite{Guo:2022,EHTC:2022a}
\begin{equation}
\displaystyle
\lim_{Q \to M} \tau_{\mathrm{(RN)}} = 4\pi M.
\end{equation}
Numerically, this corresponds to
\begin{equation}
\displaystyle
\tau_{\mathrm{(Sch)}} \approx 16.324\,M, 
\qquad 
\tau_{\mathrm{(ext)}} \approx 12.566\,M,
\end{equation}
indicating that $\tau_{\mathrm{(RN)}}$ decreases monotonically with increasing $Q$,
\begin{equation}
\displaystyle
\frac{d\tau_{\mathrm{(RN)}}}{dQ} < 0, 
\qquad 0 \le Q \le M.
\end{equation}
Thus, the half-period is a monotonically decreasing function throughout the range $0 \le Q \le M$.
\par
Finally, the orbital half-period $\tau$ encapsulates the fundamental dynamical timescales characterizing motion near the photon sphere. This may directly connect between the local spacetime geometry $(r_{0}, b_{0})$ and interferometric observables such as the sharpness, timing, and separation of photon rings.

\subsubsection{Angular Lyapunov Exponent}
\par
Building on the analysis of circular photon orbits, the \emph{angular Lyapunov exponent} $\lambda_{\rm L}$ quantifies the rate at which photons diverge from unstable circular trajectories under small radial perturbations. Its magnitude is determined by the second radial derivative of the effective potential evaluated at the photon-sphere radius, thereby characterizing the local instability of the orbit. This behavior governs observable properties of photon rings, including their sharpness and thickness on the image plane~\cite{Cardoso:2016rao,EHTC:2022a,Yang:2023,Kumara:2024}. By characterizing orbital instability, $\lambda_{\rm L}$ extends the analysis based on the orbital half-period $\tau$ and provides an additional diagnostic of photon-sphere dynamics with direct implications for high-resolution imaging.
\par
To formalize this instability, let us consider a photon near the circular orbit with a small radial perturbation $\delta r(\phi)$ along the azimuthal angle $\phi$:
\begin{equation}
\displaystyle
r(\phi) = r_{0} + \delta r(\phi), \qquad |\delta r| \ll r_{0}.
\end{equation}
Linearizing the radial equation of motion given in Eq.~\eqref{eom}, together with the conditions $V_{\mathrm{eff}}(r_0)=1/b_{0}^{2}$ and $V_{\mathrm{eff}}'(r_{0})=0$, yields
\begin{equation}
\displaystyle
\pi \frac{d}{d\phi} \delta r \simeq \lambda_{\rm L} \delta r,
\label{eomangular}
\end{equation}
with the angular Lyapunov exponent defined as
\begin{equation}
\displaystyle
\lambda_{\rm L} \equiv \pi r^{2}_{0} \sqrt{-\frac{1}{2} \frac{d^{2}}{dr^{2}} V_{\rm eff}(r_{0})}.
\end{equation}
Solving Eq.~\eqref{eomangular} yields
\begin{equation}
\displaystyle
\delta r(\phi) \simeq \delta r_{0} \, e^{\lambda_{\rm L} \phi},
\end{equation}
so that $\lambda_{\rm L}>0$ indicates exponential instability.
\par
From an observational standpoint, $\lambda_{\rm L}$ controls photon-ring morphology: larger values lead to faster trajectory divergence and sharper rings, whereas smaller values increase photon dwell time and produce thicker rings. Thus, $\lambda_{\rm L}$ provides a natural extension of the temporal diagnostic $\tau$ to the spatial domain, linking geodesic instability with image-plane structure.
\par
By construction, $\lambda_{\rm L} \geq 0$, reflecting the convexity of $V_{\rm eff}(r)$ at the photon sphere. Generic photon spheres yield $\lambda_{\rm L} > 0$, consistent with the universal instability of null circular orbits, while $\lambda_{\rm L} = 0$ arises only in marginal or finely tuned cases. Thus, $\lambda_{\rm L}$ provides a robust measure of geodesic instability that complements $\tau$ by encoding spatial rather than temporal ring properties.  
\par
For RN black holes, one finds
\begin{equation}
\displaystyle
\lambda_{\rm L(RN)} = \pi \sqrt{2 + \frac{3M (\sqrt{9M^{2} - 8Q^{2}}-3M)}{4Q^{2}}}, \qquad 0 \le Q \le M,
\end{equation}
which reduces to $\pi$ in the Sch limit $Q \to 0$ and to $\pi/\sqrt{2}$ in the extremal limit $Q \to M$. Increasing $Q$ lowers $\lambda_{\rm L}$, reflecting weaker orbital instability and producing broader photon rings, thereby linking black hole charge to observable morphology.  
\par
Furthermore, the photon image characterized by the impact parameter $b$ is closely related to the Lyapunov exponent $\lambda_{\rm L}$. Consider a photon orbiting near the photon sphere before eventually escaping to infinity. Such trajectories produce a sequence of sub-rings in the image, labeled by an integer $n$.
\par
The instability of photon-sphere dynamics causes photon trajectories to deviate from the critical impact parameter $b_{0}$. Since this instability is governed by the Lyapunov exponent $\lambda_{\rm L}$, the impact parameter of the $n$-th sub-ring, denoted by $b_n$, approaches the critical value exponentially. We define the deviation from the critical value as
\begin{equation}
\displaystyle
\Delta b_{n} \equiv b_{n} - b_{0}.
\end{equation}
Expanding $b=b(r)$ around $r=r_{0}$ with $r=r_{0}+\delta r$ gives
\begin{equation}
\displaystyle
\Delta b \simeq \frac{db}{dr} \Big|_{r=r_{0}} \delta r.
\end{equation}
For a photon trajectory producing the $n$-th sub-ring to remain near the photon sphere for $n$ orbits, the initial perturbation must scale as $\delta r_n \sim e^{-2\pi n \lambda_{\rm L}}$. Consequently, the deviation of the $n$-th sub-ring satisfies
\begin{equation}
\displaystyle
\Delta b_{n} \propto e^{-2 \pi n \lambda_{\rm L}}.
\end{equation}
\par
Moreover, the ratio of two successive sub-rings can be written as
\begin{equation}
\displaystyle
\frac{\Delta b_{n+1}}{\Delta b_{n}} = e^{-2\pi \lambda_{\rm L}}
\quad \Rightarrow \quad
\lambda_{\rm L} = -\frac{1}{2\pi}\ln \left| \frac{\Delta b_{n+1}}{\Delta b_{n}} \right|.
\label{db}
\end{equation}
This relation shows that the exponential clustering of photon sub-rings directly reflects the orbital instability of the photon sphere. Thus, the Lyapunov exponent provides a geometric measure of this instability through the self-similar structure of the photon rings.
\par
Physically, photons executing $n+1$ half-orbits around the photon sphere remain near the unstable orbit longer than those completing only $n$ half-orbits, and therefore require increasingly fine-tuned impact parameters. In order to have $\lambda_{L} > 0$, the ratio in Eq.~\eqref{db} must be less than unity, $\Delta b_{n+1}/\Delta b_{n} < 1$. A larger $\lambda_{\rm L}$ compresses the sequence of sub-rings, while a smaller $\lambda_{\rm L}$ spreads them farther apart, providing a direct probe of photon-sphere stability and its dependence on black hole parameters. Consequently, the hierarchical structure of photon sub-rings encodes the dynamical instability of null geodesics near the photon sphere.

\subsubsection{Temporal Lyapunov Exponent}
\par
Extending the analysis of circular photon orbits, we now introduce the \emph{temporal Lyapunov exponent} $\gamma_{\rm L}$, which measures the exponential growth of small radial perturbations with respect to the coordinate time $t$ as observed at infinity. Thus, $\gamma_{\rm L}$ captures the instability in the time domain and sets the characteristic decay timescale of photon orbits. This temporal diagnostic is particularly relevant for phenomena such as photon-ring echoes, quasi-normal mode (QNM) damping, and the timing of successive photon loops~\cite{Mashhoon:1985, Cardoso:2008bp,Cardoso:2016rao,Konoplya:2011,Yang:2023}.
\par
Consider a null geodesic near the photon sphere at $r_{0}$ perturbed slightly in the radial direction:
\begin{equation}
r(t)=r_{0}+\delta r(t), \qquad |\delta r|\ll r_{0}.
\end{equation}
Linearizing the radial equation of motion given in Eq.~\eqref{eom}, together with the conditions 
$V_{\mathrm{eff}}(r_0)=1/b_{0}^{2}$ and $V_{\mathrm{eff}}'(r_{0})=0$, yields
\begin{equation}
\frac{d}{dt}\delta r \simeq \gamma_{\rm L}\,\delta r,
\qquad
\gamma_{\rm L}\equiv
\sqrt{-\frac12 r_0^{2}g(r_0)\frac{d^2V_{\rm eff}}{dr^2}(r_0)},
\end{equation}
where factor $g(r_0)$ encodes the effect of gravitational redshift.
\par
The solution
\begin{equation}
\delta r(t)\simeq \delta r_0 e^{\gamma_{\rm L}t},
\end{equation}
shows that radial perturbations grow exponentially, defining a characteristic instability timescale
\begin{equation}
\tau_{\rm inst}\sim \gamma_{\rm L}^{-1}.
\end{equation}
\par
Larger $\gamma_{\rm L}$ implies faster orbital decay and shorter-lived sub-ring structures, whereas smaller values correspond to slower leakage and more persistent rings. In the eikonal limit, $\gamma_{\rm L}$ also governs the imaginary part of QNM frequencies, thus linking photon-sphere instability to gravitational-wave ring-down.  
\par
For RN black holes, the temporal Lyapunov exponent evaluates to
\begin{equation}
\displaystyle
\gamma_{\rm L(RN)} =
\frac{4 \sqrt{ \big(M (\sqrt{9 M^{2}-8 Q^{2}}+3 M)-2 Q^{2} \big) 
\big( 3 M (\sqrt{9 M^{2}-8 Q^{2}}+3 M)-8 Q^{2} \big) }}
{(\sqrt{9 M^{2}-8 Q^{2}}+3 M)^{3}},
\end{equation}
which decreases monotonically with $Q$, interpolating between $\gamma_{\rm L} = 1/(3\sqrt{3} M)$ in the Sch limit ($Q \rightarrow 0$) and $\gamma_{\rm L} = 1/(4\sqrt{2} M)$ in the extremal case ($Q \rightarrow M$). 
\par
The relation between $\gamma_{\rm L}$ and $\lambda_{\rm L}$ follows from the orbital angular velocity
$\Omega_0 \equiv d\phi/dt = \pi/\tau$.
Using
\begin{equation}
\frac{d}{d\phi}=\frac{1}{\Omega_0}\frac{d}{dt},
\end{equation}
one finds
\begin{equation}
\gamma_{\rm L} = \Omega_0 \lambda_{\rm L} = \frac{\pi}{\tau}\lambda_{\rm L}.
\end{equation}
Thus $\lambda_{\rm L}$ measures the exponential divergence per unit angular advance,
while $\gamma_{\rm L}$ characterizes the divergence per unit time. Moreover, the hierarchy of photon sub-rings, $(\Delta b_{n+1}/\Delta b_{n}) \sim e^{-2\pi \lambda_{\rm L}}$, translates into the temporal domain as
\begin{equation}
\gamma_{\rm L}
=
-\frac{1}{2\tau}
\ln\left|\frac{\Delta b_{n+1}}{\Delta b_{n}}\right|.
\end{equation}
\par
Taken together, the three quantities $(\tau, \lambda_{\rm L}, \gamma_{\rm L})$ describe the dynamics of the photon sphere: $\tau$ sets the orbital timescale, $\lambda_{\rm L}$ measures the spatial instability per orbit, and $\gamma_{\rm L}$ characterizes the temporal decay rate. This relation connects the geometry of the black hole with the structure of photon rings and time-domain observables.

\subsection{Optical Diagnostics of Black-Hole Phase Transitions}
\par
In this part, we establish a connection between black hole thermodynamics and optical observables, namely the orbital half-period $\tau$ and the Lyapunov exponents $(\lambda_{\rm L}, \gamma_{\rm L})$, to investigate the black hole phase transition from the perspective of optical characteristics. Within the framework of non-extensive statistics, the deformation parameter in the regime $\eta<0$ gives rise to a richer phase structure characterized by the SBH, IBH, and LBH branches. The emergence of these phases suggests that the associated phase transitions may also leave signatures in optical quantities related to photon orbits.
\par
Since the phase structure of the black hole can be directly diagnosed from the behavior of the temperature, we express the optical characteristics in terms of the $q$-generalized Hawking temperature. In particular, we consider
\begin{equation}
\label{eq:TH-optical}
\displaystyle
\tau = \tau(T_{\eta, \rm H}), 
\qquad
\lambda_{\rm L} = \lambda_{\rm L}(T_{\eta, \rm H}),
\qquad
\gamma_{\rm L} = \gamma_{\rm L}(T_{\eta, \rm H}).
\end{equation}
\par
The first optical parameter is the orbital half-period $\tau_{\mathrm{(RN)}}$, 
which can be expressed in terms of the charge $Q$ and the electric potential 
$\Phi_{\mathrm{(RN)}}$ as
\begin{equation}
\label{tauRN}
\displaystyle
\tau_{\mathrm{(RN)}}(Q, \Phi_{\mathrm{(RN)}}) 
= \frac{\pi Q \,\bigl(3 \Phi_{\mathrm{(RN)}}^{2} + \Delta + 3\bigr)^{2}}
{4 \sqrt{2} \,\Phi_{\mathrm{(RN)}} \, \sqrt{3 \Phi_{\mathrm{(RN)}}^{4} + (\Delta - 2)\Phi_{\mathrm{(RN)}}^{2} + \Delta + 3}},
\end{equation}
where
\begin{equation}
\label{Delta}
\displaystyle
\Delta \equiv \sqrt{9 \Phi_{\mathrm{(RN)}}^{4} - 14 \Phi_{\mathrm{(RN)}}^{2} + 9}.
\end{equation}
Both the Hawking temperature and the orbital half-period depend on the same 
thermodynamic variables $(Q,\Phi_{\mathrm{(RN)}})$, namely
\begin{equation}
\displaystyle
T_{\eta, \mathrm{H(RN)}} = T_{\eta, \mathrm{H(RN)}}(\Phi_{\mathrm{(RN)}}, Q), 
\qquad 
\tau_{\mathrm{(RN)}} = \tau_{\mathrm{(RN)}}(\Phi_{\mathrm{(RN)}}, Q).
\end{equation}
Consequently, signatures of thermodynamic phase transitions may also appear in 
the behavior of the optical observable $\tau_{\mathrm{(RN)}}$.
\par
For fixed charge $Q$ and varying $\Phi_{\mathrm{(RN)}}$, the thermodynamic
extrema are determined by the condition
\begin{equation}
\displaystyle
\left(\frac{\partial T_{\eta, \mathrm{H(RN)}}}{\partial \Phi_{\mathrm{(RN)}}}\right)_{Q}=0.
\end{equation}
Using the chain rule, the derivative with respect to the orbital half-period
can be written as
\begin{equation}
\displaystyle
\left(\frac{\partial T_{\eta, \mathrm{H(RN)}}}{\partial \tau_{\mathrm{(RN)}}}\right)_{Q}
=
\frac{\left(\partial T_{\eta, \mathrm{H(RN)}} / \partial \Phi_{\mathrm{(RN)}}\right)_{Q}}
{\left(\partial \tau_{\mathrm{(RN)}} / \partial \Phi_{\mathrm{(RN)}}\right)_{Q}}.
\end{equation}
Therefore, whenever 
$\left(\partial T_{\eta, \mathrm{H(RN)}} / \partial \Phi_{\mathrm{(RN)}}\right)_{Q}=0$
with
$\left(\partial \tau_{\mathrm{(RN)}} / \partial \Phi_{\mathrm{(RN)}}\right)_{Q}\neq0$,
the derivative with respect to $\tau_{\mathrm{(RN)}}$ also vanishes.
Consequently, the extrema in the $\tau_{\mathrm{(RN)}}$--$T_{\eta,\mathrm{H(RN)}}$
diagram can be identified through the behavior of the temperature as a function
of $\Phi_{\mathrm{(RN)}}$, in close analogy with the analysis of black hole
thermodynamics.
\par
This result implies that, locally, the orbital half-period can be parametrically
expressed as a function of the temperature,
\begin{equation}
\displaystyle
\tau_{\mathrm{(RN)}}=\tau_{\mathrm{(RN)}}\bigl(T_{\eta,\mathrm{H(RN)}}\bigr).
\end{equation}
Each thermodynamic extremum $\Phi_{\mathrm{ex}}$ therefore corresponds to a
specific optical value
\begin{equation}
\displaystyle
\tau_{\mathrm{ex}} = \tau_{\mathrm{(RN)}} (T_{\eta, \rm H(\rm RN)}(\Phi_{\rm ex})) = \tau_{\mathrm{(RN)}}(\Phi_{\mathrm{ex}}),
\end{equation}
establishing a direct mapping between thermodynamic extrema and the
corresponding optical extrema points.
\begin{figure}[ht]
\centering
\includegraphics[width=6cm]{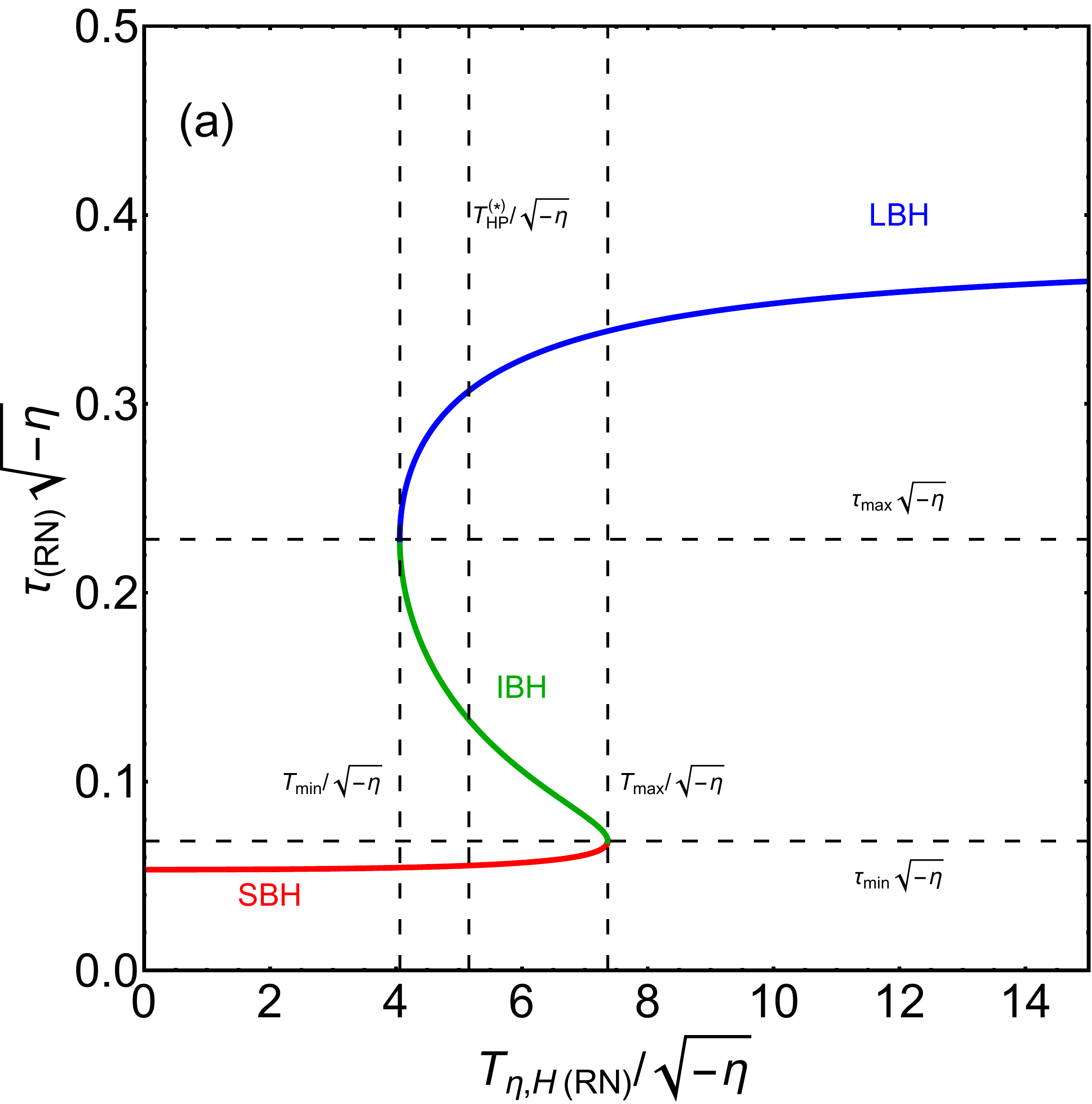}
\quad
\includegraphics[width=6cm]{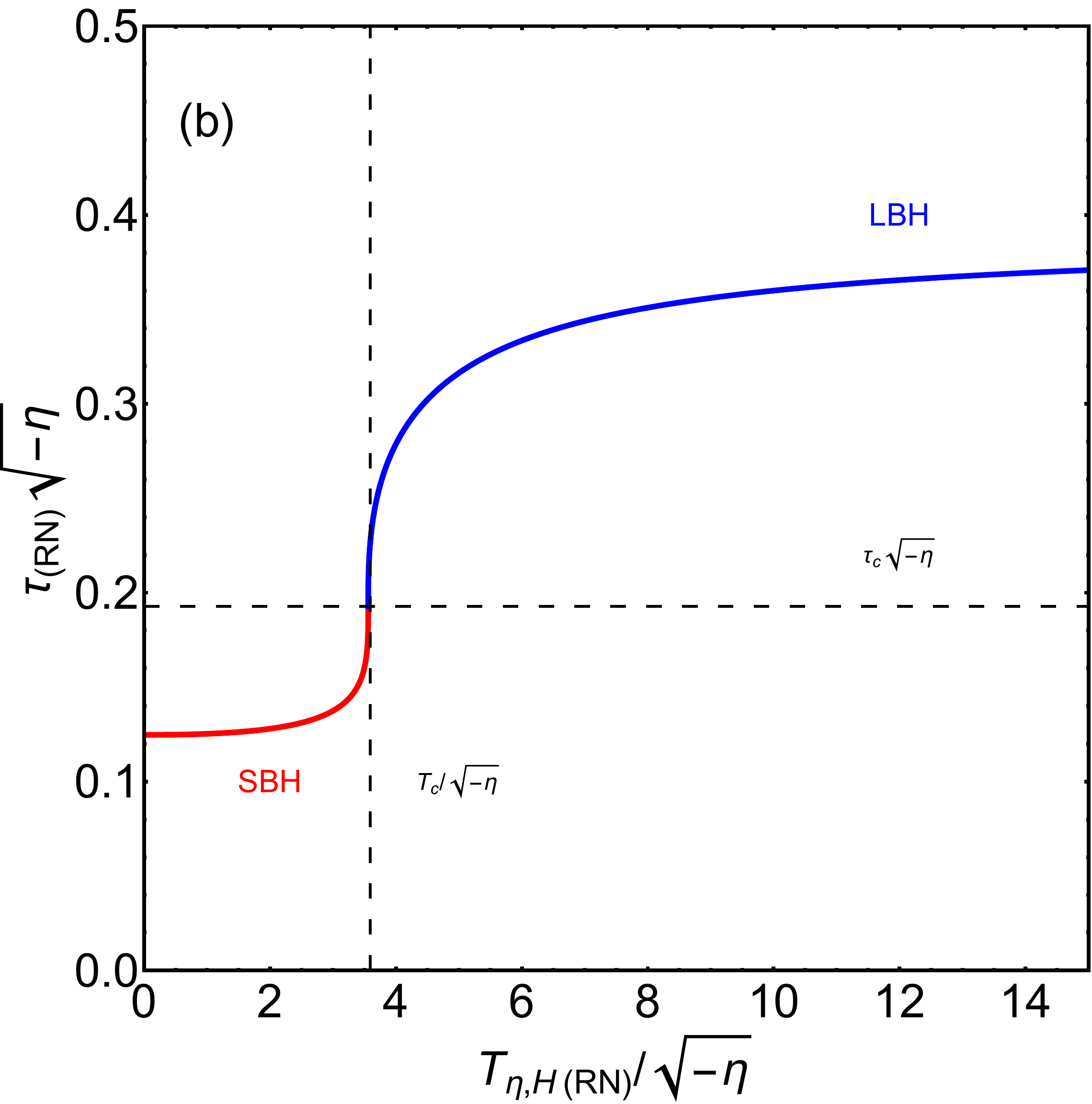}
\quad
\includegraphics[width=6cm]{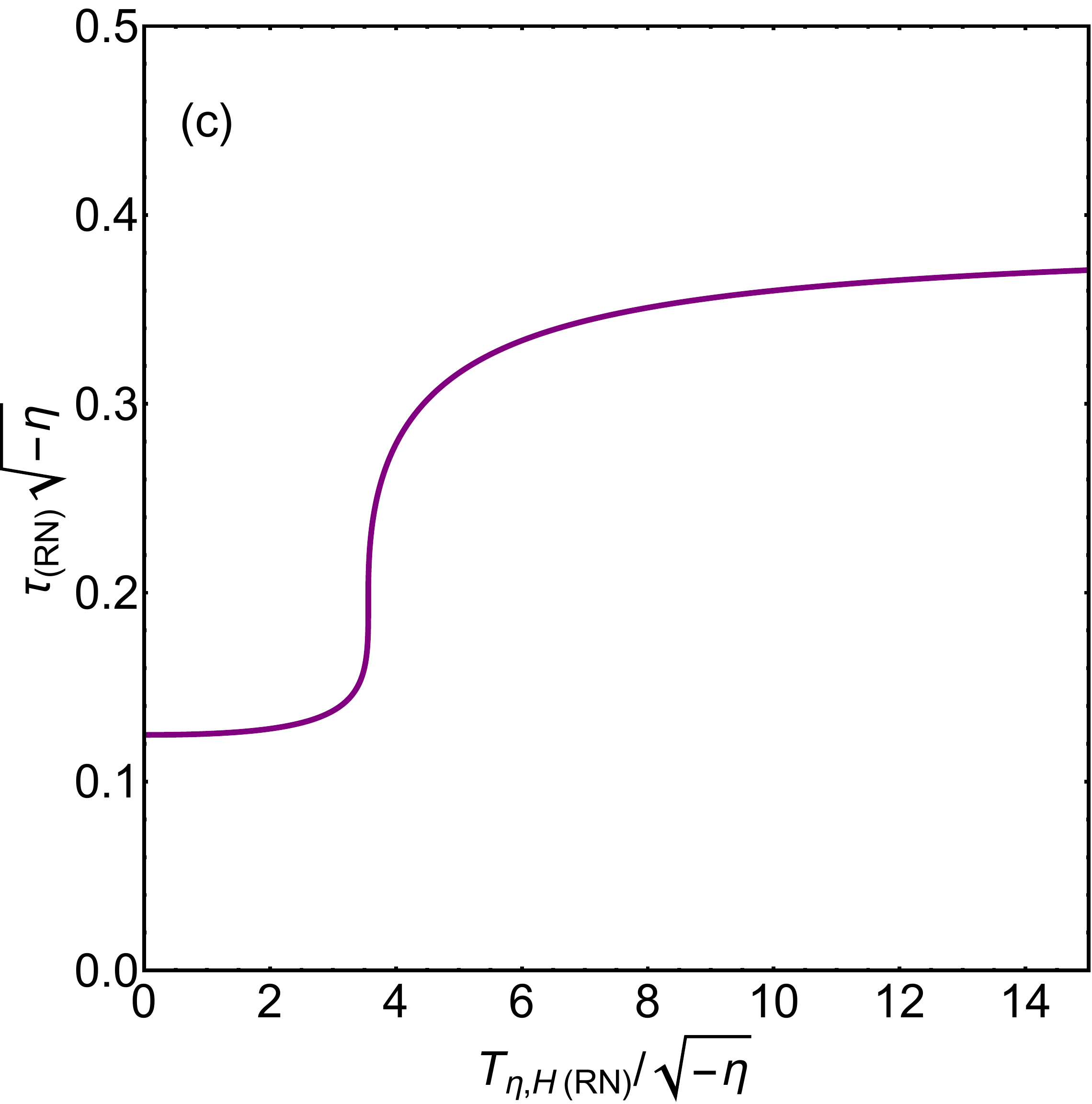}
\caption{Isobaric curves of the orbital half-period $\tau_{\mathrm{(RN)}}$ vs.\ Hawking temperature $T_{\eta,\mathrm{H(RN)}}$ for $\eta=-0.002$. 
(a) Subcritical $Q=0.095<Q_c$ showing SBH–LBH coexistence. 
(b) Critical $Q=Q_c$ with a single inflection point. 
(c) Supercritical $Q=0.3>Q_c$ showing smooth single-phase behavior.}
\label{fig:orbital_half_period}
\end{figure}
\par
The behavior of the orbital half-period and the Hawking temperature is
illustrated in the $\tau_{\mathrm{(RN)}}$--$T_{\eta,\mathrm{H(RN)}}$ diagram
shown in Fig.~\ref{fig:orbital_half_period}. In the subcritical regime $Q<Q_{c}$, the temperature curve exhibits two
turning points in the $\tau_{\mathrm{(RN)}}$--$T_{\eta,\mathrm{H(RN)}}$ plane,
which correspond to the thermodynamic critical points
$A_{C_{Q(-)}}$ and $A_{C_{Q(+)}}$ identified in Sec.~\ref{thermo}. These turning points divide the system into three branches in $\tau$-space, corresponding to the SBH, IBH, and LBH phases. The SBH and LBH branches are thermodynamically stable, while the IBH branch is unstable. At the Hawking--Page temperature $T_{\rm HP}^{(*)}$, the parameter $\tau_{\mathrm{(RN)}}$ becomes discontinuous between the SBH and LBH branches, signaling the first-order phase transition. At the critical charge $Q=Q_{c}$, the two turning points merge into a single
inflection point, and the orbital half-period varies continuously with the
temperature, corresponding to the second-order phase transition.
For the supercritical case $Q>Q_{c}$, the turning points disappear and the
curve becomes smooth and monotonic, indicating that the system remains in a
single thermodynamic phase. These results suggest that the orbital half-period can serve as an optical probe of black hole thermodynamics, providing an indirect way to infer the underlying phase structure through observable photon dynamics in this model.
\begin{figure}[ht]
\centering
\includegraphics[width=6cm]{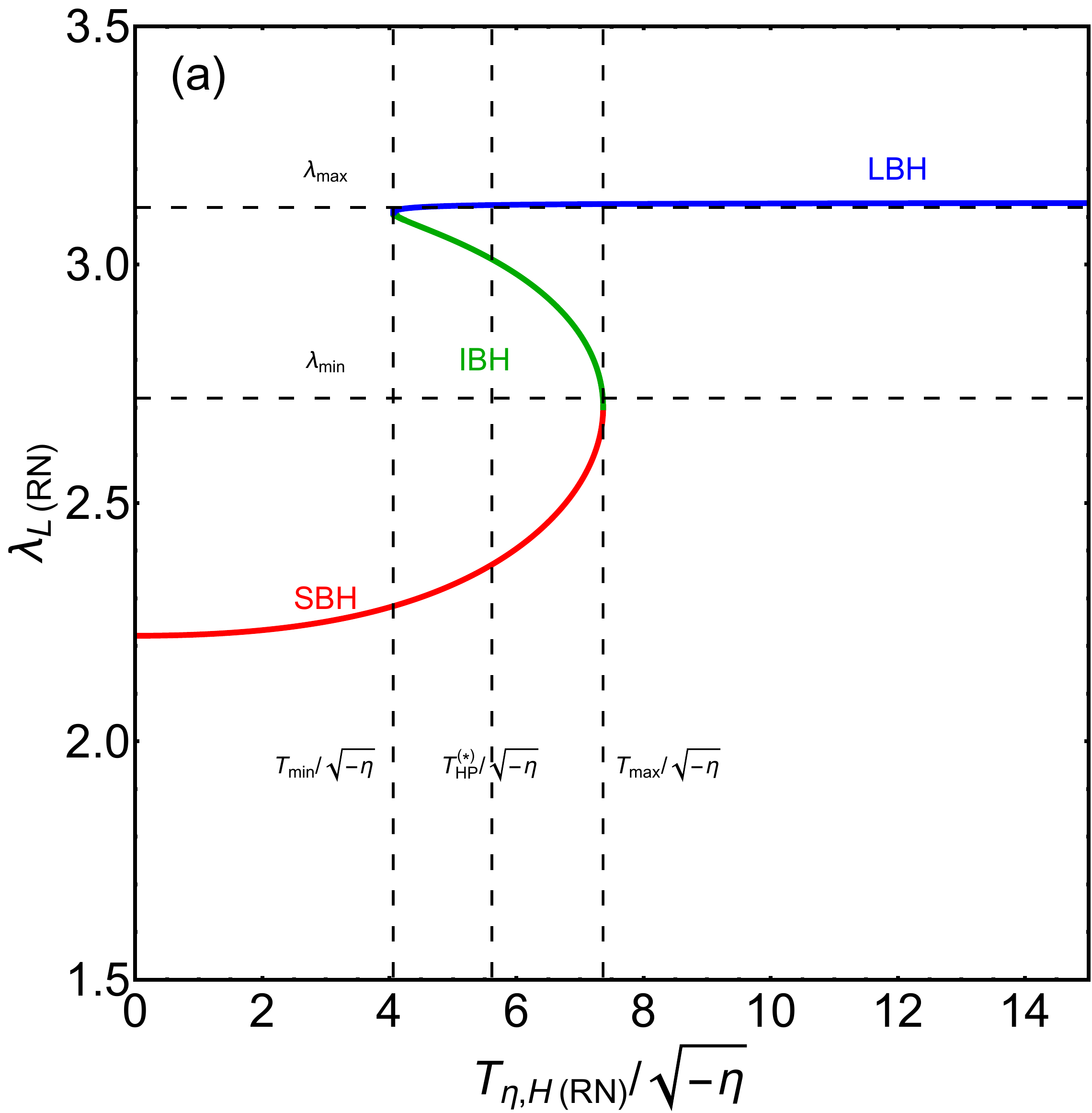}
\quad
\includegraphics[width=6cm]{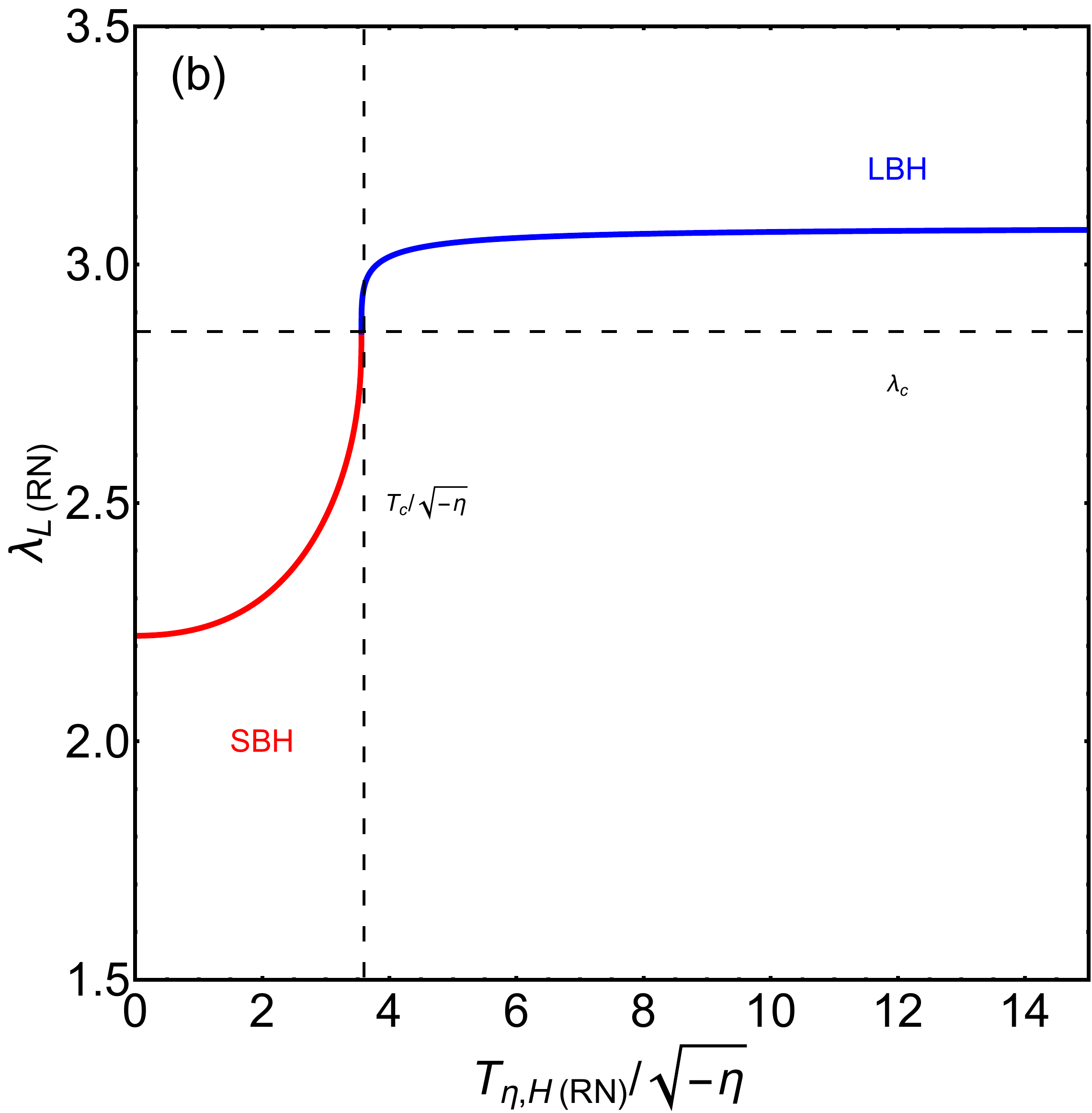}
\quad
\includegraphics[width=6cm]{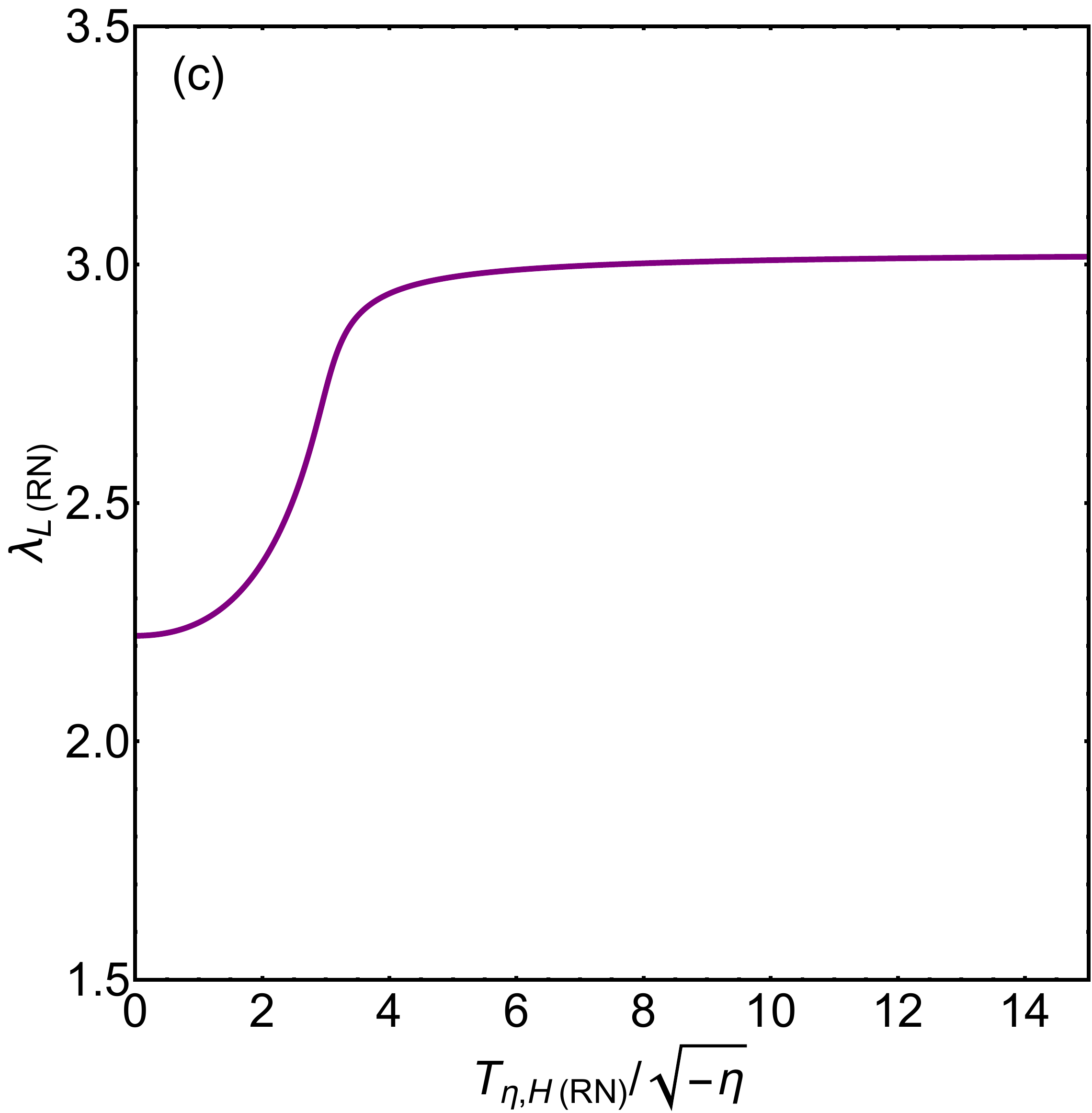}
\caption{Isobaric curves of the angular Lyapunov exponent $\lambda_{\rm L(RN)}$ vs.\ Hawking temperature $T_{\eta,\mathrm{H(RN)}}$ for $\eta=-0.002$. 
(a) Subcritical $Q=0.095<Q_{c}$ showing divergences from SBH–LBH coexistence. 
(b) Critical $Q=Q_{c}$ with a vertical tangent at the inflection point. 
(c) Supercritical $Q=0.3>Q_{c}$ showing smooth, single-phase behavior.}
\label{angular Lyapunov}
\end{figure}
\par
Furthermore, two additional optical parameters are the angular and temporal Lyapunov exponents, $\lambda_{\mathrm{L(RN)}}$ and $\gamma_{\mathrm{L(RN)}}$, respectively, which can be expressed in terms of $Q$ and $\Phi_{\mathrm{(RN)}}$ as
\begin{figure}[ht] 
\centering 
\includegraphics[width=6cm]{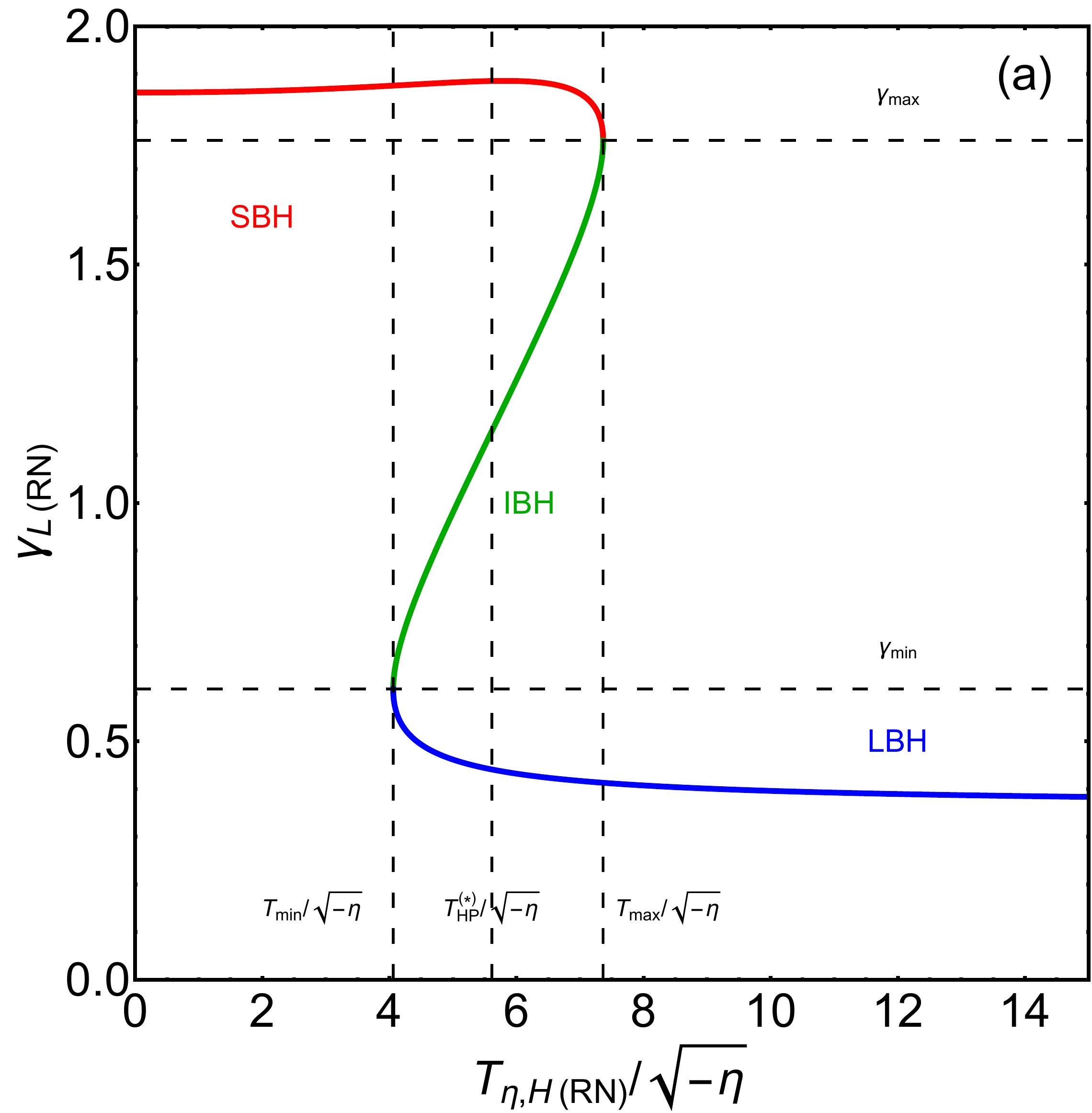}
\quad 
\includegraphics[width=6cm]{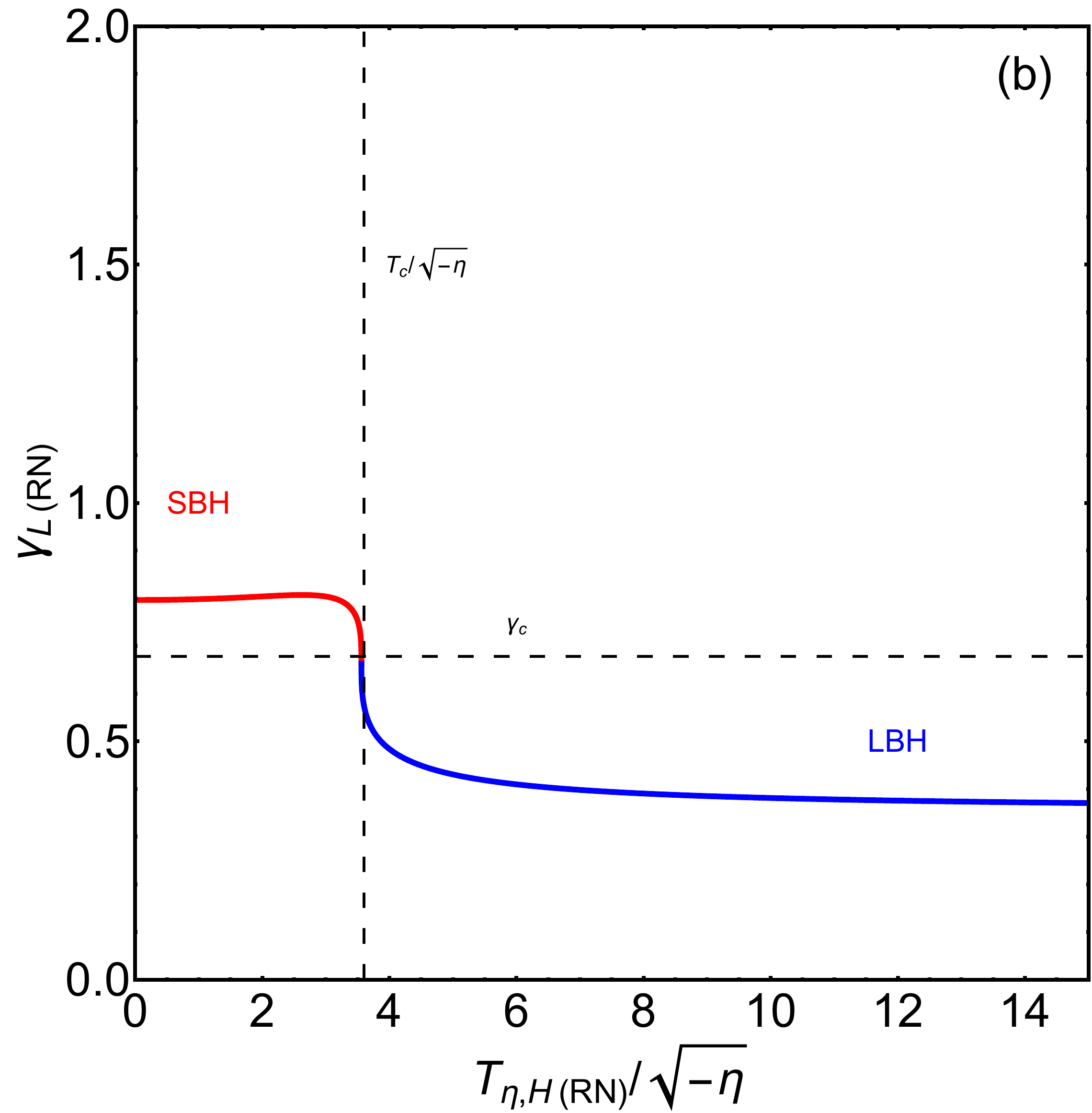}
\quad 
\includegraphics[width=6cm]{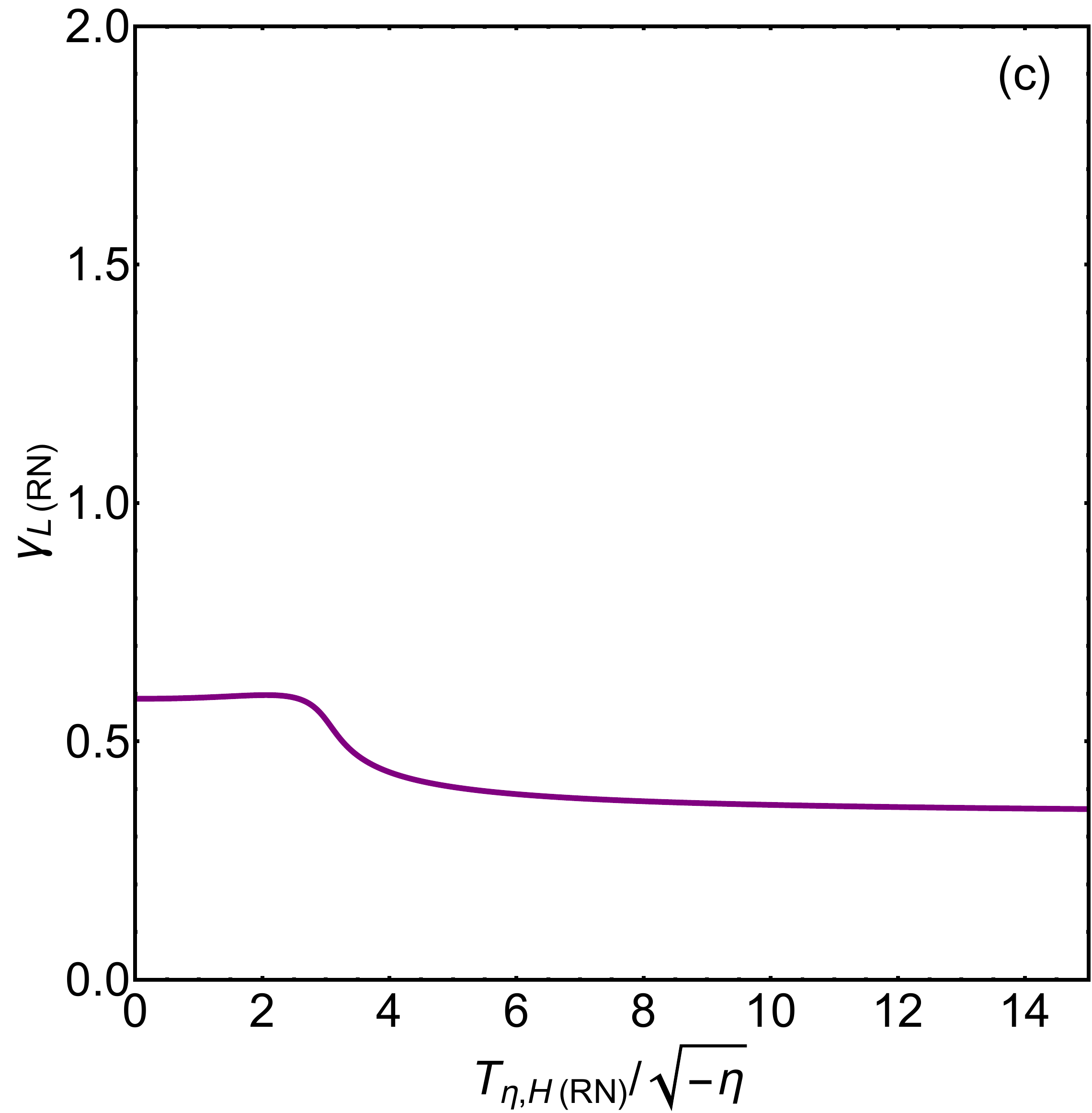} 
\caption{Isobaric curves of the temporal Lyapunov exponent $\gamma_{\rm L(RN)}$ vs.\ Hawking temperature $T_{\eta,\rm H(\rm RN)}$ for $\eta=-0.002$. 
(a) Subcritical $Q=0.095<Q_{c}$ showing SBH–LBH coexistence. 
(b) Critical $Q=Q_{c}$ with a single vertical tangent. 
(c) Supercritical $Q=0.3>Q_{c}$ showing smooth, single-phase behavior.}
\label{temporal_Lyapunov} 
\end{figure}
\begin{equation}
\displaystyle
\lambda_{\mathrm{L(RN)}} =
\frac{\sqrt{2}\,\pi\,\sqrt{\Delta}}
{\sqrt{3 \Phi_{\mathrm{(RN)}}^{2} + \Delta + 3}},
\end{equation}
and
\begin{equation}
\displaystyle
\gamma_{\mathrm{L(RN)}} =
\frac{8 \, \Phi_{\mathrm{(RN)}} \, \sqrt{
\bigl(3 \Phi_{\mathrm{(RN)}}^{4} + (\Delta - 2)\Phi_{\mathrm{(RN)}}^{2} + \Delta + 3\bigr)
\bigl(9 \Phi_{\mathrm{(RN)}}^{4} + (3 \Delta - 14) \Phi_{\mathrm{(RN)}}^{2} + 3(\Delta + 3)\bigr)}}
{Q \, \bigl(3 + 3 \Phi_{\mathrm{(RN)}}^{2} + \Delta\bigr)^{3}} .
\end{equation}
\par
In the same manner as we have done for the case of $\tau_{(\rm RN)}$, the angular and the temporal Lyapunov exponents can be parametrically written as functions of the temperature,
\begin{equation}
\displaystyle
\lambda_{\rm L(\rm RN)} = \lambda_{\rm L(\rm RN)}(T_{\eta, \rm H(\rm RN)}),
\qquad
\gamma_{\rm L(\rm RN)} = \gamma_{\rm L(\rm RN)} (T_{\eta, \rm H(\rm RN)}).
\end{equation}
These parameters can be mapped to the extremum point $\Phi_{\rm ex}$ of the temperature through the relation
\begin{eqnarray}
\displaystyle
\lambda_{\rm ex} &=& \lambda_{\rm L(\rm RN)}(T_{\eta, \rm H(\rm RN)}(\Phi_{\rm ex})) = \lambda_{\rm L(\rm RN)} (\Phi_{\rm ex}),  \\
\displaystyle
\gamma_{\rm ex} &=& \gamma_{\rm L(\rm RN)} (T_{\eta, \rm H(\rm RN)}(\Phi_{\rm ex})) = \gamma_{\rm L(\rm RN)} (\Phi_{\rm ex}).
\end{eqnarray}
Accordingly, these optical parameters establish a direct one-to-one mapping with the thermodynamic extremum point. 
\par
The behaviors of the angular and the temporal Lyapunov exponents are illustrated in Figs.~\ref{angular Lyapunov} and~\ref{temporal_Lyapunov}, respectively. For the subcritical regime $Q<Q_{c}$, the $\lambda_{\rm L(\rm RN)}-T_{\eta, \rm H(\rm RN)}$ and the $\gamma_{\rm L(\rm RN)}-T_{\eta, \rm H(\rm RN)}$ curves develop two extrema, namely $\lambda_{\rm min}$ and $\lambda_{\rm max}$, and $\gamma_{\rm min}$ and $\gamma_{\rm max}$, marking the coexistence of SBH and LBH phases. At the coexistence phase associated with the temperature $T_{\rm HP}^{(*)}$, these optical parameters change discontinuously between the SBH and the LBH branches. This coexistence of black hole thermodynamic phases can be detected through the optical parameters. If a discontinuous change in these optical quantities is observed, this may indicate that the charged black hole undergoes a first-order phase transition characterized by non-extensive effects, thereby allowing the parameter $\eta$ to be constrained. Moreover, for the cases $Q=Q_{c}$ and $Q>Q_{c}$, the optical parameters change continuously and smoothly, respectively, as shown in panels~(b) and~(c) of both figures.

\subsection{Optical Signatures of Criticality in Black Hole Thermodynamics}
\par
Here we demonstrate that optical observables associated with photon motion exhibit critical behavior analogous to that of black hole thermodynamic phase transitions. The motion of photons around a black hole carries detailed information about the underlying spacetime geometry, and consequently about the thermodynamic state of the system. Optical observables therefore provide a possible way to probe the thermodynamic phase structure, linking photon-sphere dynamics with the phase transition of charged black holes described by Tsallis statistical mechanics.
\par
Focusing on three complementary diagnostics—the orbital half-period $\tau$, the angular Lyapunov exponent $\lambda_{\rm L}$, and the temporal Lyapunov exponent $\gamma_{\rm L}$—we analyze how these quantities respond near the thermodynamic critical point. These optical parameters exhibit pronounced changes close to criticality, forming a robust optical--thermodynamic correspondence that maps photon dynamics to phase-transition phenomena. Near the critical regime, these observables display universal scaling behavior characteristic of second-order phase transitions, paralleling conventional thermodynamic order parameters.
\par
Along the coexistence line separating the SBH and LBH phases, each optical observable takes two distinct values corresponding to the two branches. This difference naturally motivates the definition of an optical order parameter, analogous to the difference between thermodynamic order parameters in coexisting phases of ordinary fluids.
\par
To formalize this idea, we introduce optical variables $\mathcal{V}_{i}$ $(i=1,2,3)$ corresponding to the quantities $\tau$, $\lambda_{\rm L}$, and $\gamma_{\rm L}$, respectively. Since these observables can be expressed in terms of thermodynamic quantities, for example $\mathcal{V}_{i}(Q,\Phi)$ or $\mathcal{V}_{i}(T_{\eta,\mathrm{H}})$, we define the \textit{reduced optical order parameter} as
\begin{equation}
\displaystyle
\frac{\Delta \mathcal{V}_{i}}{\mathcal{V}_{i(c)}} :=
\frac{|\mathcal{V}_{i(\rm SBH)}-\mathcal{V}_{i(\rm LBH)}|}{\mathcal{V}_{i(c)}},
\label{optical parameter}
\end{equation}
where $\mathcal{V}_{i(\rm SBH)}$ and $\mathcal{V}_{i(\rm LBH)}$ denote the values of the optical quantities along the SBH and LBH branches, respectively, and $\mathcal{V}_{i(c)}$ represents the corresponding critical value.
\par
To determine the near-critical scaling behavior, we expand $\mathcal{V}_{i}$ around the critical potential $\Phi_c$ along the coexistence line $Q = Q(\Phi)$,
\begin{align}
\displaystyle
\mathcal{V}_{i} (\Phi_{\rm LBH})
&=
\mathcal{V}_{i}(\Phi_{c})
+
\frac{\partial \mathcal{V}_{i}}{\partial \Phi}
\Big|_{\Phi_{c}}
(\Phi_{\rm LBH}-\Phi_{c})
+
\cdots, \\
\displaystyle
\mathcal{V}_{i} (\Phi_{\rm SBH})
&=
\mathcal{V}_{i}(\Phi_{c})
+
\frac{\partial \mathcal{V}_{i}}{\partial \Phi}
\Big|_{\Phi_{c}}
(\Phi_{\rm SBH}-\Phi_{c})
+
\cdots.
\end{align}
In the near-critical regime $|\Phi_{\rm SBH}-\Phi_{\rm LBH}| \to 0$, the optical order parameter can therefore be approximated as
\begin{equation}
\displaystyle
\frac{\Delta \mathcal{V}_{i}}{\mathcal{V}_{i(c)}}
\simeq
\frac{1}{\mathcal{V}_{i(c)}}
\frac{\partial \mathcal{V}_{i}}{\partial \Phi}
\Big|_{\Phi_{c}}
(\Phi_{\rm SBH}-\Phi_{\rm LBH}).
\label{eq:deltaVi_linear}
\end{equation}
\par
Next, expanding the black hole equation of state along the coexistence line, where $\partial_{\bar{\Phi}}\bar{T}\simeq0$, yields
\begin{equation}
\displaystyle
(\Phi_{\rm SBH} - \Phi_{\rm LBH})
\simeq
\pm 2\Phi_{c}
\left[
-\frac{2\,\partial_{\bar{Q}}\dot{\bar{T}}(0,0)}
{\dddot{\bar{T}}(0,0)\,\partial_{\bar{Q}}\bar{T}(0,0)}
\right]^{1/2}
(1-t^{(*)})^{1/2},
\qquad
t^{(*)}\equiv\frac{T^{(*)}_{\rm HP}}{T_c}.
\end{equation}
Substituting this result into Eq.~\eqref{eq:deltaVi_linear}, we obtain the universal near-critical scaling
\begin{equation}
\displaystyle
\frac{\Delta\mathcal{V}_{i}}{\mathcal{V}_{i(c)}}
\simeq
\tilde{\alpha}_{i}
(1-t^{(*)})^{1/2},
\qquad
i\in\{\tau,\lambda_{\rm L},\gamma_{\rm L}\},
\label{optical order}
\end{equation}
where the positive prefactors are given by
\begin{equation}
\displaystyle
\tilde{\alpha}_{i}
\equiv
\left|
\frac{2\Phi_{c}}{\mathcal{V}_{i(c)}}
\frac{\partial \mathcal{V}_{i}}{\partial \Phi}
\Big|_{\Phi_{c}}
\right|
\left[
-\frac{2\,\partial_{\bar{Q}}\dot{\bar{T}}(0,0)}
{\dddot{\bar{T}}(0,0)\,\partial_{\bar{Q}}\bar{T}(0,0)}
\right]^{1/2}.
\label{optical order 2}
\end{equation}
\begin{figure}[ht]
\centering
\includegraphics[width=6cm]{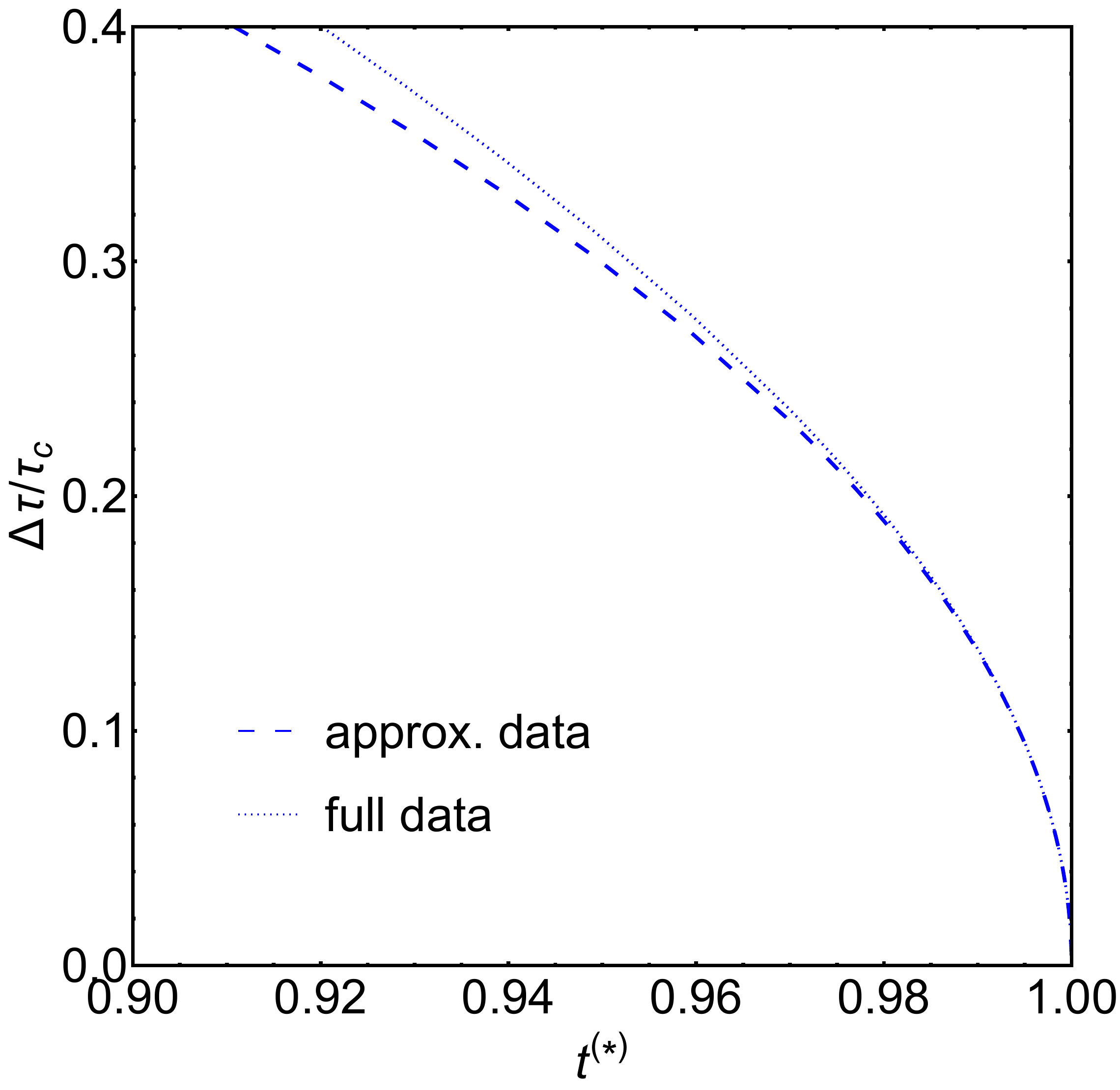}
\quad
\includegraphics[width=6cm]{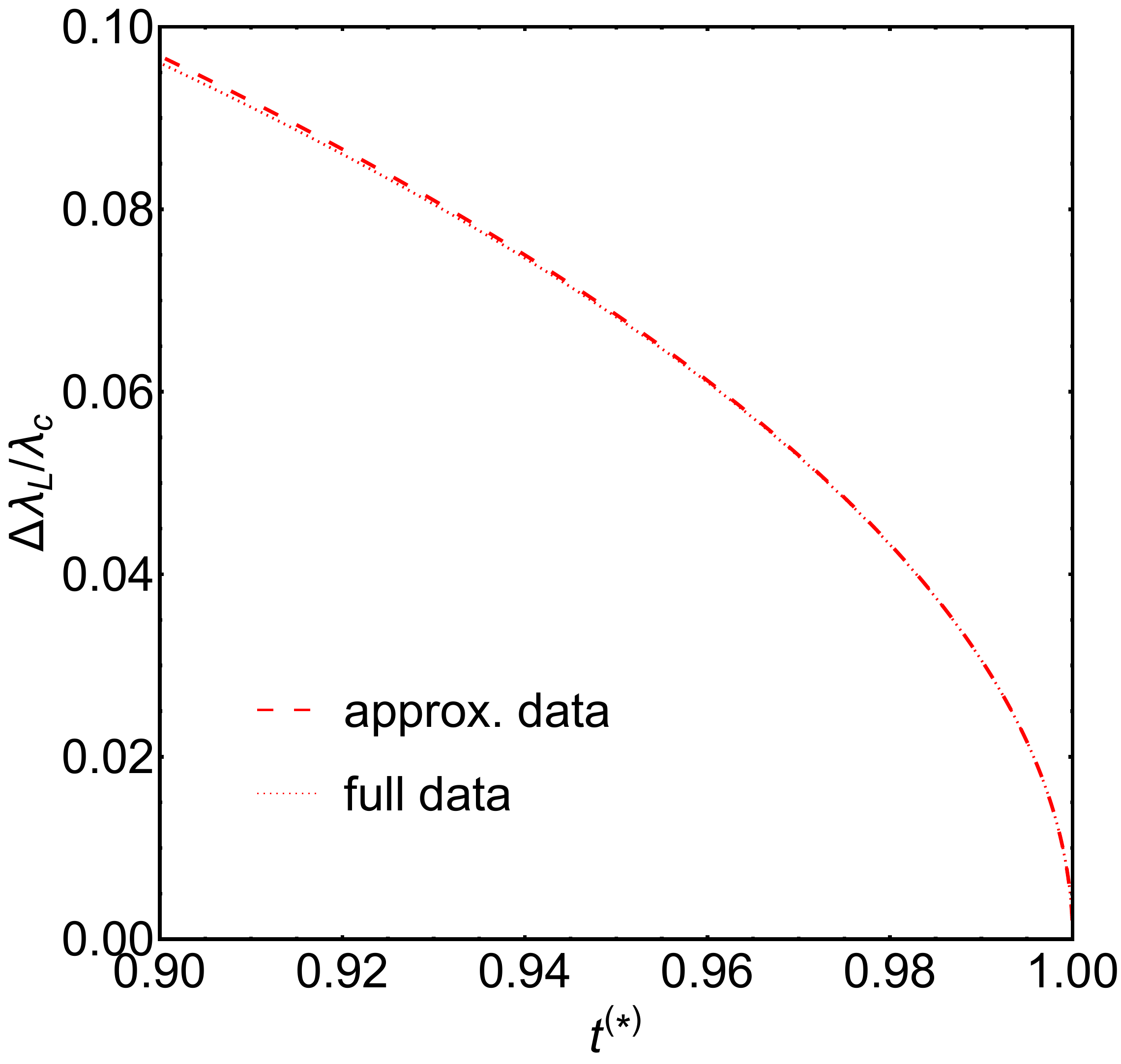}
\quad
\includegraphics[width=6cm]{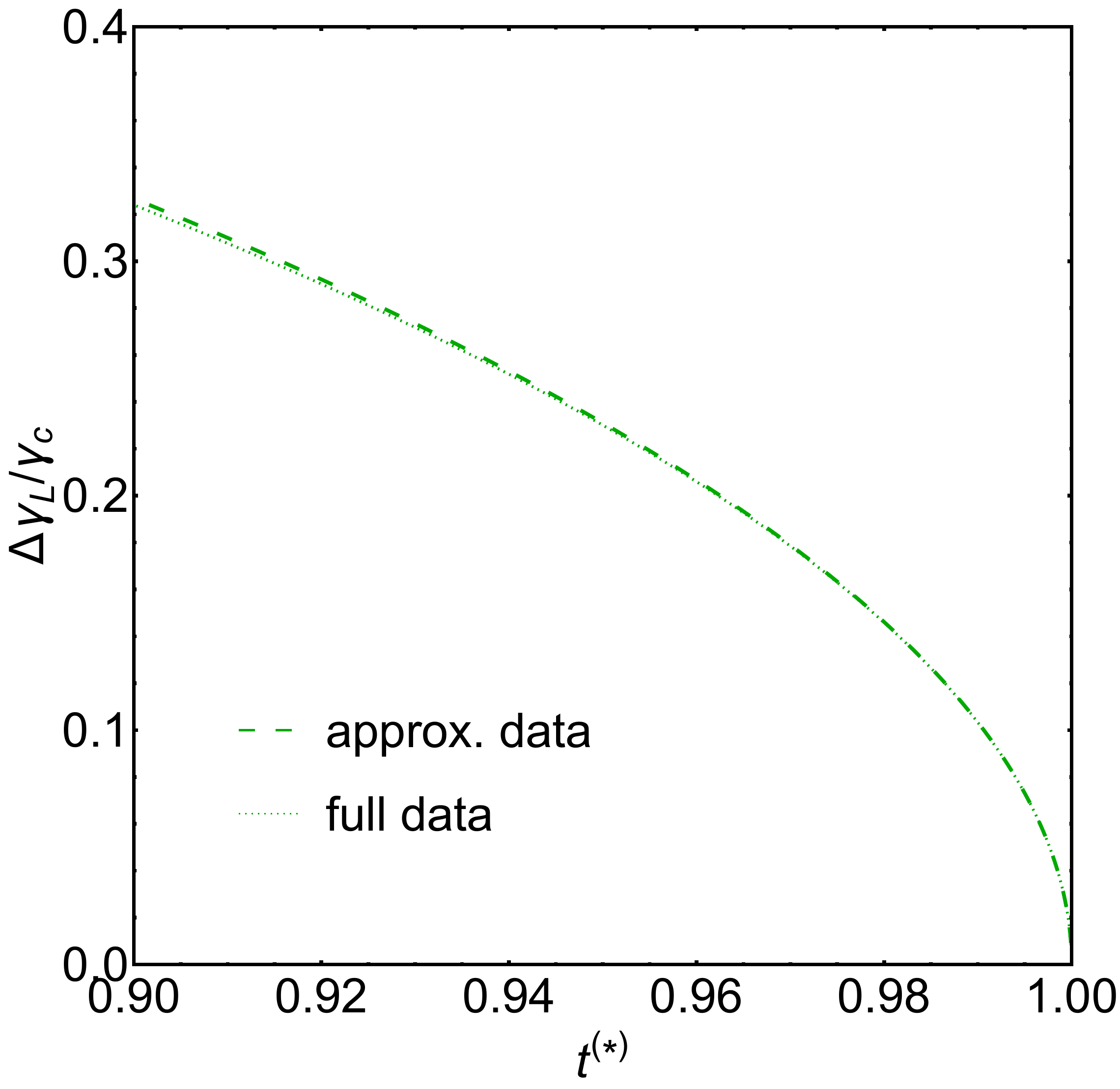}
\caption{Reduced optical order parameters $\Delta \mathcal{V}_{i}/\mathcal{V}_{i(c)}$ as functions of $t^{(*)}$ for the SBH and LBH branches at the first-order phase transition ($\eta=-0.002$). The dashed lines denote the approximate scaling $\sim(1-t^{(*)})^{1/2}$ obtained from the near-critical expansion, while the dots correspond to the full photon-sphere expressions. Their agreement near the critical point confirms the universal square-root scaling behavior.}
\label{order parameter}
\end{figure}
\par
In particular, by evaluating the prefactors $\tilde{\alpha}_i$ at the critical point using Eqs.~\eqref{optical order} and~\eqref{optical order 2}, we obtain
\begin{align}
\displaystyle
\frac{\Delta \tau}{\tau_c} &\simeq 1.33936 (1 - t^{(*)})^{1/2}, \\
\frac{\Delta \lambda_{\rm L}}{\lambda_c} &\simeq 0.30613 (1 - t^{(*)})^{1/2}, \\
\frac{\Delta \gamma_{\rm L}}{\gamma_c} &\simeq 1.03323 (1 - t^{(*)})^{1/2}.
\end{align}
These relations exhibit the universal square-root scaling characterized by the mean-field critical exponent $\beta = 1/2$. Consequently, the optical order parameters vanish as the reduced temperature approaches the critical value,
\begin{equation}
\displaystyle
\lim_{t^{(*)}\to 1} \frac{\Delta \tau}{\tau_c} = 0, \qquad 
\lim_{t^{(*)}\to 1} \frac{\Delta \lambda_{\rm L}}{\lambda_c} = 0, 
\qquad 
\lim_{t^{(*)}\to 1} \frac{\Delta \gamma_{\rm L}}{\gamma_c} = 0.
\end{equation}
\par
To validate the analytical approximation, we compare the optical order parameters obtained from the approximate analysis with those derived from the full photon-sphere expressions in the near-critical regime, as shown in Fig.~\ref{order parameter}. The close agreement confirms that the reduced optical order parameters from the full expressions follow the near-critical scaling behavior given in Eq.~\eqref{optical order}.
\par
These results establish a direct connection between black hole thermodynamic phase transitions and optical characteristics. In this framework, photon-sphere observables provide an effective probe of the underlying phase structure. In particular, the optical signatures associated with different phases encode information about the black hole charge and offer a potential avenue for testing Tsallis-deformed statistics, as well as constraining the non-extensive parameter $\eta$ through observational data.

\section{Conclusion and Discussion}\label{conclusion}
\par
In this work, we have investigated black hole thermodynamics within the framework of Tsallis statistical mechanics, in which the non-extensive formalism is designed to capture the intrinsically non-extensive nature of black hole entropy. By employing a near-horizon photon-gas model, we derive a generalized entropy as a function of the horizon area. Using this entropic expression, we analyze the thermodynamic properties of charged black holes, including stability, phase transitions, and critical behavior. Furthermore, we demonstrate that an emergent phase transition, appearing within an appropriate range of the non-extensive parameter, can be probed through optical characteristics of the spacetime. This establishes a coherent connection between phase transitions, statistical descriptions, and potentially observable photon-sphere dynamics.
\par
In the first part of this paper, we construct black hole entropy from a near-horizon photon-gas model using statistical mechanics based on both the Gibbs-Boltzmann (GB) entropy and the Tsallis entropy. Within the GB framework, we obtain the black hole entropy as a function of the horizon area, $S_{\rm BH} = k_{B} A_{h}/(4l_{P}^{2})$, reproducing the conventional Bekenstein--Hawking expression. We then extend the analysis to Tsallis statistical mechanics, which leads to the generalized Bekenstein--Hawking entropy given in Eq.~\eqref{used Sq}. Interestingly, the entropy expressions derived in both statistical frameworks match those obtained from a classical ideal-gas model. Since a distant external observer cannot distinguish the accessible information between the horizon and the near-horizon gas, the observer has access only to the horizon data, which is entirely determined by the black hole spacetime geometry. This physical insight leads to explanation why the mathematical structures of black hole entropy derived from different near-horizon gas models are equivalent.
\par
It is interesting to emphasize that the entropy functional associated with the Tsallis framework is distinct from the well-known results obtained in quantum gravity. We consider an expansion around $q=1$, corresponding to a small deviation from extensivity. The first term reproduces the conventional Bekenstein--Hawking entropy derived from the geometric approach, while the sub-leading term takes the form of a power-law correction of the horizon area, i.e.,
\(
(1-q) k_{B} \left(A_{h}/l_{P}^{2}\right)^{2}.
\)
By contrast, in quantum gravity, the sub-leading corrections typically contain a logarithmic term of the horizon area, i.e.,
\(
\alpha k_{B} \ln \left|A_{h}/l_{P}^{2}\right|,
\)
where $\alpha$ is a dimensionless constant. We find that the sub-leading term in the Tsallis entropy originates from non-extensive statistics rather than from geometric features as in quantum gravity. Moreover, the black hole entropy becomes smaller than the standard Bekenstein--Hawking entropy due to the sub-extensive parameter $q>1$, which is qualitatively similar to the well-known results obtained in many quantum gravity studies where $\alpha = -3/2$~\cite{Kaul2000, Domagala:2004}. This feature opens the possibility that non-extensive effects could contribute to the understanding of black hole microstates, alongside quantum gravitational approaches.
\par
With the construction of black hole entropy within the Tsallis statistical framework, thermodynamic quantities such as the temperature and response functions can be consistently defined so that they satisfy the first law of thermodynamics together with a generalized Smarr relation. Using these quantities and their properties, we investigate the thermodynamic behavior of a charged black hole. We find that an emergent new phase appears in the non-extensive parameter range $\eta<0$, which is absent in the conventional Bekenstein--Hawking description. In contrast, for $\eta>0$ the system exhibits only two distinct phases, identical to those of the standard charged black hole with Bekenstein--Hawking entropy. For the sub-extensive regime, $\eta<0$, the long-range correlations enhance the number of accessible microstates of the black hole system. This enlargement of microscopic configurations modifies the statistical organization of the system and consequently affects its macroscopic thermodynamic behavior. As a result, the phase structure becomes richer, giving rise to three distinct phases: small, intermediate, and large black-hole branches.
\par
Due to the emergence of the new phase when $\eta<0$, the charged black hole associated with the Tsallis statistical description can be reinterpreted as a Van der Waals (VdW) fluid-like system, and its physical implications can be systematically described using the language of conventional fluid dynamics. We first compute the corresponding fluid quantities, such as the thermal compressibility, which characterize the mechanical stability of black holes. Moreover, the VdW fluid-like behavior allows us to determine the criticality of black holes by identifying the critical point in the $T_{\eta, \rm H}$--$S_{\eta, \rm BH}$ diagram, as illustrated in Fig.~\ref{Maxwell area 1}, from which the critical temperature and associated response functions are obtained.
\par
In analogy with ordinary fluids, the multi-valued branch structure of the temperature as a function of entropy leads to a first-order phase transition between small and large black holes. The coexistence of these phases is established through the Maxwell equal-area construction in the $T_{\eta, \rm H}$--$S_{\eta, \rm BH}$ plane, where the area under the curve has a direct thermodynamic interpretation in terms of exchanged heat. This construction guarantees that the Gibbs free energies of the coexisting phases are equal and that the transition occurs at a fixed Hawking--Page temperature. As a consequence, the Tsallis framework not only predicts the existence of an emergent intermediate phase but also provides a consistent thermodynamic mechanism governing its stability and phase transition toward the small and large black-hole branches.
\par
Furthermore, we analyze the behavior of the critical phenomena of black holes in analogy with ordinary fluids. We discover that the critical exponents of the Tsallis-deformed charged black hole for $\eta<0$ take the same values as those of the conventional VdW fluid system, implying that the charged black hole with Tsallis-modified entropy belongs to the same universality class as the classical VdW fluid. The small black-hole and large black-hole branches closely parallel the liquid--gas phase transition observed in ordinary fluids. Although the microscopic degrees of freedom of black holes remain largely unknown, these universal features---which do not depend on the detailed microscopic structure of the system---govern their macroscopic critical behavior.
\par
In addition to developing the theoretical framework of black hole thermodynamics under Tsallis deformation, we also establish a possible way to connect between these theoretical predictions and observational quantities through optical characteristics, such as the orbital half-period and the Lyapunov exponents. By relating black hole thermodynamic quantities to optical observables, we find that the optical features faithfully mirror the underlying black hole phase transitions, as illustrated in Figs.~\ref{fig:orbital_half_period}, \ref{angular Lyapunov}, and~\ref{temporal_Lyapunov}. Furthermore, we determine the critical behavior of the optical signatures corresponding to the black hole phase transitions. At criticality, the optical order parameters scale with the temperature as a power law of exponent $1/2$, consistent with the mean-field critical exponent. These analyses therefore indicate that the emergence of new phases induced by Tsallis deformation with $\eta<0$ can indeed be captured via optical characteristics. Such investigations open promising avenues for probing theoretical descriptions of black hole thermodynamics through potentially observable photon-sphere dynamics, even though direct measurements remain challenging at present.
\par
Our investigation demonstrates that black hole thermodynamics can be consistently formulated within Tsallis-deformed statistical mechanics, particularly in the description of horizon thermodynamics. The generalized Bekenstein--Hawking entropy provides the foundation for constructing thermodynamic quantities in this non-extensive framework. The stability of charged black holes is reinterpreted through a fluid-like analogy, and we establish an optical--thermodynamic correspondence that connects the theoretical framework to observable quantities. In this context, photon-sphere dynamics act as a proxy for black-hole phase transitions, offering an experimentally accessible probe. 
\par
We further find that the connection between black hole thermodynamics and optical features allows the phase structure of the system to be probed through photon-sphere observables. In this way, the optical signatures associated with the emergent phases can indirectly indicate the presence of charge in black holes and provide observational support for the Tsallis-based model, potentially allowing the non-extensive parameter $\eta$ to be constrained from optical measurements. Finally, this approach can be extended to more general gravity theories with deformed entropy, opening new avenues to explore black hole thermodynamics through photon-sphere measurements, which we leave for future work.

\appendix
\newpage

\section{\texorpdfstring{$q$}{q}-Deformed Integral in Non-Extensive Quantum Statistical Mechanics}\label{Integral}
\par
In this part, we derive the result in Eq.~\eqref{partition 5 q}. We begin by taking the logarithm of the partition function in Eq.~\eqref{Zq final}, originally expressed as a discrete sum over photon modes, and approximate it by an integral representation valid for a large number of microstates:
\begin{equation}
\displaystyle
\ln Z_{q} \simeq - \int_{0}^{\infty} d\mathcal{V} \, \ln \left| 1 - \exp_{q}(-\beta_{q} E) \right|,
\label{partition 2 q}
\end{equation}
At high effective temperatures ($T_q \to \infty$ or $\beta_{q} \to 0$), the argument of the $q$-exponential satisfies $-\beta_{q} E \ll 1$, justifying the factorization approximation over the integration domain. It is important to emphasize that the phase-space volume in Eq.~\eqref{phase3} remains geometrically determined, particularly through the near-horizon behavior of the horizon function $g(r)$. Therefore, the non-extensive deformation does not modify the integration measure, and $\ln Z_{q}$ can be evaluated using the same local phase-space structure as in the extensive case.
\par
From the fact that the phase-space volume in Eq.~\eqref{phase3} remains geometrically determined, particularly through the near-horizon behavior of the horizon function $g(r)$, the non-extensive deformation does not modify the integration measure. Therefore, $\ln Z_{q}$ can be evaluated using the same local phase-space structure as in the extensive case. We then follow the same strategy as in Sec.~\ref{subsec: S GB}. The integration over the phase-space volume near the black hole horizon can be approximated as
\begin{equation}
\displaystyle
\ln Z_{q} \simeq -\frac{2 \pi c^{3} A_{h}}{h^{3} \kappa^{3} l_{\text{loc}}^{2}} \int_{0}^{E_{\text{max}}} dE \, E^{2} \ln \left|1 - \exp_{q}(-\beta_{q} E) \right|,
\label{partition 3 q}
\end{equation}
where the upper limit $E_{\text{max}}$ is determined by the domain of the $q$-exponential: for $q>1$, $E_{\text{max}} \to \infty$, while for $q<1$, $E_{\text{max}} = 1/[(1-q)\beta_q]$. In the high-temperature limit ($\beta_q \to 0$), the cutoff becomes arbitrarily large, allowing $E_{\text{max}} \to \infty$ in both cases.
\par
The integral in Eq.~\eqref{partition 3 q} can be written as
\begin{equation}
\label{integral q}
\displaystyle
\mathcal{I}(q) = \int_{0}^{\infty} dx_{q}\, x_{q}^{2} \ln \left|1 - \exp_{q}(-x_{q})\right|,
\end{equation}
where $x_{q} = \beta_{q} E$. To evaluate Eq.~\eqref{integral q}, we expand the logarithm using the Mercator series,
\begin{equation}
\label{series log}
\displaystyle
\ln |1 - u| = -\sum_{k=1}^{\infty} \frac{u^{k}}{k}, \qquad |u|<1,
\end{equation}
with the identification $u = \exp_{q}(-x_{q})$. The convergence condition $|\exp_{q}(-x_q)| < 1$ depends on the deformation regime, which we now analyze.
\par
\begin{itemize}
\item \textbf{Case $q < 1$} ($1-q>0$): 
\[
\exp_{q}(-x_{q}) = [1 - (1-q)x_{q}]^{1/(1-q)}, \quad 0 < x_{q} < \frac{1}{1-q}.
\]  
For $x_{q} \ge \frac{1}{1-q}$, $\exp_{q}(-x_{q}) = 0$, producing a UV cutoff in the integral.
\item \textbf{Case $q > 1$} ($1-q<0$): 
\[
\exp_{q}(-x_{q}) = [1 + (q-1)x_{q}]^{-1/(q-1)}, \quad x_{q} \in [0, \infty),
\]  
so that the series converges over the entire positive domain, leading to power-law tails characteristic of sub-extensive statistics.
\end{itemize}
Substituting $u = \exp_{q}(-x_{q})$ into Eq.~\eqref{series log}, we obtain
\begin{equation}
\label{series expansion step}
\displaystyle
\mathcal{I}(q) = - \int_{0}^{\infty} dx_{q}\, x_{q}^{2} \sum_{k=1}^{\infty} \frac{[\exp_{q}(-x_{q})]^{k}}{k}.
\end{equation}
Assuming uniform convergence, we interchange the order of summation and integration:
\begin{equation}
\label{integral q 2}
\displaystyle
\mathcal{I}(q) = - \sum_{k=1}^{\infty} \frac{1}{k} \int_{0}^{\infty} dx_{q}\, x_{q}^{2} [\exp_{q}(-x_{q})]^{k}.
\end{equation}
\par
To proceed further, we consider the weakly deformed regime ($q \to 1$), or equivalently the high-temperature limit ($x_{q} \ll 1$), in which the approximation
\begin{equation}
\label{exp approx}
\displaystyle
[\exp_{q}(-x_{q})]^{k} \simeq \exp_{q}(-k x_{q})
\end{equation}
becomes valid for both $q<1$ and $q>1$. As a result, the integral in Eq.~\eqref{integral q 2} becomes
\begin{equation}
\label{integral q 3}
\displaystyle
\mathcal{I}(q) \simeq - \sum_{k=1}^{\infty} \frac{1}{k^{4}} \int_{0}^{\infty} d(kx_{q})\, (kx_{q})^{2} \exp_{q}(-kx_{q}).
\end{equation}
The remaining integral is identified as the $q$-gamma function (see Eq.~\eqref{def q Gamma} with $s=3$), yielding a $k$-independent quantity. The remaining summation is recognized as the Riemann zeta function,
\begin{equation}
\label{zeta function}
\displaystyle
\zeta(4) = \sum_{k=1}^{\infty} \frac{1}{k^{4}} = \frac{\pi^{4}}{90}.
\end{equation}
Combining these results, we obtain a generalization of the standard BE integral in the non-extensive setting,
\begin{equation}
\label{final q integral}
\displaystyle
\mathcal{I}(q) \simeq - \Gamma_{q}(3) \zeta(4) = - \frac{\pi^{4}}{90} \Gamma_{q}(3).
\end{equation}
Applying this result, together with the definition of volume~\eqref{V loc}, to Eq.~\eqref{partition 3 q}, we obtain the $q$-partition function in Eq.~\eqref{partition 5 q}.

\section{Verification of the First Law of Thermodynamics}\label{check 1st law}
\par
We now verify that the $q$-generalized thermodynamic quantities derived in non-extensive statistical mechanics satisfy the $q$-generalized first law of thermodynamics. This verification ensures that the modified framework is not only formally consistent but also physically meaningful. Specifically, we demonstrate the differential identity found in Eq.~\eqref{1st law of q}
\begin{equation}
\label{eq:first_law_q}
\displaystyle
dU_{q, \mathrm{loc}} = \frac{1}{k_{B} \beta_{q, \mathrm{loc}}} \, dS_{q, \mathrm{loc}} - P_{q, \mathrm{loc}} \, dV_{\mathrm{loc}},
\end{equation}
where we have assumed all thermodynamic variables are smooth and continuously differentiable. This establishes the connection between the generalized entropy and the dynamical behavior of energy and pressure.
\par
The $q$-generalized local internal energy of a photon gas is given by
\begin{equation}
\label{eq:Uq_def}
\displaystyle
U_{q, \mathrm{loc}} = C V_{\mathrm{loc}} \beta_{q, \mathrm{loc}}^{-4} \exp\!\left((1-q)A\right),
\end{equation}
where we have introduced the auxiliary quantities
\begin{equation}
\label{eq:A_C_def}
\displaystyle
A := \frac{4\pi^{5}}{45 h^{3} c^{3}} \frac{V_{\mathrm{loc}}}{\beta_{q, \mathrm{loc}}^{3}} \Gamma_{q}(3), 
\qquad
C := \frac{4\pi^{5}}{15 h^{3} c^{3}} \Gamma_{q}(3),
\end{equation}
to simplify subsequent calculations. Differentiating Eq.~\eqref{eq:Uq_def} with respect to $\beta_{q,\mathrm{loc}}$ and $V_{\mathrm{loc}}$ gives the total differential
\begin{equation}
\label{eq:dU_total}
\displaystyle
dU_{q, \mathrm{loc}} = \left( \frac{\partial U_{q, \mathrm{loc}}}{\partial \beta_{q, \mathrm{loc}}} \right)_{V_{\mathrm{loc}}} d\beta_{q, \mathrm{loc}} 
+ \left( \frac{\partial U_{q, \mathrm{loc}}}{\partial V_{\mathrm{loc}}} \right)_{\beta_{q, \mathrm{loc}}} dV_{\mathrm{loc}}.
\end{equation}
Explicit differentiation yields
\begin{align}
\displaystyle
\left( \frac{\partial U_{q, \mathrm{loc}}}{\partial \beta_{q, \mathrm{loc}}} \right)_{V_{\mathrm{loc}}}
&= C V_{\mathrm{loc}} \left[ -4 \beta_{q, \mathrm{loc}}^{-5} + (1-q)\beta_{q, \mathrm{loc}}^{-4} \frac{\partial A}{\partial \beta_{q, \mathrm{loc}}} \right] e^{(1-q)A}, \label{dUbeta} \\
\left( \frac{\partial U_{q, \mathrm{loc}}}{\partial V_{\mathrm{loc}}} \right)_{\beta_{q, \mathrm{loc}}}
&= C \beta_{q, \mathrm{loc}}^{-4} \left[ 1 + (1-q) V_{\mathrm{loc}} \frac{\partial A}{\partial V_{\mathrm{loc}}} \right] e^{(1-q)A}, \label{dUV}
\end{align}
with
\begin{equation}
\displaystyle
\frac{\partial A}{\partial \beta_{q, \mathrm{loc}}} = -3 D V_{\mathrm{loc}} \beta_{q, \mathrm{loc}}^{-4}, 
\qquad
\frac{\partial A}{\partial V_{\mathrm{loc}}} = D \beta_{q, \mathrm{loc}}^{-3},
\label{AD}
\end{equation}
where
\begin{equation}
\displaystyle
D := \frac{C}{3} = \frac{4\pi^{5}}{45 h^{3} c^{3}} \Gamma_q(3),
\end{equation}
so that $A = D V_{\mathrm{loc}} \beta_{q, \mathrm{loc}}^{-3}$, simplifying differentiation. Substituting these results into Eqs.~\eqref{dUbeta}, \eqref{dUV}, and~\eqref{AD} in Eq.~\eqref{eq:dU_total}, we obtain 
\begin{align}
\label{eq:dUq_final}
\displaystyle
dU_{q, \mathrm{loc}} &= -3 D V_{\mathrm{loc}} \beta_{q, \mathrm{loc}}^{-5} (4 + 3(1-q)A) e^{(1-q)A} \, d\beta_{q, \mathrm{loc}} \nonumber \\
&\quad + 3 D \beta_{q, \mathrm{loc}}^{-4} (1 + (1-q)A) e^{(1-q)A} \, dV_{\mathrm{loc}}.
\end{align}
Next, we consider the $q$-generalized local entropy, which can be expressed as
\begin{equation}
\displaystyle
S_{q, \mathrm{loc}} = \frac{k_{B}}{1-q} f(A), 
\qquad 
f(A) = \bigl[ 1 + 3(1-q) A \bigr] e^{(1-q)A} - 1.
\end{equation}
Its differential form is
\begin{equation}
\displaystyle
dS_{q, \mathrm{loc}} = \frac{k_{B}}{1-q} \frac{df}{dA} \, dA,
\label{dS}
\end{equation}
with
\begin{equation}
\displaystyle
\frac{df}{dA} = (1-q)(4 + 3(1-q)A) e^{(1-q)A},
\qquad 
dA = -3 D V_{\mathrm{loc}} \beta_{q, \mathrm{loc}}^{-4} d\beta_{q, \mathrm{loc}} + D \beta_{q, \mathrm{loc}}^{-3} dV_{\mathrm{loc}}.
\label{dA}
\end{equation}
Substituting Eq.~\eqref{dA} into Eq.~\eqref{dS}, we obtain
\begin{align}
\displaystyle
dS_{q, \mathrm{loc}} &= k_{B} D (4 + 3(1-q)A) \left[ -3 V_{\mathrm{loc}} \beta_{q, \mathrm{loc}}^{-4} d\beta_{q, \mathrm{loc}} + \beta_{q, \mathrm{loc}}^{-3} dV_{\mathrm{loc}} \right] e^{(1-q)A}, \\
\displaystyle
\frac{1}{k_{B} \beta_{q, \mathrm{loc}}} dS_{q, \mathrm{loc}} &= D (4 + 3(1-q)A) \left[ -3 V_{\mathrm{loc}} \beta_{q, \mathrm{loc}}^{-5} d\beta_{q, \mathrm{loc}} + \beta_{q, \mathrm{loc}}^{-4} dV_{\mathrm{loc}} \right] e^{(1-q)A}.
\label{eq:dSoverT}
\end{align}
Similarly, the $q$-generalized local pressure reads
\begin{equation}
\displaystyle
P_{q, \mathrm{loc}} = D \beta_{q, \mathrm{loc}}^{-4} e^{(1-q)A}, 
\qquad
-P_{q, \mathrm{loc}} dV_{\mathrm{loc}} = - D \beta_{q, \mathrm{loc}}^{-4} e^{(1-q)A} dV_{\mathrm{loc}}.
\label{eq:work_term}
\end{equation}
\par
By combining Eqs.~\eqref{eq:dSoverT} and~\eqref{eq:work_term}, and then comparing with Eq.~\eqref{eq:dUq_final}, the $q$-generalized first law of thermodynamics is satisfied:
\begin{equation}
\displaystyle
dU_{q, \mathrm{loc}} = \frac{1}{k_{B} \beta_{q, \mathrm{loc}}} dS_{q, \mathrm{loc}} - P_{q, \mathrm{loc}} dV_{\mathrm{loc}},
\end{equation}
confirming the internal consistency of the non-extensive thermodynamic framework, which supports the broader applicability of $q$-generalized entropy, including in black hole thermodynamics.

\section*{Acknowledgement}
\par
This research has received funding support from the NSRF via the Program Management Unit for Human Resources \& Institutional Development, Research and Innovation [grant number B39G690007]. The authors acknowledge the use of AI-assisted language tools to improve the readability and clarity of the manuscript.

\nocite{*}
\bibliography{ref.bib}

\end{document}